\documentclass[journal]{IEEEtran}
\usepackage[edges]{forest}
\usetikzlibrary{shadows}
\usepackage{amsmath,amssymb,amsmath,amsfonts}
\usepackage{bm}
\usepackage{bbm}
\usepackage{graphicx}
\usepackage{cite}
\usepackage{epstopdf}
\usepackage{balance}
\usepackage{caption}
\usepackage{gensymb}
\usepackage{color}
\usepackage{lipsum}
\usepackage{float}
\usepackage{epstopdf}
\usepackage[mathcal]{eucal}
\usepackage{subfiles}
\usepackage[T1]{fontenc}
\usepackage{siunitx}
\usepackage{tabulary}
\usepackage{array}
\usepackage{adjustbox}
\usepackage{booktabs}
\usepackage{cite}
\usepackage{times}
\usepackage{placeins}
\usepackage{textcomp}
\usepackage[breaklinks=true]{hyperref}
\usepackage{tabularx}
\usepackage{bm}
\usepackage{enumerate}
\usepackage{tikz}
\usepackage{pgfplots}
\usepackage{enumerate}
\usepackage{epstopdf}
\usetikzlibrary{shapes,arrows}
\usepackage[utf8]{inputenc}
\usepackage{mathtools}
\usepackage[utf8]{inputenc}
\usepackage{tikz}
\usepackage{multicol}
\usepackage{multirow}
\usepackage{caption}
\usepackage{subcaption}

\usepackage{pifont}
\newcommand{\cmark}{\ding{51}}%
\usepackage{tikz}
\usetikzlibrary{trees,positioning,shapes,shadows,arrows}

\usepackage{enumitem}
\usepackage{tabulary}

\newcommand{\black}[1]{\textcolor{black}{#1}}

\usepackage[normalem]{ulem}

\makeatletter

 \let\oldforeign@language\foreign@language
 \DeclareRobustCommand{\foreign@language}[1]{%
   \lowercase{\oldforeign@language{#1}}}

\usepackage{dblfloatfix}

\makeatother

\IEEEoverridecommandlockouts

\begin{document}
%
\title{Evolution of Non-Terrestrial Networks \\ from 5G to 6G: A Survey}


\author{M.~Mahdi~Azari, Sourabh Solanki, Symeon~Chatzinotas,  \\  Oltjon Kodheli, Hazem Sallouha, Achiel Colpaert, Jesus Fabian Mendoza Montoya, \\ Sofie Pollin, Alireza Haqiqatnejad,  Arsham Mostaani, Eva Lagunas, Bjorn Ottersten  
 \thanks{M.~M.~Azari, S.~Solanki, S.~Chatzinotas, J.~F.~Mendoza~Montoya, A.~Mostaani,  E.~Lagunas, and B.~Ottersten are with SnT, University of Luxembourg. O.~Kodheli is with SES, Luxembourg. A.~Haqiqatnejad is with OQ Technology, Luxembourg. H.~Sallouha, A.~Colpaert, and S.~Pollin are with KU Leuven, Belgium.}
}

\maketitle

\begin{abstract}

 Non-terrestrial networks (NTNs) traditionally have certain limited applications. However, the recent technological advancements and manufacturing cost reduction opened up myriad applications of NTNs for 5G and beyond networks, especially when integrated into terrestrial networks (TNs). \black{ This article comprehensively surveys the evolution of NTNs highlighting their relevance to 5G networks and essentially, how it will play a pivotal role in the development of 6G ecosystem. We discuss important features of NTNs integration into TNs and the synergies by delving into the new range of services and use cases, various architectures, technological enablers, and higher layer aspects pertinent to NTNs integration. Moreover, we review the corresponding challenges arising  from  the  technical  peculiarities and the new approaches being adopted to develop efficient integrated ground-air-space (GAS) networks. Our survey further includes the major progress and outcomes from academic research as well as industrial efforts representing the main industrial trends, field trials, and prototyping towards the 6G networks.}
 
\end{abstract}
\begin{IEEEkeywords}
Non-terrestrial network (NTN), satellite, unmanned aerial vehicle (UAV), drone, high altitude platform (HAP), terrestrial network (TN), 5G, mmWave, Internet of Things (IoT), localization, mobile edge computing (MEC), machine learning (ML), artificial intelligence (AI), reinforcement learning (RL), 6G, cell free, mega constellation, intelligent Reconfigurable Surfaces (IRS), terahertz communications, 3GPP, MAC, NET, network slicing, virtualization, open-RAN, QUIC.
\end{IEEEkeywords}

\IEEEpeerreviewmaketitle


\section{Introduction
}\label{sec:intro}

\subsection{Motivation
}

Non-terrestrial networks (NTNs), which include unmanned aerial vehicles (UAVs), high altitude platforms (HAPs), and satellite networks, are traditionally used for certain applications such as disaster management, navigation, television broadcasting, and remote sensing. However, recent tremendous developments of aerial/space technologies coupled with reduced cost of their manufacturing and launching have enabled more advanced applications of NTNs, when integrated with terrestrial communication networks. In this context, various new use cases and applications have been envisioned mostly focusing on providing continuous, ubiquitous, and high-capacity connectivity across the globe \cite{3GPP22822}. 
Inherent limitations of ground infrastructure as well as economic rationales may prohibit terrestrial networks (TNs) deployment in remote or unreachable regions, such as rural areas, deserts, and oceans. As a result, the user equipments (UEs) cannot access the terrestrial services within these un-served areas. \black{The use of NTNs in conjunction with the existing terrestrial infrastructure could provide a feasible and cost-effective solution for continuous and ubiquitous wireless coverage and thereby enable network scalability \cite{3GPP22822}}. In such cases, NTNs act as access nodes to augment the performance of existing terrestrial networks in terms of capacity, coverage, and delay. It can also address the shortfalls of terrestrial infrastructure in satisfying the required levels of reliability and widespread presence for future wireless applications. 

Following the benefits offered by NTNs and also to capitalize on the economies of scale, the 3rd generation partnership project (3GPP) study items include the satellite access integration into 5G ecosystem \cite{3GPP38811,3GPP22822}. In addition, the development of the 5G ecosystem and its specifications, particularly, in providing enhanced mobile broadband (eMBB) and ultra-reliable low-latency communication (uRLLC), enables a seamless integration of UAVs into the terrestrial networks. 5G technology is capable of satisfying critical features of UAVs for control and communication and therefore, can facilitate reliable beyond visual line-of-sight (BVLoS) drones connectivity \cite{3GPP22829,AzaRosPol2019}, which is essential for several applications \cite{commag22}.

The emerging applications such as autonomous vehicles, precision agriculture, and many more yet-to-discover use cases drive extensive research towards the 6G and beyond to provide global connectivity. The need for global coverage and the imminent plans of large low earth orbit (LEO) satellite constellations would lead to a more important roles of NTNs for future networks. Unarguably, fully integrated NTNs can provide anywhere, anything, at anytime connectivity with remarkable socio-economic impact \black{\cite{3GPP38811,3GPP22822}}. 
Therefore, it becomes crucial to provide an in-depth overview of the NTNs integration into the current and future Gs of wireless networks.
\begin{table*}[h!] 
\centering
\caption{Survey structure and content comparison with existing surveys and tutorials (partially covered: $\partial$, covered: \cmark).} 
\begin{tabular}{ |l||c|c|c|c|c|c|c|c|} 
 \hline\hline
 \hspace{15mm} Reference $\rightarrow$    & \cite{geraci2021will} & \cite{dao2021survey}  & \cite{rinaldi2020non} &  \cite{kodheli2020satellite} &  \cite{zhang2020survey} & \cite{wu20205g}  & \cite{vinogradov2019tutorial} & \cite{liu2018space}  \\
 Topics  $\downarrow$     &   &   &   &   &  & & &   \\
 \hline\hline
 Sec.\ \ref{sec:NTN_5G}: NTNs Integration In 5G Ecosystem    &  &   &   &   &   & & &  \\
 $-$ Satellite  &  & $\partial$  & \cmark  & \cmark  &  \cmark & & & \\
 $-$ UAV     & \cmark &  $\partial$ &   &   &  $\partial$  & \cmark & $\partial$ & \\
 \hline
Sec.\ \ref{sec:NTN_mmWave}: NTNs Operation in mmWave    &  &   &   &   &   & & &  \\
 $-$ Satellite  &  &   &   &   &   & & &  \\
 $-$ UAV   & $\partial$  &   &   &   &   & &  $\partial$ &  \\
 $-$ Multi-Segment    &  &   &   &   &   & & &  \\
 \hline
Sec.\ \ref{sec:NTN_IoT}: NTNs Integration In IoT   &  &   &   &   &   & & &  \\
$-$ Satellite  &  &   &   &   &   & & &  \\
 $-$ UAV   &  &  $\partial$ &   &   &   & & $\partial$ &  \\
 $-$ Multi-Segment    &  &   &   &   &   & & & $\partial$ \\
 \hline
 Sec.\ \ref{sec:NTN_MEC}: NTNs Integration In MEC Networks   &  &   &   &   &   & & &  \\
$-$ Satellite  &  &   &   &   &   & & &  \\
 $-$ UAV   &  & $\partial$  &   &   &   & $\partial$ & &  \\
 $-$ Multi-Segment    &  &   &   &   &   & & &  \\
 \hline
 Sec.\ \ref{sec:NTN_ML}: ML-Empowered NTNs   &  &   &   &   &   & & &  \\
$-$ Satellite  &  &   &   & $\partial$  &   & & &  \\
 $-$ UAV   & $\partial$ &   &   &   &   & $\partial$ & &  \\
 $-$ Multi-Segment    &  &   &   &   &   & & &  \\
 \hline
 Sec.\ \ref{sec:NTN_higherlayer}: Higher Layer Advancements   &  &   &   &   &   & & &  \\
 $-$ Network Virtualization  &  &   &   & $\partial$  &   & & & $\partial$ \\
 $-$ C-RAN Architecture and Cloud/Edge Computing   &  &   & $\partial$  &   &   & & &  \\
 $-$ Transmission Control Protocol    &  &   &   &   &   & & & \cmark \\
 $-$ Smart Gateway Diversity and Aerial Links   &  &   &   &   &   & & &  \\
 \hline
 Sec.\ \ref{sec:NTN_FieldTrials}: NTNs Field Trials \& Industrial Efforts   &  &   &   &   &   & & &  \\
$-$ Satellite  &  &   &   & $\partial$  & $\partial$   & & &  \\
 $-$ UAV   &  &   &   &   &   & & &  \\
 \hline
Sec.\ \ref{sec:NTN_6G}: NTNs Integration In  6G and Beyond    &  &   &   &   &   & & &  \\
 $-$ Use Cases  &  &   &   &   &   & & & \\
 ~~ * Satellite  &  &   &  $\partial$ &   &   & & & \\
 ~~ * UAV  & $\partial$ &   &   &   &   & & & \\
 $-$ Architectures  &  &   &   &   &   & & &  \\
 ~~ * Open-RAN  &  &   &   &   &   & & &  \\
 ~~ * Multi-Segments  &  &   &   &   &   & & & $\partial$ \\
 ~~ * 3D Cell-free   & $\partial$ &   &   &   &   & & &  \\
  ~~ * Mega LEO Constellation  &  &   &   &   &   & & &  \\
 $-$ Technological Enablers    &  &   &   &   &   & & &\\
  ~~ * X-Communication Co-Design  &  &   &   &   &   & & &\\
 ~~ * Intelligent Reconfigurable Surfaces  & $\partial$ & \cmark  &   &   &   & $\partial$ & & \\
 ~~ * Multi-Mode Communication    &  &   &   &   &   & & &\\
  ~~ * Dynamic Spectrum Access  &  &   &   &   &   & & & \\
 ~~ * THz Communication & $\partial$ & $\partial$  &   &   &   & & &\\
 ~~ * AI-Empowered Networks   & $\partial$ &   &   &   &   & & & \\
 ~~ * Task-Oriented Communications   &  &   &   &   &   &  & & \\
 ~~ * Quantum Satellite Networks &  &    &   &   &   & & &\\
 $-$ Higher Layer Aspects  &  &   &   &   &   & & & \\
  ~~ * Software-Defined Satellite  &  &   &   &   &   & & & \\
 ~~ * Quick UDP Internet Connections  &  &   &   &   &   & & & \\
  ~~ * Network Slicing and Virtualization  &  &   &  $\partial$ & $\partial$  &   & & & \\
 ~~ * Highly distributed RAN &  &   &   &   &   & & & \\
 \hline
 \hline
\end{tabular}
\label{tab:ContentComparison}
\end{table*}

\begin{table*}[]
	\caption{List of Abbreviations.}
	\aboverulesep = .25ex
	\belowrulesep = .25ex
	\centering
	\begin{tabular}{lc|lc}
		\toprule
		\bf Abbreviation & \bf Definition                                 & \bf Abbreviation & \bf Definition                                           \\
		\midrule\midrule
		3GPP    & 3rd generation partnership   project           & ML  & Machine learning                        \\
		4G      & 4th generation mobile communications   technology  & mMTC    & Massive machine type communication\\
		5G      & 5th generation mobile communications   technology  & mmWave  & Millimeter wave                   \\
		5G NR   & 5G new radio& MPLS    & Multi-protocol label switching             \\
		A2A    & Air-to-air  & NB-IoT  & Narrowband Internet of things\\
		A2G    & Air-to-ground   & NFV     & Network function virtualization\\
		AI      & Artificial intelligence                            & NGSO    & Non-geostationary orbit                                  \\
		AoI     & Age of information                                 & NLoS    & Non-line-of-sight                                    \\
		API     & Application programming interface                  & NOMA    & Non-orthogonal multiple access                       \\
		BH      & Beam hopping      & nSAT & Nano satellite\\
		BS      & Base station                                       & NT, NTN     & Non-terrestrial, non-terrestrial network                              \\
		BVLoS   & Beyond visual line-of-sight                                       & OFDM    & Orthogonal frequency division multiplexing           \\
		CNN     & Convolutional neural network                       & OVSDB   & Open vSwitch database management protocol            \\
		CQI     & Channel quality indicator                          & PAPR    & Peak-to-average power ratio                          \\
		C-RAN   & Cloud radio access network                         & PEP     & Performance enhancing proxies                        \\
		CWN     & Crowdsourced wireless network                      & PHY     & Physical layer                                       \\
		D2D     & Device-to-device       & POMDP  & Partially observable Markov decision process\\
		DDQN  & Double deep Q-network                                   & QoE     & Quality-of-experience                                \\
		DNN   & Deep neural network & QoS     & Quality-of-service \\
				DRL   & Deep reinforcement learning & RAN  & Radio access network  \\ E2E     & End-to-end   & RF      & Radio frequency            \\ 
				eMBB    & Enhanced mobile broadband & RL      & Reinforcement learning                                                        \\
		eMBMS   & Enhanced multimedia broadcast and multicast system & RSS     & Received signal strength \\
		GAS     & Ground-air-space  & RTT     & Round-trip time   \\
		GEO     & Geostationary earth orbit & SAGIN   & Space-air-ground integrated network      \\
		gNB     & Next generation NodeB & SDN     & Software-defined networking    \\
		GPS     & Global positioning system & SDR & Software-defined radio \\     
		GW      & Gateway   & SLA     & Service level agreement  \\                  
		HAP     & High altitude platform & SLAM &   Simultaneous  localization  and  mapping                            \\
		HAPS    & High altitude platform station  & SMBS    & Super macro base station                                                 \\
		HARQ    & Hybrid automatic repeat request   & SVM     & Support vector machine                                               \\
		HIBS    & High altitude platform station as IMT base station & SWIPT   & Simultaneous wireless information and power transfer  \\
		HO      & Handover & TCP     & Transmission control protocol \\              
		IAB     & Integrated access and backhaul& TD      & Temporal difference   \\
		IMT     & International mobile telecommunication & TDMA    & Time division multiple access                                     \\
		IoT     & Internet of things     & TDoA    & Time difference of arrival                                                        \\
		IP      & Internet protocol & TN & Terrestrial network  \\        
		IRS   & Intelligent  Reconfigurable  Surface & TTI     & Transmission time interval                              \\
		ISL     & Inter-satellite link      & UAS     & Unmanned aircraft system                                                       \\
		KPI    & Key performance indicator   & UAV     & Unmanned aerial vehicle                            \\
		LEO     & Low earth orbit   & UDP     & User datagram protocol                                                                       \\
		LoS     & Line-of-sight    & UE      & User equipment                                 \\
		LPWAN   & Low-power wide area network  & UL & Unsupervised learning         \\
		LTE     & Long-term evolution       & uRLLC   & Ultra-reliable low-latency communication                                                      \\
		M2M     & Machine-to-machine        & UTM & UAV traffic management                                      \\
		MAB    & Multi-armed bandit  & V2V     & Vehicle-to-vehicle                          \\
		 MAC     & Media access control      & VLAN    & Virtual local area network                                                            \\
		MDP     & Markov decision process      & VN      & Virtual network                                                     \\
		MEC     & Multi-access edge computing   & VPN     & Virtual private network                                               \\
		MEO     & Medium earth orbit     & VSAT    & Very small aperture terminal                                                         \\
		MIMO    & Multiple-input multiple-output        & WPT     & Wireless power transfer             \\
		\bottomrule                                                 
	\end{tabular}
	\label{abbrev}
\end{table*}

\subsection{\black{Related Work}} 
A detailed content comparison between this survey and relevant tutorial/survey articles is provided in Table \ref{tab:ContentComparison}. For instance, the tutorial paper \cite{geraci2021will} focused on UAV cellular communication, mostly where UAVs act as end-users, and provides important takeaways for the corresponding networks. The authors review the use cases, communication requirements, challenges, and potential solutions. In \cite{dao2021survey}, the authors focus on the design of aerial access nodes in 6G from the analytical perspective. The 3GPP standardization for satellite integration into 5G is reviewed in \cite{rinaldi2020non}. The survey articles \cite{kodheli2020satellite,zhang2020survey} considered satellite networks and overview the important 5G applications along with the system designs and prevalent challenges. The authors in \cite{wu20205g,vinogradov2019tutorial} reviewed UAVs related communication and its networking aspects. Finally, the integrated ground-air-space (GAS) networks were the focus of \cite{liu2018space} where specific challenges in network design and resource allocation are discussed. Other surveys, tutorials, and short review articles with little relevance to this survey can be found in \cite{commag22,mozaffari2019tutorial,saeed2020point,yao2018space,yaacoub2020key,shakeri2019design,frew2008airborne,gupta2015survey,giordani2020non,AbdMar2020,shakhatreh2019unmanned,liu2020cell,zhang2019research,naqvi2018drone,zhang2017software,xie2020satellite,hong2020space,hosseini2019uav,motlagh2016low,zeng2016wireless,shi2018drone,bor20195g,zhang2019cooperation,mishra2020survey,sharma2020communication,zeng2019accessing,lin2020integrated,boero2018satellite}. 
\subsection{Contributions}
In this survey article, we provide an in-depth review of the major trends in both partially and fully integrated GAS networks in the context of 5G/6G. Unlike related studies \cite{geraci2021will,dao2021survey,rinaldi2020non,kodheli2020satellite,zhang2020survey,wu20205g,vinogradov2019tutorial,liu2018space}, we provide a comprehensive survey addressing satellite, UAV, and multi-segment networks, industrial efforts, higher layer aspects, and 6G vision.  This survey includes the following axes:
\begin{itemize}
    \item \emph{Time Evolution:} We first review the fundamental features of NTNs independent of Gs development; then we review the integration of NTNs into the 5G ecosystem by pointing out important use cases, architectures, enablers, challenges, and proposed solutions as well as existing studies; finally we review the integration of NTNs into 6G and beyond, how NTNs assist with developing 6G networks, and how NTNs benefit from 6G TNs. 
    \item \emph{Layers:} We review general architecture options of an integrated ground-air-space (GAS) network. We further discuss various use cases of UAVs and satellites corresponding to 5G/6G such as in IoT and MEC networks where synergies between NTNs and 5G/6G ecosystems are specified. The use of mmWave and THz frequencies as well as ML approach in NTNs integration into 5G/6G are addressed. We further elaborate on candidate 6G technologies and advancements on media access control (MAC) and network (NET) layers through the discussion on network programmability and virtualization, to name a few.
    \item \emph{Approach:} We review the academic research results as well as the prevailing efforts across the industries and academia for standardization, prototyping, and field trials.
    \item \emph{Segments Contribution:} We review the following inter- and intra-connected architectures in each section: ground-air, air-air, ground-space, space-space, air-space, and multi-segment connectivity.
\end{itemize}
\subsection{Organization}
Section \ref{sec:NTN_5G} presents the preliminaries of NTNs, their certain features, and the integration of NTNs into the 5G ecosystem. The NTNs operation in mmWave, integration into the terrestrial networks, and the co-existence networks at mmWave are reviewed in Section \ref{sec:NTN_mmWave}. In Section \ref{sec:NTN_IoT}, the NTNs integration into Internet of Things (IoT) networks are discussed. Mobile edge computing (MEC) with NTNs is the topic of Section \ref{sec:NTN_MEC} followed by Section \ref{sec:NTN_ML} where machine learning (ML) powered non-terrestrial communications and networking integration into TNs is presented. Within all these sections, we classify the references based on the background studies, use cases, architectures, design opportunities and challenges. Section \ref{sec:NTN_higherlayer} provides 5G related advances in higher layers. The field trials and major industrial and academic players on aerial and space communications are highlighted in Section \ref{sec:NTN_FieldTrials}. Next, in Section \ref{sec:NTN_6G}, the vision of NTNs integration into 6G and beyond is presented wherein several selected applications, new architectures, technological enablers, and higher layer aspects are discussed. Finally, the conclusion is drawn in Section \ref{sec:conclusion}. \black{ A more detailed organization overview can be found in Table \ref{tab:ContentComparison} and the acronyms used are given in Table~\ref{abbrev}.} 

\begin{figure*}
\centering
\begin{subfigure}{\columnwidth}
\centering
\includegraphics[width=\columnwidth]{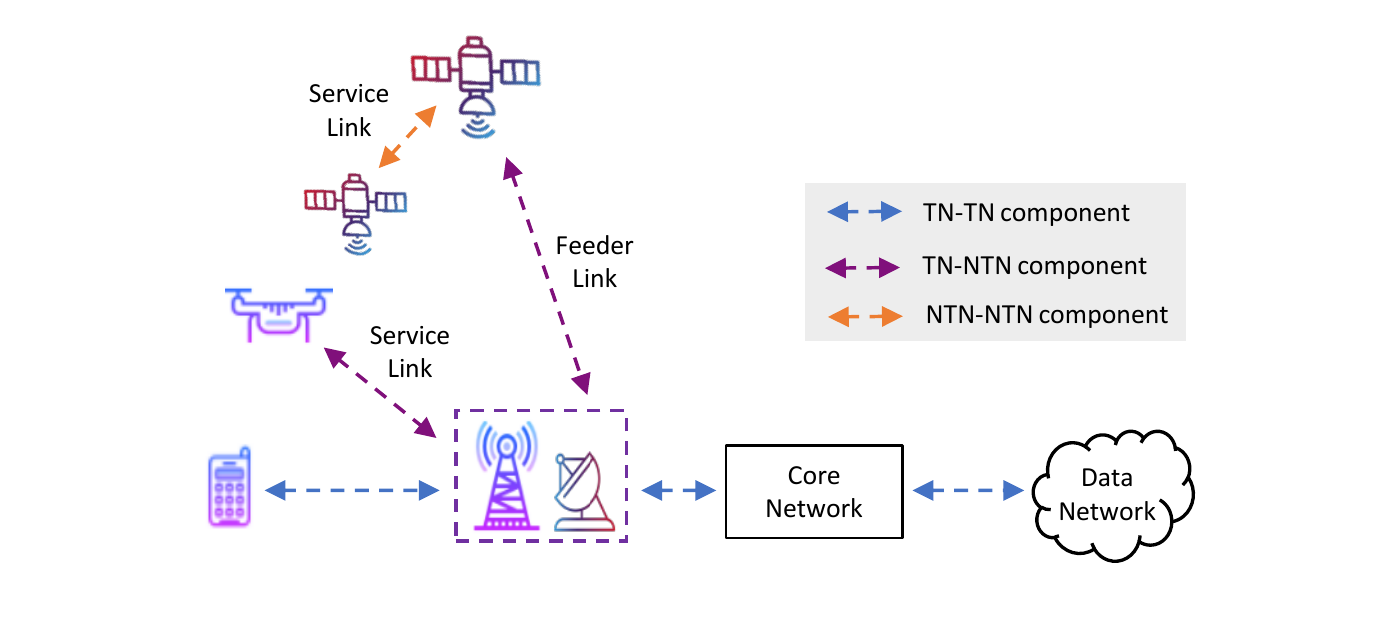}
	\caption{NT platform as a user. 
	}
	\label{fig:ntnUser}
\end{subfigure}
\begin{subfigure}{\columnwidth}
\centering
\includegraphics[width=\columnwidth]{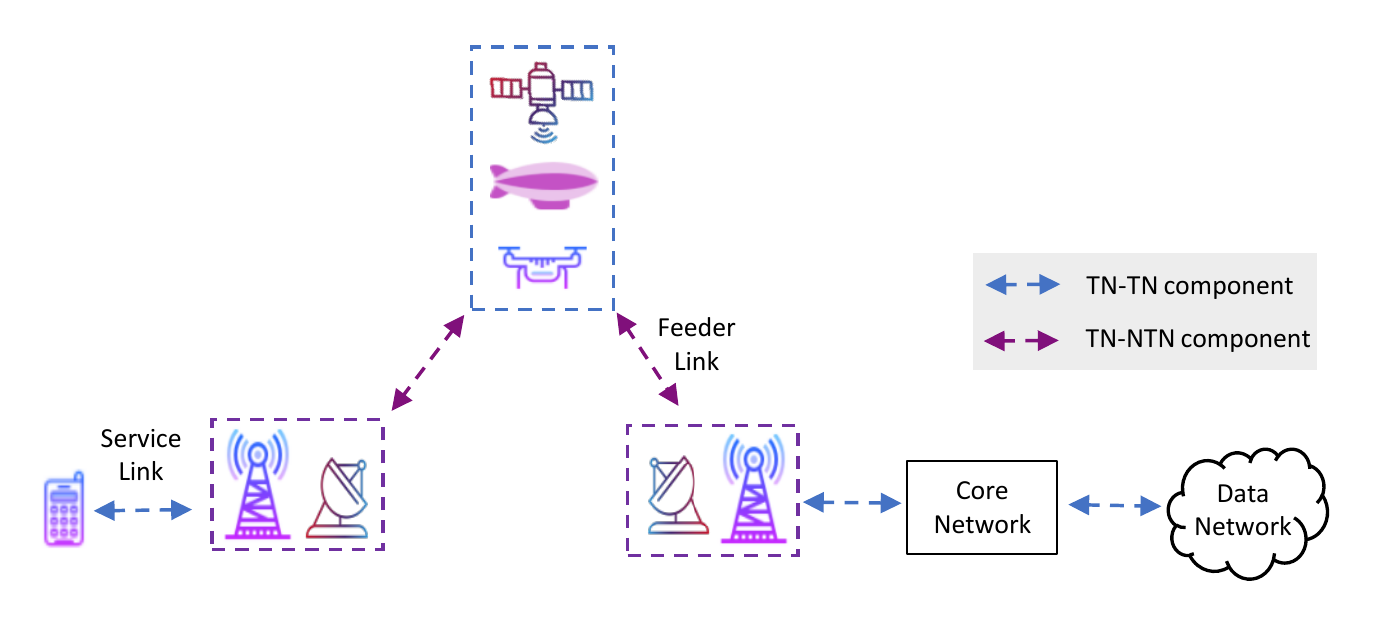}
	\caption{NT platform as a relay for backhauling.}
	\label{fig:ntnRelay1}
\end{subfigure}
\newline
\begin{subfigure}{\columnwidth}
\centering
\includegraphics[width=\columnwidth]{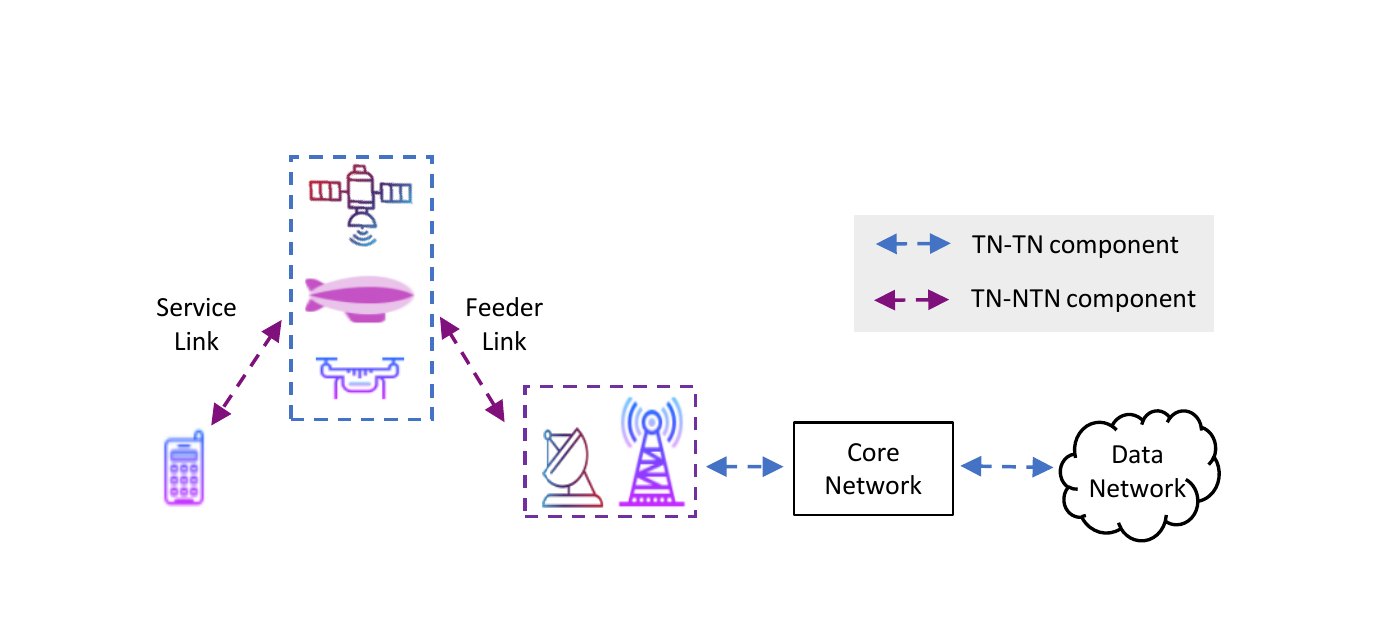}
	\caption{NT platform as a relay for end-users.}
	\label{fig:ntnRelay2}
\end{subfigure}
\begin{subfigure}{\columnwidth}
\centering
\includegraphics[width=\columnwidth]{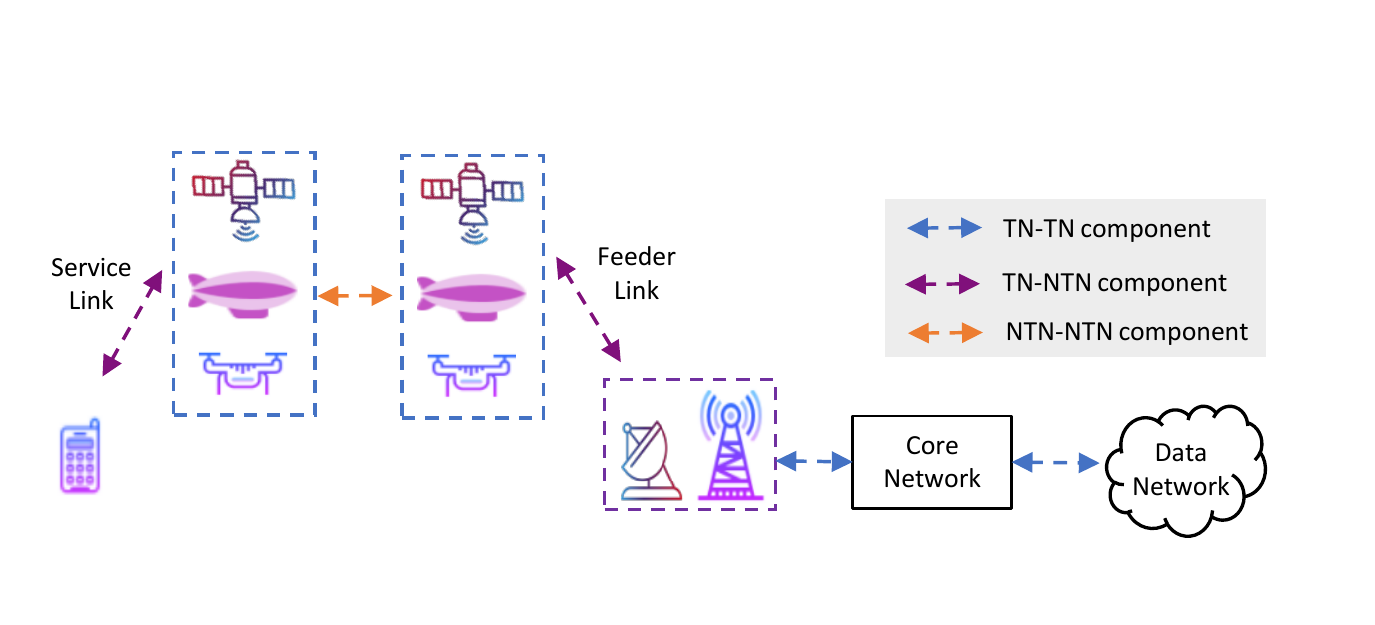}
	\caption{NT platform as a BS.}
	\label{fig:ntnBS}
\end{subfigure}
\caption{\textcolor{black}{NTN architecture options classified by the role of NT component in the overall communication chain of a terrestrial network. Detailed descriptions are provided in Section \ref{sec:general_architecture_options}.}}
\label{fig:NTNarchitecture}
\end{figure*}
\section{NTNs Integration in 5G Ecosystem}\label{sec:NTN_5G}

In this section, we review general assumptions, architectures, key features, and challenges of NTNs and their roles in 5G use cases.

\subsection{Preliminaries}
\subsubsection{NTNs} \label{sec2a}

As defined by the 3GPP \cite{3GPP38811}, an NTN refers to a network which partially/fully operates for communication purposes through a spaceborne vehicle i.e., Geostationary earth orbit (GEO), Medium earth orbit (MEO), and LEO satellites, or an airborne vehicle (i.e., HAPs and UAVs). The most important feature that makes NTNs unique is their capability to provide connectivity in unreachable areas for a terrestrial network (i.e., vessels and airplanes), or remote areas (i.e., rural areas) where a huge investment is required  to build a terrestrial infrastructure. 



In the spaceborne category, the most traditional ones are the GEO satellites which have a circular and equatorial orbit at an altitude of 35786 km. They appear stationary from a ground observer perspective and cover a large portion of the Earth’s surface. Their typical beam footprint size varies from 200 km to 3500 km \cite{3GPP38821}. Satellites operating at lower altitudes are the MEO and LEO ones, which are also known as non-geostationary (NGSO). This is because their rotational speed overcomes the rotational speed of the Earth and they appear as moving objects from a ground observer perspective. They have a circular or elliptical orbit with an altitude that varies between 7000-25000 km for MEO and 300-1500 km for LEO. Their typical beam footprint size ranges from 100 km to 1000 km \cite{3GPP38821}.

The airborne category includes unmanned aircraft system (UAS) platforms, where the most common ones contemplated for communication are HAPs and UAVs. The altitude in which the HAPs operate is 20 km \cite{rinaldi2020non}, whereas, UAVs fly at lower altitudes (e.g., few hundred meters). The footprint size of UASs can reach up to \black{few hundred km} on the ground. 

\subsubsection{General Architecture Options} \label{sec:general_architecture_options}

A terrestrial cellular radio access network (RAN) typically features the following system elements: a) the \textit{user equipment} (UE) which can be a handheld or an IoT device; b) the \textit{base station} (BS) which provides the access (service) link to the on-ground UEs; c) the \textit{core network} which has multiple functionalities including mobility management, authentication, session management application of different quality of services, etc.; and d) the \textit{data network}
which handles the transferring of data from one network access point to another. Integration of NTN components in existing terrestrial architecture can play a significant role in the typical communication chain. Such integration can be done at the physical (PHY) layer or at the NET layer. The PHY layer integration means that the NTN component has to utilize the same radio access technology as the terrestrial one \cite{3GPP38811, 5gdirectaccess}. Whereas, the integration at the NET layer allows for different radio access technologies, e.g., digital video broadcasting (DVB) satellite network with a 3GPP terrestrial core network \cite{3GPP24502}. Depending on the non-terrestrial (NT) platform placement, following different architecture options may exist:


\begin{itemize}
    \item NT platform as a user: 
    In this architecture option, the terrestrial infrastructure of a terrestrial network is used to serve the NTN platform (see Figure~\ref{fig:ntnUser}). The 3GPP has already concluded a study item \cite{3GPP36777} to explore the challenges and opportunities for serving the UAVs as a new type of UE \cite{azari2017coexistence,azari2018reshaping,AzaRosPol2019}. 
    \black{Furthermore, satellites can also be seen as users in space being served by other satellites at higher altitudes. This is an interesting architecture model that releases the need for a network of ground stations to gather/send data to the satellite directly from space \cite{satasuser}.}
    \item NT platform as a relay:
    In this architecture, there may be two options as demonstrated in Figure~\ref{fig:ntnRelay1} and Figure~\ref{fig:ntnRelay2}. The NTN platform can be used in the link between the BS and the core network, which usually is ensured by fiber optics, thus providing backhauling services. Also, the NTN platform can act as a relay in the link between the on-ground users and BS, providing a direct access connectivity. For these architectures, a transparent payload at the NTN platform is sufficient to relay the signal coming from a terrestrial component. 
    \item NT platform as a BS:
    In case of an NTN platform with a regenerative payload with enough processing capabilities, the BS functionalities can be incorporated on the fly, as illustrated in Figure~\ref{fig:ntnBS}.
    \item  Mixed architectures:
    Starting from the three architecture models explained above, other ones may exist combining various NT platforms playing a different role. For example, a satellite with BS functionalities may serve the on-ground users through the help of LEO/UAV platforms acting as a relay in the access link \cite{zhang2019multiple, bai2018multi}, or may directly serve the UAVs which will act as aerial users.
    Obviously, one architecture option can be more favourable than another depending on the particular use cases, which we will describe in the upcoming sections.
\end{itemize}

\begin{table*}[b]
\color{black}
	\caption{A general technical comparison between TNs and NTNs.}
	\centering
	\begin{tabular}{|l || l | l |}
		\toprule
		\textbf{Technical feature} & \textbf{Terrestrial}  &  \textbf{Non-terrestrial} \\
		\hline
		\hline
		\textbf{Coverage on earth} & Up to 100 km & Up to 3500 km (GEO satellite) \\ \hline
		\textbf{Propagation delay} & Up to 0.67 ms (100 km cell) & Up to 540 ms (GEO satellite with transparent payload) \\ \hline
		\textbf{Propagation path loss} & $\approx$ 138 dB (100 km cell \& 2 GHz $f_c$) & $\approx$ 190 dB (GEO satellite \& 2 GHz $f_c$) \\ \hline
		\textbf{Doppler shift} & $\approx$ 1 KHz (high speed train \& 2 GHz $f_c$) & $\approx$ 48 KHz (600 km altitude LEO satellite \& 2 GHz $f_c$) \\ \hline
		\textbf{Handovers} & Triggered when users move from one cell to another & Periodic HO due to NGSO satellite movement \\ \hline
		\textbf{Network deployment} & Long-term deployment & Temporary or long-term depending on the NTN platform \\
		\bottomrule
	\end{tabular}
	\label{tabletnvsntn}
\end{table*}

\subsubsection{Key Features and Challenges of NTNs}
Several characteristics of the NTNs are different from the terrestrial systems, mainly caused by the altitude and movement of the NT platforms. These features are briefly described in the sequel in a comparative manner:
\begin{itemize}
    \item Propagation delay and path-loss:
    The altitude of the NT platforms can cause extra delays in the communication link, especially in the case of GEO satellites reaching round trip latency of 270 ms \cite{3GPP38821}. This may create a bottleneck for specific services and applications which require ultra-low latency. Longer propagation delay could impact protocol layers in terms of retransmissions and response time and result in an outdated channel quality indicator (CQI) measurement. Not only this, but the propagation path losses would also be higher since the signal has to propagate over larger distances. For UAVs and HAPs, the propagation delays and losses can be in the range of terrestrial communication as their altitude is much lower compared to the satellites. Furthermore, satellite users will experience different propagation delays in different regions of the their cells due to the large coverage areas. This will affect the initial access and synchronization of users at the cell center and cell edge. 
    \item Doppler effect: 
    The movement of the NT platforms may also cause extra Doppler effects in the communication link. Of course, the Doppler effects also occur in the terrestrial networks due to relative motion between the users (e.g., a UE inside a high speed train) and the fixed ground BS. However, in an integrated GAS network, the movement of the NT platform may drastically increase the Doppler effects, especially in case of a LEO satellite. Assuming a communication link over a LEO satellite at 600 km altitude and 2 GHz carrier frequency ($f_c$), the Doppler shift reaches peak values of 48 kHz \cite{lin20195g}. This is around 10 times higher compared to the one experienced by a user inside a high speed train, thus an important feature to be addressed. Large frequency shifts due to Doppler effects must be considered when determining the subcarrier spacing of orthogonal frequency division multiplexing (OFDM) systems. 
    The reference signals, used to monitor signal strengths for example, should be adapted to take into account Doppler and possible specific multipath effects \cite{3GPP38821}.
    \item Coverage, throughput, and handover (HO):
    The higher the altitude of the NT platform, the wider is the beam footprint size. For a GEO satellite, the coverage can reach up to 3500 km, whereas, for MEO and LEO, it goes up to 1000 km \cite{rinaldi2020non}. Although, the specific value depends on the exact altitude and supported elevation angles of communication between ground users and the satellite. The movement of NGSO satellites results in a varying coverage in time and space, while the coverage of GEO ones is quite static. Frequent handover procedures are required to route the traffic from one NGSO satellite to another and to ensure continuity of services to ground users. On the other hand, though the coverage of HAPs and UAVs is smaller and the risk of handover might be higher \cite{rinaldi2020non}, they can provide higher throughput links to the ground users in a more flexible manner. They can be deployed quickly to targeted areas, or can be re-directed depending on the network demand. Last but not the least, an important aspect worth mentioning is the increased risk of interference caused by the larger footprint of the NTN platforms on the ground. Mitigating the interference coming from different NT platforms placed at various orbital ranges is crucial in a multi-segment 
    communication scenario (mixed-architecture option). To avoid the co-channel intereference, orthogonal frequency allocation may not be a possible solution in any case due to the spectrum scarcity, raising the need for spectrum co-existence \cite{spectrumcoex}.
    \item Deployment:
    UAVs and HAPs offer a cheap, quick, flexible, and temporary deployment solution compared to satellites which, in general, are meant for long-term deployment. Aerial BSs can be deployed for operation within hours, whereas, satellites require months (and sometimes years, i.e., GEO) until they are ready for operation. 
    UAVs/HAPs can directly complement the existing infrastructure in case of failure/congestion. In addition, the cost of deploying aerial platforms in a network is typically much smaller compared to satellites. 
    However, once deployed, satellite can provide wide area coverage in contrast to UAVs/HAPs.
    \item Space weather effects: Studying space weather is crucial for the satellite operations. The earth space is traversed by large and variable radiation flux and also by small pieces of matter. Specifically, the radiation field arises from the cosmic radiation, solar radiation, and trapped radiation belts (also known as Van Allen belts). Amongst these, the radiation belts, which are low-energy particle radiation, must be considered for satellites that spend a large amount of time in MEO orbits \cite{AUSBureau}. The orbits below the 1500 km (LEO satellites) are mostly below the radiation belts, whereas geosynchronous orbits (GEO satellites) lie above them. Within the outer belt, the most intense radiation occurs between 9,000 and 12,000 miles above earth's surface.
The accelerated particles in the Van Allen Radiation belt can pose danger to the spacecrafts. Nevertheless, a greater understanding of the radiation belts can aid in the building of radiation-resistant satellites \cite{nasadotgov}. To this end, NASA in 2012 launched the Van Allen Probes spacecraft to evaluate the situations that can disrupt the satellite operations. These findings help engineers with precise data to develop the better technologies and hardware that can withstand radiation. 
\end{itemize}
\textcolor{black}{Table \ref{tabletnvsntn} provides a summary of the above-mentioned features in a comparative manner between TNs and NTNs.}

\begin{figure*}[!h]
\centering
\includegraphics[width=2\columnwidth]{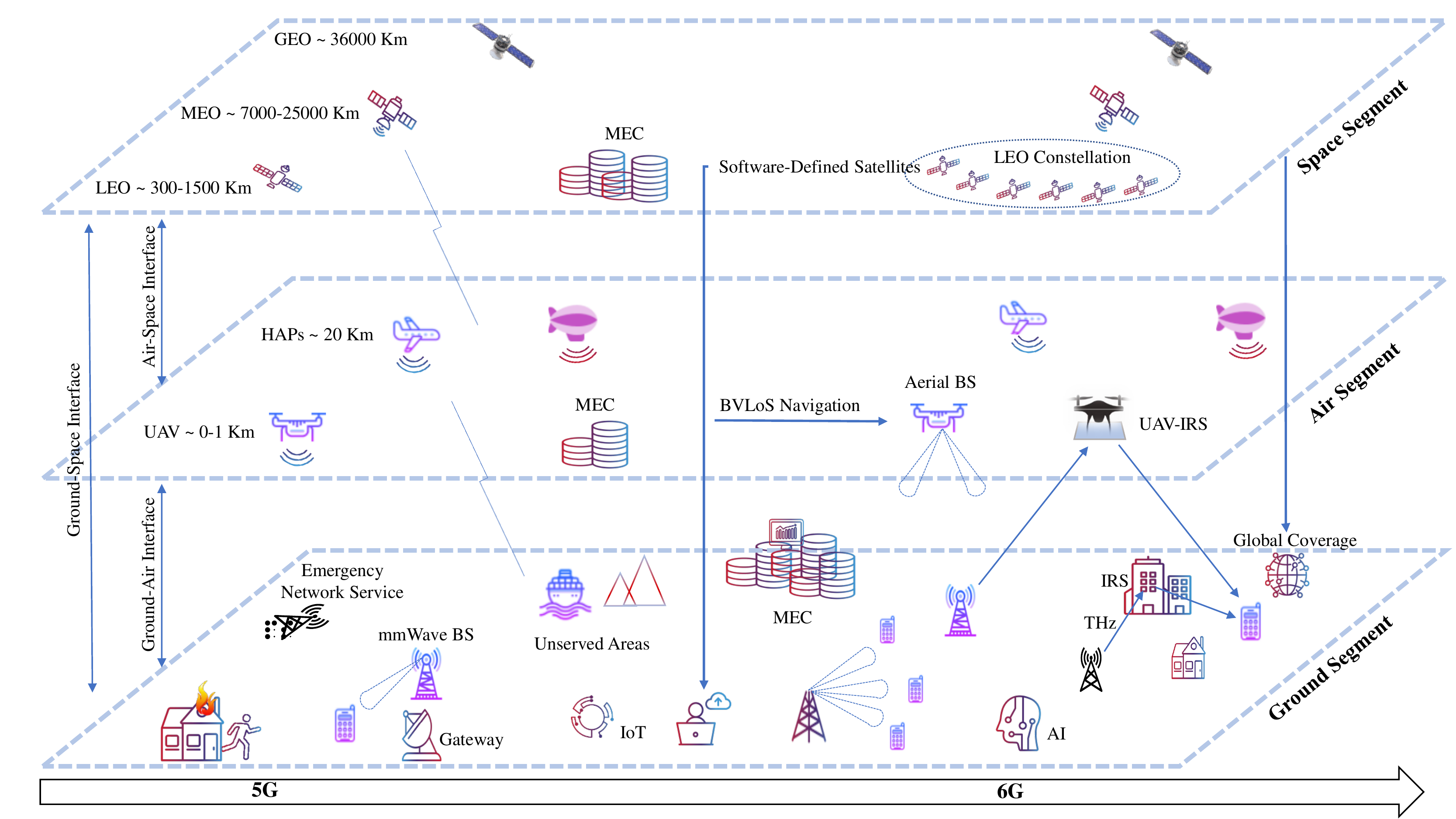}
	\caption{The evolution of integrated GAS networks towards 6G and beyond.}
	\label{general_fig}
\end{figure*}

\subsection{Key Drivers for NTNs Integration in 5G}

Thanks to the NTNs capabilities in providing a widespread coverage and their reduced vulnerability to physical attacks and natural disasters, they can play important roles in supporting the 5G terrestrial networks \cite{3GPP38811,3GPP22822}:
\begin{itemize}
    \item NTNs are able to improve the limited performance of 5G terrestrial networks in un-served and underserved areas or in disaster-hit regions where the terrestrial networks are destructed or in outage, by offering \textit{ubiquitous} services from the space/sky in a cost-effective manner.
    \item NTNs improve the reliability of 5G network by ensuring \textit{continuous} service for IoT devices and mobile on-board passengers such as in ships, aircrafts, and trains.
    \item NTNs also enable \textit{scalable} service through efficient broadcasting allowing streaming content to large areas and offloading popular content to the network edge caching.
\end{itemize}
In such NTN-assisted terrestrial communication scenarios, the NTN component can be deployed according to architecture options shown in Fig. \eqref{fig:NTNarchitecture}. 

Among three 5G service enablers, i.e., eMBB, uRLLC, and massive machine type communications (mMTC), the roles of NT \textit{access} networks fit mostly within eMBB and mMTC use cases as specified in \cite[Sec. 4.2]{3GPP38811} through several examples. The high propagation delay of satellite systems are the major barrier in using them for uRLLC purpose, however, the UAVs can still remarkably contribute into this aspect of 5G use cases too. For such purposes, UAVs acting as aerial BSs (see Fig. \eqref{fig:ntnBS}) are flexible platforms that can adjust their 3D locations and employ 5G communication technologies to serve the target users \cite{azari2017ultra,arani2020learning}. Finally, \cite{rinaldi2020non} provides an overview of the 3GPP NTN features, uncovering their potential to satisfy consumer expectations in 5G networks.

In addition to use cases corresponding to NT access networks, UAVs acting as aerial users (see Fig. \eqref{fig:ntnUser}) can significantly benefit from the development of 5G terrestrial networks for various applications \cite{3GPP22.125,3GPP22829,AzaGerGar2020}. For instance, 5G eMBB and uRLLC service enablers allow an efficient UAV traffic management (UTM) and reliable control of drones. Selected 5G UAV use cases, where UAV acts as end-user, along with their requirements are discussed in \cite{geraci2021will}. \black{ Figure~\ref{general_fig} provides an illustration of how integrated GAS networks are evolving from 5G and progressing towards 6G. Especially, it attempts to highlight the present use-cases and opportunities while showcasing the prospective applications of the NTN in future.}

\vspace{0.1in}
\noindent\textit{\black{{\textbf{Key Takeaways} -- The above-mentioned use cases are a clear indicator that 5G and beyond systems will rely more and more on non-terrestrial components to offer their services globally. This is due to their unique capabilities in extending coverage in areas where a terrestrial infrastructure is impossible or cost-inefficient to reach, as well as their complementary role in offloading an important part of the traffic especially in highly congested areas. Nevertheless, the technical peculiarities coming from the presence of the non-terrestrial channel, which is quite different from a terrestrial one, brings various challenges to be solved and raises the need for novel solutions.}}}

\section{NTNs Operation in mmWave 
} \label{sec:NTN_mmWave}

\tikzset{
  level/.style   = { line width=2pt, black },
  level2/.style   = { ultra thick, black},
  connect1/.style = { thick, blue },
  connect2/.style = { thick, dashed, blue },
  notice/.style  = { draw, rectangle callout, callout relative pointer={#1} },
  label/.style   = { text width=2cm }
}

\begin{figure*}
\begin{tikzpicture}
  \draw[level] (-1,0) -- node[above] {mmWave} (1,0);

  \draw[connect1] (-1,0)  -- (-1.5,2) (-1,0) -- (-1.5,-2) (1,0) -- (1.5,2) (1,0) -- (1.5,-2);
  \draw[level2]   (-1.5,2)  -- node[above] {Background} node[below] {} (-3,2);
  \draw[level2]   (-1.5,-2) -- node[above] {UAV} node[below] {} (-3,-2);
  \draw[level2]   (1.5,2) -- node[above] {Satellite} node[below] {} (3,2);
  \draw[level2]   (1.5,-2) -- node[above] {Multi} node[below] {} (3,-2);

  \draw[connect2] (-3,2) -- (-3.5,2);
  \draw[level]   (-3.5,2)  node[left] {\footnotesize Terrestrial Networks \cite{3GPP38104,chiaraviglio2017bringing,shafi2018microwave,rappaport2017overview}};
  
  \draw[connect2] (3,2) -- (3.5,2) (3,2) -- (3.5,1.5) (3,2) -- (3.5,1) (3,2) -- (3.5,0.5) (3,2) -- (3.5,0) ;
  \draw[level]   (3.5,2)  node[right] {\footnotesize Architecture \cite{babich2019nanosatellite,kourogiorgas2017cooperative,mudonhi2018SDN,artiga2018shared, ziaragkas2017sansa, zhang2020multicast, di2019ultra}};
  \draw[level]   (3.5,1.5) node[right] {\footnotesize Channel Modeling \cite{xiao2017millimeter,khan2011mmWave,wenzhen2001ka,chun1998land,kelmendi2021alphasat}};
  \draw[level]   (3.5,1) node[right] {\footnotesize Resource Allocation \cite{kodheli2020satellite,choi2005optimum,lei2011multibeam,kibria2020carrier,lagunas2017carrier,lagunas2018fair, shaat2018integrated,kawamoto2020flexible}};
  \draw[level]   (3.5,0.5) node[right] {\footnotesize Performance Evaluation \cite{Liang2018outage, an2019hybrid, hraishawi2021scheduling}};
   \draw[level]   (3.5,0) node[right] {\footnotesize Co-existence \cite{guidolin2015study,icolari2016genetic,lagunas2015resource,lagunas2015power, lagunas2016power, sharma20153d,an2017outage,zhang2020coexistence, peng2020hybrid}};
 
  \draw[connect2] (-3,-2) -- (-3.5,-2) (-3,-2) -- (-3.5,-1.5)  (-3,-2) -- (-3.5,-1) (-3,-2) -- (-3.5,-0.5) ;
  \draw[level]    (-3.5,-0.5)  node[left] {\footnotesize  Architecture \cite{xiao2016enabling,zhang2019survey,xiao2020uav,feng2018spectrum}};
  \draw[level]    (-3.5,-1)  node[left] {\footnotesize Channel Modeling \cite{geraci2021will,zhang2019research, khawaja2017uav,dabiri2020analytical,zhang2019survey,polese2020experimental}};
  \draw[level]    (-3.5,-1.5) node[left] {\footnotesize Beamforming \cite{geraci2021will,colpaert2018aerial,kalamkar2020beam,colpaert20203d,song2020beam,zhou2019beam,heimann2018potential,zhang2019fast}};
  \draw[level]    (-3.5,-2) node[left] {\footnotesize Performance Evaluation \cite{AzaRosPol2019,AzaGerGar2020,geraci2021will,yang2020performance,gapeyenko2018flexible, wang2019multiple,khosravi2018performance,rahmati2019energy,zhao2018channel}};
  
  \draw[connect2] (3,-2) -- (3.5,-2);
  \draw[level]   (3.5,-2)  node[right] { \footnotesize  Architecture \cite{li2019extensible,kodheli2020satellite,li2020forecast}};

\end{tikzpicture}
\caption{NTNs integration in mmWave.}
    \label{fig:mmWaveRev}
\end{figure*}
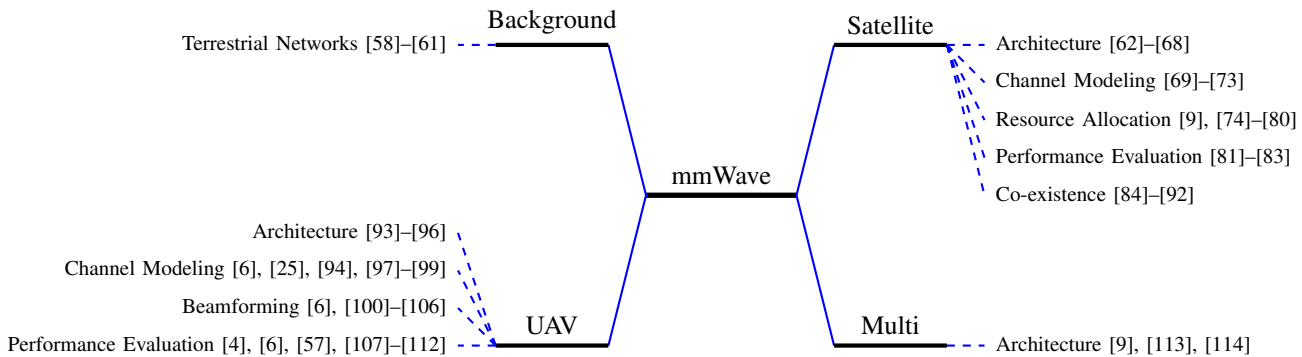
The promise of 5G new radio (NR) to provide widespread broadband access to everyone has thus far been difficult to realize. To fulfill this requirement, 5G new radio opens up a new part of the spectrum with its frequency band 2, ranging from 24.25 GHz
to 52.60 GHz, to effectively realize this broadband access to all the users \cite{3GPP38104}.
The use of mmWave frequencies is envisioned to be effective in many parts of the 5G NR network. For example, in both the fronthaul and backhaul of a BS, mmWave can offer deployment with the same data rates when compared to fiber. Another example is densifying the network by cell shrinking, in this case the large propagation losses and directivity in combination with large bandwidths will prove most useful. 
While urban regions can be equipped with sufficiently dense mmWave access points supported by fixed high data rate backhaul links, it would be ineffective or not cost efficient for rural or remote areas \cite{chiaraviglio2017bringing}. Thus, deploying NTNs using mmWave communication to provide broadband has lately attracted a lot of research attention. While broadcasting services at mmWave frequencies are not new in satellite communications, the complexity of integrating SatComs into a broadband access network will have additional challenges related, for example, to system architecture, resource allocation, co-existence, etc.

Although the advantages seem promising, the use of mmWave frequencies will impact the design choices to be made in NTN. The main difference being that mmWave signals have different propagation behaviour when compared to lower frequencies and therefore fundamental large scale propagation and channel models have to be re-investigated. For example, path loss, penetration loss, and shadow fading behave differently \cite{rappaport2017overview,shafi2018microwave}.  A list of the open challenges for mmWave satellites are given in \cite{giordani2020non}. The work provided a multitude of challenges including but not limited to the lack of second order aerial mmWave channel statistics, such as time and spatial correlation. 

Another issue is the fact that fast moving stations in the air or in space will create noncontinuous coverage zones. These problems can be solved by managing the constellations of satellites. Note that planning a large scale satellite network might be more straightforward than providing full terrestrial mmWave connectivity in a dense urban environment. Another challenge is mobility which requires very precise beamforming and tracking, although the use of directional antennas might be somewhat helpful. Mobility also implies the need for efficient handovers. In this section, we outline an overview of the literature focusing on NTN integration in the mmWave communications networks. A schematic of the discussed works can be found in Fig.~\ref{fig:mmWaveRev}. 
\black{The main differences for mmWave deployment in terrestrial, aerial, and space segments can be found in Table \ref{tab:mmWave}.}

\begin{table*}[t]
\color{black}
	\caption{mmWave operation for terrestrial, aerial and space nodes}
	\centering
	\begin{tabular}{ | l || l | l | l |}
		\toprule
		\textbf{Aspect} & \textbf{Terrestrial}\cite{rappaport2017overview} &  \textbf{Aerial} \cite{xiao2020uav,zhang2019survey} & \textbf{Space} \cite{giordani2020non,giordani2020satellite} \\
		\hline
		\hline
		\textbf{Propagation} & Blockage dominated &  LOS dominated & Atmospheric attenuation \\
		\hline
		\textbf{Channel modeling}  & Well investigated  & Open problem &  Older models \\ 
		\hline
		\textbf{Station deployment} & Static & Highly dynamic & Dynamic\\
		\hline
		\textbf{Coverage area} & Small & Dynamic & Large\\
		\hline
		\textbf{Synchronisation} & Easier due to static deployment & Difficult due to large Doppler shifts & Difficult due to atmospheric effects\\
		\bottomrule
	\end{tabular}
	\label{tab:mmWave}
\end{table*}

\subsection{Satellite Operation in mmWave}
The use of satellites utilizing mmWave frequencies promises a high intrinsic capacity while simultaneously supporting global coverage \cite{giordani2020satellite}. Satellite communication at mmWave is not new, as the Ka-band has been used for fixed satellite services for years. However, integrating satellites into the cellular network over mmWave links is a recent concept. In this context, 3GPP released 16 envisions satellite links to offer broadband services over mmWave frequencies because of their large bandwidths and high spatial resolution \cite{3GPP38821}. These 3GPP envisions concluded that NR functionalities form a sound basis for supporting NTN scenarios. However, they identified several issues due to considerable propagation delays, significant Doppler effects, and moving cells in an NTN. Therefore, they proposed to specifically focus on the following: timing enhancements, UL time and frequency synchronization, and enhancements on the PRACH sequence in the case the UE itself does not perform timing and frequency offset correction.

\subsubsection{Architecture}
Several different architectures involving satellite networks operating in the Ka-band are presented in literature.
In \cite{di2019ultra}, Di \textit{et al.} proposed a ground-space network, integrating ultra-dense LEO and terrestrial networks where the terrestrial nodes operate in the C-band and the satellites in the Ka-band. They considered a dynamic backhaul and proposed a scheme for data offloading.
Authors of \cite{kourogiorgas2017cooperative} presented a hybrid ground-space cooperative system for backhaul applications where a relay node repeats the satellite signal to a destination ground station, which in turn utilizes both the relayed and the direct signals in a diversity combiner to compensate for the large propagation losses.
In \cite{babich2019nanosatellite}, a complete network architecture for an integrated nanonsatellite (nSAT)-5G mmWave system is envisioned, where a constellation of nSATs communicating through mmWave links operates as gateways between several isolated terrestrial BSs. A detailed architecture is presented and a comprehensive analysis of the system is performed by implementing the system in NS3.

Several works proposed and evaluated different software defined networks (SDNs) architectures where the integration with satellite mmWave links are present and are used to improve the performance of terrestrial networks, to improve resilience \cite{mudonhi2018SDN} or to increase flexibility \cite{artiga2018shared}. In the SANSA project \cite{ziaragkas2017sansa}, a spectrum efficient self-organizing hybrid terrestrial-satellite backhaul network is presented enabled by smart antenna techniques at mmWave and software defined intelligent hybrid network management.
On the other hand, \cite{zhang2020multicast} studied a cloud-based ground-space network design to support high speed multimedia services. A central processing unit interconnects all the nodes and manages joint user scheduling and multicast beamforming.

\subsubsection{Channel Modeling}
It is well known that mmWave signals traveling from space undergo severe Doppler shifts and several stages of attenuation as presented by \cite{ xiao2017millimeter}. These losses can be countered by utilizing large antenna arrays to generate highly directive communication \cite{khan2011mmWave} and due to the short wavelengths the aperture of these antenna arrays will be very small. 

As such, satellite communication over mmWave has been studied extensively in the past and various measurements also exist to support the theoretical models. \black{In particular, \cite{wenzhen2001ka} studied a Ka-band land mobile satellite channel (LMS) model which also accounted for the weather impairments. Based on this model, the authors derived the statistics of the received signal and thereby analyzed bit error rate performance. They also showed that their proposed model provided more realistic results compared to other weather-affected LMS model. In \cite{chun1998land}, the authors extensively reviewed the progress of the LMS channel modelling and measurements. For the measurements, various platforms including helicopters, balloons, airplanes, and satellites were considered at varying elevation angles. Moreover, the channel characterization included different frequencies ranging from UHF-band to Ka-band. More recently, the authors in \cite{kelmendi2021alphasat} analyzed the signal attenuation due to rains based on three years of measurement data from the Alphasat satellite in Ka- and Q-bands. The investigation revealed that the number of fades and fades duration at Ka-band are notably lower in comparison to Q-band. } 
 
\subsubsection{Resource Allocation}
Since satellite resources are scarce, it is necessary to efficiently allocate them \cite{kodheli2020satellite}. Power allocation should be based on channel conditions and traffic demands for the different users \cite{choi2005optimum}. However, when beamforming is used, one also should consider inter beam interference as it will affect the operation of the whole system\cite{lei2011multibeam}.

Allocating frequencies and bandwidths to the different beams and users should be done carefully. Carrier aggregation (CA) could aid in utilizing the limited spectrum efficiently. \cite{kibria2020carrier} discussed several CA techniques and technical problems. The authors proposed an efficient multi-user aggregation scheme to enhance data rates of satellite users as well as efficient use of resources.
The work of \cite{lagunas2017carrier} considered the problem of carrier allocation in a hybrid satellite-terrestrial backhaul, proposing a sequential allocation algorithm to tackle this problem. In an additional work, an improved carrier allocation algorithm is presented \cite{lagunas2018fair}.
While, \cite{shaat2018integrated} stresses on the benefits of integrating satellite and terrestrial networks within the same band as well as the need for efficient resource allocation strategies to improve spectral efficiency.
Further, efficient frequency channel allocation method is required to optimize the capacity. As orthogonal channel techniques seem to not provide enough capacity for future NTN networks, channels should either be semi- or non-orthogonal\cite{kodheli2020satellite}. The authors of \cite{kawamoto2020flexible} presented a frequency resource allocation method with inter-beam interference for satellites equipped with a digital channelizer. A flexible channel allocation framework dependent on demand will probably be the most efficient. However, this concept makes the allocation problem much more challenging and more research is required in this direction.


\subsubsection{Performance Evaluation}
The authors in \cite{Liang2018outage} investigated the outage performance of a hybrid ground-space network utilizing terrestrial stations as relays. \black{They proposed a relay selection system utilizing rain attenuation values and derive outage expressions to evaluate the system performance. Relay selection based on rain attenuation essentially helped in reducing the computational complexity  while improving the outage performance. Additionally, both fixed and variable gain relaying protocol have been considered for their investigation.
Further, the work in \cite{an2019hybrid} considered a hybrid satellite-terrestrial network and derives the analytical expressions of the channel capacity for two adaptive transmission schemes namely, optimal power and rate adaptation and truncated channel inversion with fixed rate. It is found that the optimal power and rate adaptation scheme perform almost same as conventional optimal rate and fixed power scheme at high SNR regime. However, the former perform better in low to medium SNR regime.
In a recent work \cite{hraishawi2021scheduling}, authors investigated the performance of satellite system using CA from a link layer perspective. It is noted in the study that CA in satellite systems require efficient scheduling methods to achieve reliable communications. They proposed a load balancing algorithm for packet distribution across the carriers depending upon the channel capacities. They concluded that adding CA has a relative low complexity and it can reduce queue delays through offloading to other carriers and thereby improving the user experience in general.} 

\subsubsection{Co-existence}
As mentioned before, broadband mmWave satellite networks may co-exist with the upcoming cellular networks operating in the same frequency bands. In \cite{guidolin2015study}, Guidolin \textit{et al.}  studied the interference levels seen by Fixed-satellite services as a result of mmWave cellular networks. A number of European projects aim to study or solve these interference issues. For example, the CoRaSat project aimed to investigated and develop cognitive radio techniques in SatCom networks for spectrum sharing, targeting a better exploited spectrum through flexible spectrum usage. In this context, \cite{icolari2016genetic} focused on exploiting frequencies not used by existing fixed services for broadband downlinks toward the users. While, \cite{lagunas2015resource} proposed a spectrum exploitation framework and evaulates the system based on available deployment data of a cognitive satellite link.

Several works\cite{lagunas2015power, lagunas2016power} discussed different power allocation algorithms for the cognitive uplink for satellite-terrestrial co-existence. Researchers in \cite{sharma20153d} explored the use of a 3D beamforming antenna at the satellite to further mitigate the interference to terrestrial networks.
Several other works
investigated the outage performance of these type of co-existing networks, e.g., \cite{zhang2020coexistence, an2017outage}. A general framework for the co-existence of these networks and analytical expressions for the outage probability are provided in \cite{an2017outage}, while \cite{zhang2020coexistence} evaluated a more advanced scenario where the satellite simultaneously serves multiple users and is equipped with a uniform planar antenna array. Furthermore, they also provided an analytical expression for the beamforming weights. On the other hand, \cite{peng2020hybrid} considered an integrated satellite network utilizing massive multiple-input multiple-output (MIMO) but evaluates intra-system interference between the satellite and terrestrial systems. They proposed a hybrid precoding algorithm to mitigate this interference.

\black{Since 5G mainly targets the use of the Ka-band, ranging from 26.5 GHz to 40 GHz, this band is already being used by older satellite technologies such as weather forecasting and very small aperture terminals (VSAT) and hence spectrum sharing becomes challenging. There has been discussion in the past on how terrestrial and space services will share frequency bands above 1GHz. At the World Radio communication Conference (WRC), some regulatory advice regarding spectrum sharing has been agreed on. For example, receiving stations should avoid directing their antennas towards geostationary-satellite orbit in specific mmWave bands. They are recommended to maintain a minimum separation angle. Other guidelines covered maximum transmitted powers for different bands as well as limits of power-flux densities from space stations\cite{wrc15,wrc19}. They also agreed that more research is required to verify the coordination between these new and old technologies.}

\subsection{UAV Operation in mmWave}

Introducing mmWave communication in UAV cellular networks has proven to have great potential such as in a video monitoring scenario where large amounts of traffic data are sent back to a control station for fast response\cite{xiao2016enabling} or in general to improve coverage for aerial vehicles by reducing the interference levels \cite{colpaert2018aerial}.

\subsubsection{Architecture}
The works in \cite{zhang2019survey} and \cite{xiao2020uav} presented current achievements of 5G mmWave in UAV-assisted wireless networks and highlighted three potential applications. First of all, aerial access, where a UAV provides access to the ground users by operating as an aerial BS. Secondly, aerial relay, where a UAV acts as a relay in a multi-hop fashion, providing indirect access to ground users. Finally, aerial backhaul links to provide Gigabit data transmissions in a flexible manner. Both works gave an overview on the advantages and challenges of mmWave communications for UAV-assisted networks. The three main advantages mentioned are availability of large bandwidth communication reaching peak data rates up to 10 Gbit/s, short wavelengths allowing for design of small and lightweight antenna arrays supporting large flexibility and lastly the use of directive transmission significantly reducing the interference levels and prevents eavesdropping. On the other hand several challenges still remain. The first challenges being the increased free space path-loss at higher frequencies and the propagation issues due to atmospheric losses and blockages. Efficient beam alignment is another requirement for effective mmWave UAV wireless networks. Another challenge in context of 5G mmWave, is the need to share the spectrum with the existing applications in the mmWave bands, namely the Ka-band. An overview of potential spectrum management scenarios for mmWave enabled UAV swarm networks can be found in \cite{feng2018spectrum}, they proposed a novel spectrum management architecture specifically for UAV-assisted cellular networks.

\subsubsection{Channel Modeling}
UAV channel models remain a research challenge due to rapid changing channel parameters and the high mobility of UAVs. The work in \cite{zhang2019research} presented the challenges for UAV mmWave communications and arising research opportunities. It describes the key channel characteristics of UAV mmWave including new air-to-ground (A2G) and air-to-air (A2A) channel models, temporal and spatial channel variations, airframe shadowing, Doppler effect, and 3D blockages. It is also noted that the UAV mobility gives rise to two remarkable challenges, i.e., reduced channel coherence time and channel path changes. To overcome these challenges, they outlined some methods for UAV channel estimation and beam training and tracking.

Authors of \cite{khawaja2017uav} investigated A2G mmWave channel characteristics utilizing ray tracing simulations at 28 GHz and 60 GHz. Specifically, they analyzed receive signal strength (RSS) and root mean square delay spread (RMS-DS) for various UAV altitudes in different environment viz., urban, suburban, rural, and over sea. They conclude that a two-ray propagation model can be applicable in all the environments. It is also observed that the scatterers predominantly affect the RSS in urban scenario while its fluctuation rate is higher at 60 GHz with respect to distance. 
The authors in \cite{dabiri2020analytical} provided an tractable, closed-form statistical channel model for different UAV communication links: UAV-to-UAV, aerial relaying for aerial nodes, and aerial relaying for terrestrial nodes. Especially, the analytical channel model also accounts for the random vibrations and orientation fluctuations of the UAV. They also studied the effect of antenna directivity gain under different channel conditions. It is found that increasing antenna directivity does not necessarily leads to improved performance, and hence optimization of radiation pattern of antenna is crucial.

An experimental analysis of the UAV-to-UAV mmWave channel at 60 GHz is performed in \cite{polese2020experimental}. The authors proposed an empirical propagation loss model based on an extensive aerial measurement campaign using the channel sounder mounted on two UAVs. The analysis found that the path loss for A2A link is independent of the UAV altitude, primarily between 6-15 m.
In addition to the above, the work in \cite{zhang2019survey} presented an extensive overview on the different A2G and A2A channel models available in literature. 

\subsubsection{Beamforming}
\begin{figure}[t]
\centering
\input{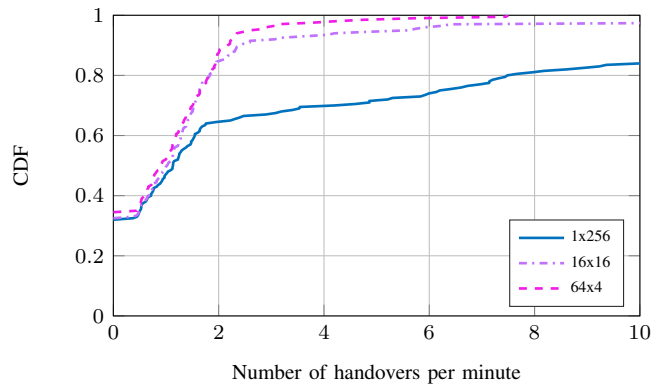}
\caption{Cumulative distribution function (CDF) plot of the number of handovers per minute for different antenna array topologies of a total of 256 elements for a UAV at an altitude of $40$~m \cite{colpaert20203d}.}
\label{fig:handovers40}
\end{figure}

One of the advantages of using mmWave is the improved security and interference reduction  by using highly directive beamforming  \cite{colpaert2018aerial}. \black{It is shown in the study that, through the use of beamforming, an aerial user can have coverage up to the altitudes of 100~m and more without experiencing too much interference as is the case in an LTE system. However, the beamforming also leads to additional challenges, for example, beam misalignment becomes more prevalent and frequent with the narrower beamwidth.} In \cite{kalamkar2020beam}, the authors provided a system-level stochastic geometry model to analyse the important aspects of mmWave beam management problems, while \cite{song2020beam} proposed a possible solution to perform consistent beam tracking from ground station to UAV in the form of a Kalman filter based tracking model. The work of \cite{zhang2019fast}, on the other hand, presented an inter-UAV beam tracking scheme, capable of overcoming slight beam misalignment. The fact that mmWave networks become more dense, will also pose problems in terms of handovers, as cell size will shrink dramatically in urban environments \cite{colpaert20203d}. \black{It is shown hereby that a drone will experience at least one handover per minute, see Fig.~\ref{fig:handovers40}, and also that care should be taken in designing the antenna array. Using large antenna arrays will result in large gains, but the beamwidth will become more narrow, resulting in very small beam misalignment error margins.} Zhou \textit{et al.} in \cite{zhou2019beam}, evaluated beam management and network self-healing in UAV mesh networks. The work of \cite{heimann2018potential} is worth to mention as they performed actual measurements with a mmWave antenna equipped on a UAV to evaluate pencil beam alignment and beam tracking algorithms.

\subsubsection{Performance Evaluation}
Regarding the performance of integrated UAV-ground networks, \black{general network and link level analytical modelings based on 3GPP and ITU reports are proposed in \cite{AzaRosPol2019, AzaGerGar2020} which can be applied to various scenarios. These papers extensively studied the impact of blockages, altitude, urban types, network density, antenna patterns, and also spectrum sharing mechanisms on the KPIs such as coverage and throughput.} Authors of \cite{gapeyenko2018flexible} analysed UAV assisted mmWave ground networks, where the UAV assists the BSs in case of link blockages. 
In the context of mesh networking, \cite{yang2020performance} formulated the performance and fairness of UAV swarms connected to cellular networks. 
In \cite{khosravi2018performance} the performance of a lamp post deployment of mmWave access points with additional coverage provided by UAVs is evaluated.
The authors of \cite{rahmati2019energy} noted that new multiple access schemes will be necessary when deploying mmWave for UAVs to efficiently distribute resources. For this reason they introduced two new multiple access schemes and evaluate their energy efficiency. Authors of \cite{wang2019multiple} concluded that existing approaches for multiple access do not suffice in a UAV mmWave context and proposed their own multiple access scheme for a system where the UAV act as relay nodes. Authors of \cite{zhao2018channel} proposed a novel channel tracking method for UAVs utilizing the on-board flight control system and positioning information in combination with a few pilot tones in an OFDM system.\\



 \subsection{Multi-Segment Operation in mmWave}
The study in \cite{li2019extensible} envisioned a multi-segment communications network to solve the problems of satellites satisfying the demands of huge capacity in ultra-dense regions as well as the frequent handovers caused by high movement speed of these satellites. \black{In particular, the multi-layer architecture comprised LEO mobile satellite systems for wide coverage including HAPs and terrestrial relays (TRs) being the access nodes for serving in hot-spot regions, and GEO satellites are introduced as routing nodes. Using HAPs and TRs can provide services to small user terminals which can not be directly connected with LEO satellites. Whereas, adopting GEO satellites for routing can reduce the complexity of routing, handover, and rerouting.
In \cite{li2020forecast}, the authors focused on the handover issue and propose a forecast based handover method for these multi-segment networks.
They studies a dynamic handover optimization framework for reducing the dropping probability while guaranteeing QoS of mobile terminals.
However, due to limited literature in this domain, additional studies are required to fully understand the behaviour of mmWave links integrated into multi-segment networks.}

\vspace{0.1in}
\noindent\textit{\black{\textbf{Key Takeaways} --  mmWave in NTNs can potentially provide services to areas where deployment of TNs will not be feasible. 
While satellite communications have long been utilizing mmWave frequencies (e.g., Ka-band) for fixed satellite services, their integration into cellular networks through mmWave links is being explored. However, this progression first requires to address various inherent challenges including large propagation delays, Doppler effects, etc. Also, satellite operations need to share its spectrum with existing services and the highly mobile stations need to be properly managed for providing ubiquitous coverage.\\
\indent Further, UAVs can augment cellular networks even further by providing on-demand and highly dynamic network infrastructure even at mmWave frequencies. Acting as base stations, relays, and mesh networks, these drones fulfill different high bandwidth requirements in a TN by conquering the skies. Nevertheless, this does not go without overcoming some underlying challenges. Blockage and beam misalignment, particularly, caused by UAV wobbling threaten the performance of a UAV mmWave link. Drones and their mobility are one of the main advantages over static ground stations, but its effects on the performance of a mmWave link is one of the main research problems that exist in the literature. 
Moreover, seamless vertical integration of the multi-segment networks require efficient network management techniques to facilitate the multi-layered architecture.\\
\indent In summary, mmWave has unarguably a great potential to revolutionize the broadband services in terrestrial segment. It can also be deployed on a global widespread scale through NTNs for delivering high throughput broadband access even in the middle of the Pacific. }}
\section{NTNs Integration in IoT}\label{sec:NTN_IoT} 

Over the last few years, the number of IoT devices connected worldwide is experiencing tremendous growth, and it is expected to reach 75 billion in 2025 \cite{marchese2019iot}. The IoT has potential to cover a wide range of applications including transportation, smart parking systems, smart lighting, health care, smart buildings, smart grid, smart wearable, to name a few. IoT networks can be broadly categorized into long-range and short-range networks, where long-range networks mainly cover outdoor applications. Among many long-range IoT technologies that have arisen in recent years, Sigfox, LoRa, and narrowband (NB)-IoT are today’s leading ones in the long-range category \cite{chaudhari2020lpwan}. In general, these technologies share the same goal of providing ubiquitous connectivity for IoT devices. However, in many cases, these devices are installed in rural environments where a terrestrial infrastructure does not exist, such as deserts, forests, mountains, and oceans \cite{iotsat1}, where the only means of communication is through NTNs. \black{As summarized in Fig.~\ref {fig:IOTstrc}, depending on the application, the integration of NTNs into IoT networks serves various purposes, detailed in the following.}

\tikzset{
  level/.style   = { line width=2pt, black },
  level2/.style   = { ultra thick, black},
  connect1/.style = { thick, blue },
  connect2/.style = { thick, dashed, blue },
  notice/.style  = { draw, rectangle callout, callout relative pointer={#1} },
  label/.style   = { text width=2cm }
}

\begin{figure*}
\begin{tikzpicture}
  \draw[level] (-1,0) -- node[above] {IoT} (1,0);

  \draw[connect1] (-1,0)  -- (-1.5,2) (-1,0) -- (-1.5,-2) (1,0) -- (1.5,2) (1,0) -- (1.5,-2);
  \draw[level2]   (-1.5,2)  -- node[above] {Background} node[below] {} (-3,2);
  \draw[level2]   (-1.5,-2) -- node[above] {UAV} node[below] {} (-3,-2);
  \draw[level2]   (1.5,2) -- node[above] {Satellite} node[below] {} (3,2);
  \draw[level2]   (1.5,-2) -- node[above] {Multi} node[below] {} (3,-2);

  \draw[connect2] (-3,2) -- (-3.5,2);
  \draw[level2]   (-3.5,2)  node[left] {\footnotesize Terrestrial Networks \cite{chaudhari2020lpwan}};
  
  \draw[connect2] (3,2) -- (3.5,2) (3,2) -- (3.5,1);
  \draw[level]   (3.5,2)  node[right] {\footnotesize Architecture and Standardization };
  \draw[level]   (3.5,1.6)  node[right] {\footnotesize  \cite{3GPP36763, liberg2020narrowband}};
  \draw[level]   (3.5,1) node[right] {\footnotesize Satellites-Aided IoT Connectivity \cite{satiotnewrcv,kodheli2020satellite}};
 
  \draw[connect2] (3,-2) -- (3.5,-2) (3,-2) -- (3.5,-1.5);
  \draw[level]   (3.5,-2)  node[right] { \footnotesize  Resource Allocation \cite{shi2020joint,liao2021learning}};
  \draw[level]   (3.5,-1.5)  node[right] { \footnotesize Backhauling and Data Offloading \cite{dai2020uav, cheng2019space}};
  
  \draw[connect2] (-3,-2) -- (-3.5,-2)  (-3,-2) -- (-3.5,-1.5)  (-3,-2) -- (-3.5,-1) (-3,-2) -- (-3.5,-0.5);
  \draw[level]    (-3.5,-0.5)  node[left] {\footnotesize UAV-aided Data Collection \cite{zhan2019energy,li2019joint,fu2020joint,saraereh2020performance}};
  \draw[level]    (-3.5,-1)  node[left] {\footnotesize UAV-enabled WPT for IoT \cite{feng2020uav,kang2020joint,jeong2020simultaneous}};
  \draw[level]    (-3.5,-1.5) node[left] {\footnotesize UAV-aided IoT Localization \cite{sallouha2018,sallouha2018energy,ebrahimi2020}};
  \draw[level]    (-3.5,-2) node[left] {\footnotesize IoT-aided UAV Localization \cite{sallouha2019g2a, CEDAR, lashkari2018crowdsourcing}};

\end{tikzpicture}
\caption{NTNs integration in IoT.}
   \label{fig:IOTstrc}
\end{figure*}
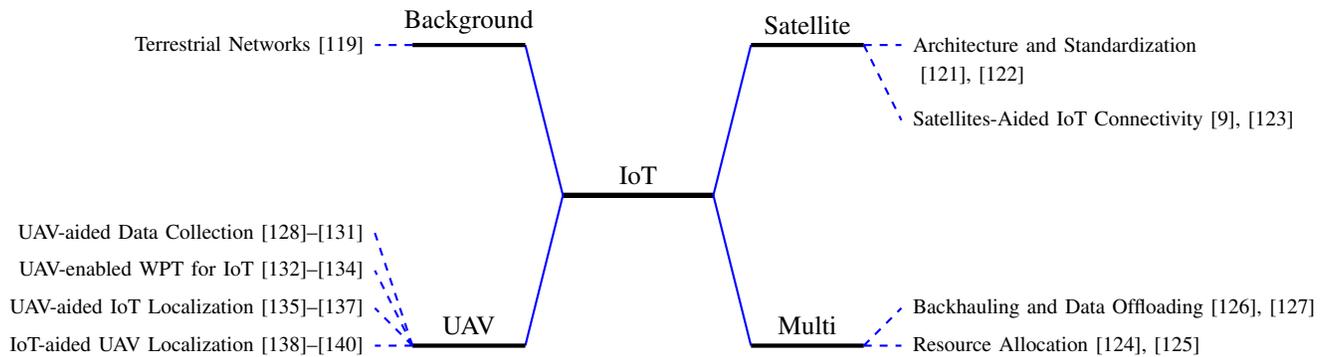





\subsection{Satellite Integration in IoT} \label{subsec:SatIOT}

    

One of the key performance indicators (KPIs) of IoT technologies, in general, is to ensure the connectivity of IoT devices throughout the globe. While UAVs offer a swift and flexible solution for IoT connectivity in areas that lack ground infrastructure, they typically have limited mission duration and coverage. GEO satellites overcome these limitations, providing global connectivity in cases of a temporary/permanent damaged terrestrial infrastructure (e.g., a natural disaster). Nevertheless, the main challenge imposed by the satellite channel is the extreme round trip delay in the communication link. The authors in \cite{iotsat6} provided a full performance analysis, including access and data phase, and propose solutions to counteract the increased delay in the GEO satellite channel. 

LEO satellites can also provide connectivity to IoT devices \cite{iotsat2}, offering smaller delays in the communication link compared to GEO and lower propagation path loses. While LEO orbits offer an advantage regarding the delay in the communication link, increased Doppler effects occur. This particular physical phenomenon arises as a result of the high-speed movement of the LEO satellite with respect to the IoT users on Earth. 
\textcolor{black}{In fact, this is also a typical impairment in a TN that occurs due to the movement of the users. Nevertheless, as it has been demonstrated in \cite{iotsat5} where the Doppler impact for NB-IoT communications over LEO have been thoroughly analysed, the Doppler shift are drastically larger.
In this context, in \cite{iotsat2}, a resource allocation technique has been proposed which reduces the differential Doppler in a LEO satellite channel down to a limit supported by the IoT devices.}

\subsubsection{Architecture and Standardization}
Different architecture options exist, as described in Section \ref{sec2a}, in order to collect the data generated by the IoT devices. A survey focusing on system architecture aspects of IoT communication via satellites under various orbits (GEO, MEO, LEO) can be found in \cite{satiotarch}. \textcolor{black}{It is worth highlighting that the architecture which is widely considered in the literature is where the satellite provides the direct access link to the terrestrial IoT users.} 

Obviously, depending on the orbital altitude of the satellite selected for operation, different impairments will be imposed in the IoT communication link which require solutions. The standardization and research effort so far has been mainly focused in modifying the existing terrestrial IoT technologies so as to be compatible even in the presence of a satellite channel. One of the main technologies that is widely addressed is the NB-IoT, which will be the one of the principal standards in the 5G to address the mMTC traffic. In fact, a study item has already started in 3GPP with the aim of studying NB-IoT via satellite communications \cite{3GPP36763}. The feasibility analysis of such an integrated system has been also performed in \cite{liberg2020narrowband}. \textcolor{black}{The main solutions proposed in the 3GPP studies in order to counteract the increased delay and Doppler effects over a NT link are based on a combination of GNSS capabilities and knowledge of the satellite ephemeris data. This allows to pre-compensate a large part of the added delay and Doppler effects by the NT channel, making them appear similar to the ones experienced over terrestrial links. } 

In parallel with the above-mentioned efforts, which mainly focus in adapting the existing NB-IoT protocol in order to counteract the satellite channel impairments, other works in the literature propose novel communication technologies. A new air interface for NB-IoT based on Turbo-frequency-shift-keying (FSK) modulation able to cope with large delays and Doppler effects has been proposed in \cite{satiotfsk}, whereas an NB-IoT receiver architecture for satellite-specific channel impairments has been analyzed in \cite{satiotnewrcv}.

\subsubsection{Satellites-Aided IoT Connectivity}
\black{The main role of satellites integrated into IoT networks, as defined by 3GPP \cite{3GPP22822,satiotnewrcv} is to realize the service ubiquity, continuity, and scalability in IoT networks. In terms of ubiquity, satellites offer ubiquitous connectivity as they can reach unserved areas or disaster-hit regions where the terrestrial networks are inapplicable, destructed, or in an outage by offering services from the space in a cost-effective manner. Moreover, satellites guarantees IoT services' continuity by providing connectivity to mobile IoT devices in areas where terrestrial infrastructure fails to reach such as in vessels, aircrafts, and trains. In terms of scalability, the vast coverage of satellites can on-demand complement ground IoT base stations when relatively high data rate is needed on a large scale, e.g., IoT nodes firmware updates. In \cite{kodheli2020satellite}, the authors categorized the use cases in which satellites serve IoT devices based on the size of the deployment area, namely wide-area and local-area services. In wide-area services, several satellites are used to provide connectivity to a group of IoT devices used in global transportation and agriculture applications. In local-area services, a satellite serves a specific set of IoT devices in applications such as a smart grid system.}



\subsection{UAV Integration in IoT}


\begin{table*}[t]
	\caption{IoT node localization with terrestrial, aerial, and space anchors.}
	\centering
	\begin{tabular}{| l || l | l | l |}
		\toprule
		\textbf{Aspect} & \textbf{Terrestrial} \cite{laoudias2018survey} &  \textbf{Aerial} \cite{sallouha2018,sallouha2018energy} & \textbf{Space (GPS)} \cite{kaplan2005understanding}\\
		\hline
		\hline
		\textbf{Localization Coverage} & Limited coverage & On-demand wide coverage &Ubiquitous coverage \\
		\hline
		\textbf{LoS Probability}  & NLoS-dominant  & Relatively high, Altitude-dependant &  LoS-dominant \\ 
		\hline
		\textbf{Accuracy} & Relatively low & Relatively high & Pinpoint accuracy\\
		\hline
		\textbf{Node's Power Consumption} & Relatively low & Relatively low & Excessive power consumption\\
		\hline
		\textbf{Node's Size and Cost} & No extra requirements & No extra requirements & Extra size and cost for the GPS radio\\
		\bottomrule
	\end{tabular}
	\label{tableLoC}
\end{table*}

The integration of UAVs into IoT networks offers substantial advantages for IoT devices as well as for UAVs. On the one hand, UAVs can provide connectivity to IoT devices in remote infrastructure-free areas, and on the other hand, the large-scale deployment of IoT devices can be exploited to extend the coverage of UAV traffic management in urban areas. \black{However, the integration of UAVs into IoT networks comes with several challenges that require careful attention. For instance, UAVs' limited battery life makes the trajectory planning an essential optimization problem for a successful UAV mission \cite{zhan2019energy,li2019joint,sallouha2018energy}. The trajectory planning problem does not only consider the energy-constrained but also the latency and reliability requirements of the corresponding IoT network. In the following, we discuss the main research works focusing on UAV integration into IoT.}

\subsubsection{UAV-Aided Data Collection}
UAVs offer swift data collection solutions due to their ability to reach remote, disaster, and poorly covered terrestrial areas on-demand, providing direct access to IoT devices. The deployment of UAVs for IoT nodes data collection is widely considered in the literature, addressing various aspects of the UAV mission such as energy efficiency \cite{zhan2019energy}, trajectory, and overall mission time \cite{li2019joint}. Depending on the IoT application, UAV data collection missions can be classified as offline and online. In offline missions, the UAV collects delay-insensitive data from IoT nodes, stores them onboard, and then delivers them to a terrestrial gateway \cite{li2019joint}. In online missions, on the other hand, the UAV acts as a mobile relay, forwarding data from IoT nodes to a terrestrial gateway in real-time \cite{fu2020joint}. Li \textit{et al.} \cite{li2019joint} introduced an offline UAV data collection mission, investigating the time minimization of a UAV data collection mission where the optimization of UAV's trajectory is studied. In particular, the authors transformed the time minimization problem to the trajectory length minimization problem. Subsequently, they decompose the problem into subproblems, namely, segment-based horizontal trajectory optimization, altitude optimization, velocity and link scheduling optimization.

An online data collection scenario is presented in \cite{fu2020joint} where multiple UAVs are used as relays between IoT devices and a terrestrial base station assumed at the center of the area of interest. A power-consumption-driven minimization problem is formulated with constraints on IoT devices' throughput. A UAV-enabled low-power wide area network (LPWAN) design for data collection from mobile IoT nodes is introduced in \cite{saraereh2020performance}. The authors consider a fleet of UAVs mesh networking with each other, promptly relaying the collected data to a centralized base station. To maintain UAVs' connectivity with the terrestrial base station a topology control algorithm is proposed.

\subsubsection{UAV-Enabled WPT for IoT} 
As UAVs facilitate realizing ubiquitous wireless connectivity to IoT devices, they can concurrently enable wireless power transfer (WPT) techniques for IoT devices by exploiting their transmitted RF signals. In \cite{feng2020uav}, a UAV with an antenna array onboard is employed to provide multiple RF beams, charging multiple users simultaneously. UAV trajectory and beam pattern optimization are considered with constraints imposed on UAV's altitude, coverage radius, and IoT nodes charging time. Subsequently, another trajectory design problem is investigated to provide wireless connectivity for the same users. Kang \textit{et al.} \cite{kang2020joint} investigated the optimization of UAV trajectory and transmit power, offering simultaneous wireless information and power transfer (SWIPT) for terrestrial terminals. A multicase scenario is proposed where users utilize a power splitting method, dividing the received signal into two parts for energy harvesting and information decoding. Multiuser UAV-enabled SWIPT is introduced in \cite{jeong2020simultaneous} where in addition to the common multicast stream, private streams for individual users are considered. The proposed SWIPT algorithm employs superposition coding and successive interference cancellation decoding, demonstrating the gains of the proposed multiuser communication. Furthermore, the authors formulated an optimization problem addressing the UAV trajectory design, power allocation, as well as power splitting ratio at IoT nodes.


\subsubsection{UAV-Aided IoT Localization}
IoT terrestrial devices are typically designed aiming at low-power consumption, low-cost and compact size. As a result, most IoT devices tend to lack a global positioning system (GPS) receiver, relying on size- and power-effective localization alternatives. Employing UAV BSs for terrestrial IoT nodes localization has been introduced in \cite{sallouha2018,sallouha2018energy,ebrahimi2020}, offering promising localization performance while meeting the IoT node's design constraints. Table \ref{tableLoC} presents a comparison between terrestrial, aerial, and space (i.e., GPS) anchors for IoT nodes localization from various aspects. 

\begin{figure}
		\centering
		\input{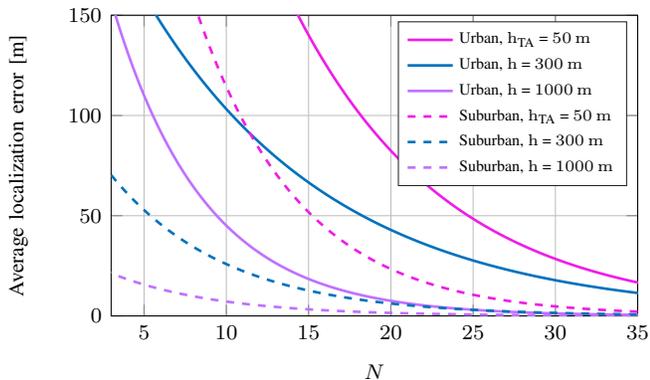}
		\caption{\black{Average localization error over uniformly distributed terrestrial nodes with aerial platform at an altitude of 1000 m versus the number of UAVs denoted by $N$ \cite{sallouha2018}.}}
		\label{AvLocErrN}
\end{figure}

A received signal strength (RSS)-based localization framework is proposed in \cite{sallouha2018} where the positioning of multiple hovering UAVs is investigated to minimize the average localization error of terrestrial IoT nodes. In particular, UAVs' altitude is identified as a novel design parameter to minimizes the localization error. \black{The reported results illustrated significant performance gains with UAV anchors flying at the optimal altitude compared to their terrestrial counterparts. 
This paper showed that in both urban and suburban environments, there exists an optimal altitude at which the average localization error is minimized. The reasoning behind this interesting trend in localization performance is two-fold. First, the localization error decreases as UAV altitude, denoted by $\mathrm{h}$, increases due to the decreasing shadowing effect's variance, which has a well-known direct relation with the localization error. Secondly, the logarithmic path loss-distance curve has a low resolution at high altitudes, making it significantly more sensitive to any relatively small variations in RSS measurements. In addition to the localization accuracy, the use of UAV anchors can significantly lower the number of anchors required to meet a given localization accuracy. Fig.~\ref{AvLocErrN} shows that the 150m accuracy that can be achieved with three UAVs at $\mathrm{h}$ = 1000\,m, requires 14 terrestrial anchors (TAs) with a typical height of $\mathrm{h_{TA}}$ = 50\,m, demonstrating the effectiveness of UAVs deployment for localization.}

The beneficial impact of UAV anchors on the localization performance is also explored in \cite{sallouha2018energy} where a single mobile UAV is considered. In addition to UAV's altitude, the optimization problem in \cite{sallouha2018energy} also considers the number of waypoints as well as the hovering time at each waypoint for an energy-constraint UAV mission. A tradeoff between UAV energy consumption and IoT nodes' localization accuracy is defined. Ebrahimi \textit{et al.} \cite{ebrahimi2020} optimized a mobile UAV mission for localizing IoT devices. They formulated a Markov decision process (MDP) while considering the overall UAV path design, including length, time, and waypoints. The proposed algorithm enables the UAV to autonomously construct its trajectory for maximum localization precision.


\subsubsection{IoT-Aided UAV Localization}
Modern IoT devices are designed to work with multiple technologies, e.g., Wi-Fi, Bluetooth, NB-IoT, SigFox, and LoRa, enabled by either separate radio-frequency (RF) front-ends or a tunable software-defined radio (SDR) with a shared RF front-end \cite{CALVOPALOMINO2020107231}. In fact, sensors ability to function over various frequency bands is a key enabler for crowdsourced-based applications. As a result, crowdsourced wireless networks (CWNs) are emerging as a new paradigm, offering unprecedented opportunities for utilizing the large-scale deployments of IoT networks in novel crowdsourced-based applications \cite{lashkari2018crowdsourcing}. UAV localization using CWNs is a particularly interesting application as it provides ubiquitous localization solutions. 

While technologies exist for UAVs to broadcast their locations \cite{minucci2020avoiding}, methods to verify the broadcast locations and complement such technologies are needed, addressing safety, anti-spoofing, and coexistence concerns. Yang \textit{et al.} \cite{CEDAR} introduced a cost-effective mobile crowdsourcing system for UAV detection and localization in urban environments. The proposed system relies on the RSS of UAV's Wi-Fi beacons captured by a crowd of terrestrial receivers to conduct a maximum likelihood estimation of the target UAV position. In \cite{sallouha2019g2a} a time difference of arrival (TDoA)-based multilateration localization method of UAVs using a non-coherent CWN is proposed. In order to achieve the time synchronization needed for TDoA, an autoregressive synchronization method in companion with a Kalman filter is employed.

\subsection{Multi-Segment Integration in IoT}

While integrating UAVs and satellites into IoT networks offers significant performance gains and interesting use-cases, as discussed earlier in this section, further improvements and applications can be realized by having a multi-segment network with integrated GAS communications \cite{shi2020joint,liao2021learning,dai2020uav,cheng2019space}. For instance, UAVs can act as a relay between IoT nodes and satellites since a direct uplink is generally power-inefficient for battery-powered IoT devices. Moreover, satellites can provide a backhaul solution for UAVs, saving crucial flight-time in cases where country- or continent-wide distributed IoT networks are considered.    

\subsubsection{Backhauling and Data Offloading}
The optimal communication performance of an integrated GAS IoT network is presented in \cite{dai2020uav}, where a UAV is deployed as a relay between satellites and terrestrial BSs to cooperatively serve terrestrial vehicles representing mobile IoT nodes. The optimization of both UAV trajectory and power consumption is investigated, providing vehicles with stable and reliable communication links. Cheng \textit{et al.} \cite{cheng2019space} proposed a flexible joint computing and communication space-air-ground IoT network, providing edge/cloud computing to remote IoT devices. An efficient computing offloading method is presented to minimize the sum-delay\footnote{The classical definition of network delay has been recently augmented by the concept of Age of Information (AoI) \cite{abd2019role}.} and energy consumption for IoT nodes. Fixed-wing UAVs are considered for edge computing services, whereas cloud services are offered through the satellite backbone network.   

\subsubsection{Resource Allocation}
Resource allocation in a GAS integrated IoT network is investigated in \cite{shi2020joint}, aiming at maximizing the spectral efficiency of UAV relaying cross-tier communication. In particular, an optimization problem for jointly gateway selection and resource allocation is formulated with the spectrum allocation among the ground-to-air, air-to-air, and air-to-space communication links taken into account. Liao \textit{et al.} \cite{liao2021learning} proposed a resource allocation and task offloading algorithms to minimize the energy consumption of IoT devices used in the smart grid oriented industrial IoT. The authors decomposed the optimization problem into three deterministic subproblems addressing: 1) task offloading; 2) resource allocation at the server-side; 3) task splitting and computational resource allocation at the device-side.\\

\noindent\textit{\black{{\textbf{Key Takeaways} -- We noted that NTNs play an indispensable role in realizing the ubiquitous connectivity envisioned for IoT networks. Satellites can guarantee the coverage and the continuous connectivity needed in IoT applications. In particular, GEO satellites can offer such global connectivity for applications that do not oppose a relatively long round trip time. Alternatively, LEO satellites provide a better delay experience on the cost of an increased Doppler effect. Resource allocation techniques can be used to address this Doppler effect. The use of UAVs in the air segment does not only address the connectivity problem in IoT networks but also provides substantial performance gains in terms of data collection, localization, and WPT. To realize these gains, the number of UAVs involved in the mission must be carefully selected, considering the deployment area and distribution of IoT devices. In addition, UAVs' altitude, trajectory, and the number of waypoints in their mission must be carefully optimized. Integrating both satellites and UAVs into IoT networks can potentially bring the advantages of both worlds. However, in such a multi-segment network, methods for backhauling and resource allocation must be introduced.}}}


\section{NTNs Integration in MEC Networks} \label{sec:NTN_MEC}

\tikzset{
  level/.style   = { line width=2pt, black },
  level2/.style   = { ultra thick, black},
  connect1/.style = { thick, blue },
  connect2/.style = { thick, dashed, blue },
  notice/.style  = { draw, rectangle callout, callout relative pointer={#1} },
  label/.style   = { text width=2cm }
}

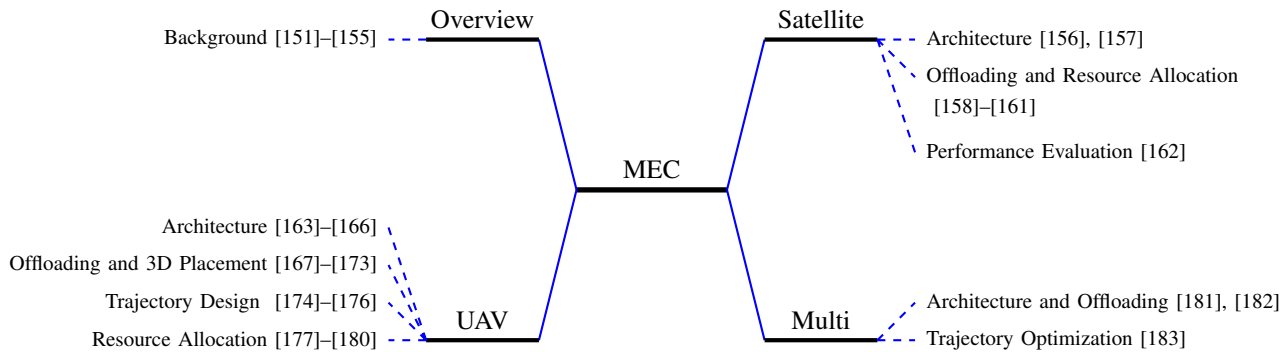
\begin{figure*}
    \centering
\begin{tikzpicture}
  \draw[level] (-1,0) -- node[above] {MEC} (1,0);

  \draw[connect1] (-1,0)  -- (-1.5,2) (-1,0) -- (-1.5,-2) (1,0) -- (1.5,2) (1,0) -- (1.5,-2);
  \draw[level2]   (-1.5,2)  -- node[above] {Overview} node[below] {} (-3,2);
  \draw[level2]   (-1.5,-2) -- node[above] {UAV} node[below] {} (-3,-2);
  \draw[level2]   (1.5,2) -- node[above] {Satellite} node[below] {} (3,2);
  \draw[level2]   (1.5,-2) -- node[above] {Multi} node[below] {} (3,-2);

  \draw[connect2] (-3,2) -- (-3.5,2);
  \draw[level]   (-3.5,2)  node[left] {\footnotesize Background \cite{mao17,taleb17,etsilight,keVTM17,JparkMEC21}};
  
  \draw[connect2] (3,2) -- (3.5,2) (3,2) -- (3.5,1.5) (3,2) -- (3.5,0.5) ;
  \draw[level]   (3.5,2)  node[right] {\footnotesize Architecture \cite{zhangNetw19, xieNetw20}};
  \draw[level]   (3.5,1.5) node[right] {\footnotesize Offloading and Resource Allocation};
  \draw[level]   (3.5,1.1) node[right] {\footnotesize \cite{wangICCS18,wangAccess20,cuiAccess20,Abderrahim20}};
  \draw[level]   (3.5,0.5) node[right] {\footnotesize Performance Evaluation \cite{kimVTC20}};

  \draw[connect2]  (-3,-2) -- (-3.5,-2)  (-3,-2) -- (-3.5,-1.5) (-3,-2) -- (-3.5,-1) (-3,-2) -- (-3.5,-0.5);
  \draw[level]    (-3.5,-0.5)  node[left] {\footnotesize  Architecture \cite{cheng18,zhouWC20,mostaani2020task, mostaani2021task}};
  \draw[level]    (-3.5,-1)  node[left] {\footnotesize  Offloading and 3D Placement \cite{zhang19,guo20,zhan20,motlagh17,messousTVT19,zhouTVT21,sunCL21}};
  \draw[level]    (-3.5,-1.5) node[left] {\footnotesize \color{black}{Trajectory Design } \cite{jeong18,cao18,tunCL21}};
  \draw[level]    (-3.5,-2.0) node[left] {\footnotesize \color{black}{Resource Allocation}  \cite{zhou18,hua18,hua19,li20}};
 
  \draw[connect2] (3,-2) -- (3.5,-2)   (3,-2) -- (3.5,-1.5);
  \draw[level]   (3.5,-1.5)  node[right] { \footnotesize  Architecture and Offloading \cite{zhouGlobecom19,zhangICC20}};
  \draw[level]   (3.5,-2)  node[right] { \footnotesize  Trajectory Optimization \cite{caoAccess19}};

\end{tikzpicture}
\caption{NTN integration in MEC.}
    \label{fig:MEC}
\end{figure*}
With the advent of IoT and evolution of 5G networks, the computing technology has witnessed a paradigm shift from centralized computing to edge computing. In this regard, MEC has been an emerging technology which aims at bringing the cloud computing facilities near the edge of a network \cite{mao17,taleb17}. Earlier, MEC has been referred to as mobile edge computing, but European telecommunications standards institute (ETSI) has revised it as multi-access edge computing to broaden its applications \cite{etsilight}. Nonetheless, mobile edge computing and multi-access edge computing are still being used interchangeably for MEC in the literature.  Owing to its proximity to the end-users and geographically distributed deployment, MEC can address the limitations of cloud computing to support the applications demanding delay-sensitive, computation-intensive, and high QoS requirements \cite{keVTM17,JparkMEC21}. Especially, uRLLC is one of the crucial KPIs of 5G system design, which can enable a plethora of new applications including autonomous driving, remote surgery, tactile internet, etc. MEC has been perceived to play a crucial role in facilitating these delay-sensitive applications \cite{mao17}. Such distinctive attributes of MEC have also driven an extensive research towards its applications in GAS networks. In what follows, we discuss various research works pursued in this direction. \black{A broader classification of these works is presented in Fig. \ref{fig:MEC}. Various aspects of MEC integration in terrestrial, aerial, and space networks have also been comparatively summarized in Table \ref{tableMEC}}.  

\subsection{Satellite Integration in MEC Networks}
\subsubsection{Architecture}
Ground-space networks have been recently explored for MEC applications to offload the heavy computational tasks of resource-limited densely distributed terrestrial terminal devices. \black{For example, a work in \cite{zhangNetw19} considered a satellite-borne offloading wherein MEC servers are deployed in LEO satellites. Such MEC servers can then enable computation offloading as well as content caching/storage for the terrestrial users. Building upon the similar idea, a research in \cite{xieNetw20} studied the architecture of the MEC assisted satellites and laid out the challenges and issues for the same.}
Although, the latency of satellite-assisted MEC can be higher compared with ground-air MEC, it may still offer significant improvement in latency performance against the remote cloud computing. Some existing works in literature studied various aspects of the ground-space MEC. 
\subsubsection{Offloading and Resource Allocation}
In \cite{wangICCS18}, the authors introduced a computation offloading with double edge computing in ground-space networks. In particular, the computation tasks are offloaded to either a LEO satellite deployed with MEC server or terrestrial MEC depending on a certain threshold mechanism. 
\black{Wang \textit{et al.} \cite{wangAccess20} proposed a game theory based approach to optimize the computation offloading from remote terrestrial mobile users to a satellite MEC server. The average response time and average energy consumption of a task have been considered as performance metrics. The developed algorithm then enables mobile device to properly allocate the tasks and also to effectively  use local and on-board resources.}
Authors in \cite{cuiAccess20} studied a MEC-assisted satellite IoT networks. \black{In particular, for a satellite IoT networks including multiple satellites and gateways, offloading decision, user association, computing and communication resource allocation have been jointly optimized for minimizing the latency and energy cost.}
In \cite{Abderrahim20}, the authors considered an integrated ground-space network in which a traffic offloading scheme is proposed such that uRLLC traffic is offloaded to the terrestrial segment, whereas, eMBB traffic is offloaded to the satellite.
\subsubsection{Performance Evaluation}
An analytic study was presented in \cite{kimVTC20} to evaluate the performance of satellite MEC network. \black{Specifically, based on constellation model with undirect graph representation, the authors considered propagation and queuing delays with uplink/downlink packet error rate for performance analysis. They investigated the average total latency and jitter with respect to the ground separation between transmitter and receiver for different satellite altitudes and offloading rate.
Following the analysis, the authors proposed guidelines for the network architecture of the satellite MEC server using the considered constellation model. This study noted that latency decreases with the offloading rate while it increases with altitude.}



\subsection{UAV Integration in MEC Networks}
\subsubsection{Architecture}
Recently, UAVs have been explored for their possible applications in MEC networks \cite{cheng18,zhouWC20}. Specifically, UAVs can be exploited as a MEC server for computation of offloaded tasks from the terrestrial nodes \cite{zhouWC20}. On the other hand, certain UAV applications such as disaster resilient networks, BVLoS navigation, and dynamic geofencing etc., require extensive storage and processing capabilities, which may increase the payload and power consumption of UAVs and eventually shorten their battery life. \textcolor{black}{To address this challenge, task-oriented communication can be considered as a candidate \cite{mostaani2020task, mostaani2021task}, where the power dedicated to the communication of IoT devices and UAVs can be optimized, refer to section \ref{subsubsect: AI-MEC} to learn more.} For handling this issue, computation intensive tasks can be offloaded to ground MEC servers for the processing \cite{cheng18}. 


\subsubsection{Offloading and 3D Placement}

 \black{For UAV-assisted MEC system, average weighted energy consumption of mobile devices and UAV was minimized in \cite{zhang19} based on a stochastic queue model. Various constraints included number of offloaded computation tasks, CPU cycle frequency of devices and UAV, and trajectory scheduling.} Guo \textit{et al.} \cite{guo20} studied a UAV-enhanced edge network wherein an UAV moves above the locations with high density IoT mobile devices. Energy consumption required for the computational tasks at mobile devices has been minimized under a binary offloading strategy. \black{Authors in \cite{zhan20} studied the joint design of offloading, resource allocation, and trajectory to minimize the task completion time and energy consumption of UAV-enabled MEC. They also unveiled a trade-off between completion time and energy consumption of the UAV.}
 In \cite{motlagh17}, the authors presented a crowd surveillance use case for UAV-based IoT networks. Specifically, they compared the performance of UAVs onboard processing of video data with the offloading to a MEC node in terms of energy consumption and processing time using an experimental setup.
  \black{The authors in \cite{messousTVT19} proposed a game theory based solution for offloading the UAVs task while ensuring a tradeoff between the delay, cost, and energy consumption. They considered offloading complex tasks to a fleet of UAVs to reduce the execution delay while optimizing the energy overhead.}
 More recently, Zhou \textit{et al.} \cite{zhouTVT21} proposed a method to jointly optimize the mobility, communication, and computation for a UAV in MEC-assisted air-ground cooperative network to maximize the UAV's energy efficiency. The work in \cite{sunCL21} aimed at minimizing the execution time required for UAVs to complete the offloaded tasks by optimizing the UAV's 3D location.

\subsubsection{Trajectory Design}
In \cite{jeong18}, the authors studied a UAV-mounted cloudlet assisted MEC system to minimize the energy consumption of the mobile users. Specifically, joint optimization of bit allocation and UAV's trajectory had been performed to demonstrate the energy savings at the mobile device compared with local computation and partial offloading. 
In \cite{cao18}, the authors considered a scenario where a mobile UAV offloads the data to MEC server at ground base station. This work focused on minimizing the mission completion time of UAV by optimizing its trajectory along with the computation offloading.
 Tun \textit{et al.} \cite{tunCL21} considered the problem of jointly minimizing the energy consumption of UAVs as well as IoT devices in UAV-aided MEC system while optimizing the offloading decisions, resource allocation, and trajectory.

\subsubsection{Resource Allocation}
The authors in \cite{zhou18} analyzed a computation rate maximization problem for the MEC-enabled wireless powered networks under two schemes, viz., partial and binary offloading. 
\black{In \cite{hua19}, the authors considered UAV as a flying BS to execute the delegated tasks from terminal devices to save their energy consumption. A one-by-one access scheme has been proposed such that one portion of bits is computed locally at the device while other is offloaded to UAV for computation. For such a framework, total energy consumption of multiple terminal devices has been minimized. In a similar line, a study in \cite{li20} considered an aerial cloudlet at UAV to carry out the computations for ground users. For this setup, the energy efficiency of UAV has been maximized while optimizing the computation offloading.}

\subsection{Multi-Segment Integration in MEC Networks}
\subsubsection{Architecture and Offloading}
Integrated GAS network forms a multidimensional heterogeneous architecture which can potentially cater the offloading services by opportunistically accessing the different network segments in a flexible manner.
A few research works have recently focused on such architecture for MEC-aided offloading services. For instance, GAS has been considered for MEC assisted offloading in IoT applications \cite{zhouGlobecom19}. Specifically, the collected task from the IoT devices is processed locally at UAV or offloaded to a satellite/ground edge server using a linear programming based scheduling. Zhang \textit{et al.} \cite{zhangICC20} introduced a air-space integrated computing architecture for disaster applications. This work considered that the tasks from the terrestrial/aerial users are either computed locally at the HAPs or offloaded for computation at LEO satellite. 
\subsubsection{Trajectory Optimization}
In \cite{caoAccess19}, a MEC-driven GAS network has been presented for the routine inspection in wind farms. Specifically, the UAVs detect the wind turbines and can optimally offload the sensory data to ground station or satellite. \black{For minimizing the total completion time, the joint optimization of UAV trajectory and computations has been performed while ensuring the data processing accuracy. } \\

\noindent\textit{\black{\textbf{Key Takeaways} -- Based on the survey in  this section, it is clear that application of MEC in NTN can offer several benefits essentially for the resource limited devices on the ground or in air. For NTN, especially aerial networks, the use of MEC is quite promising which can overcome the limitation of onboard computation for UAVs to increase their operational lifetime by reducing the energy consumption. UAVs can utilize the MEC servers on ground or in space for offloading the heavy tasks. Though in contrast to TN MEC, offloading to satellite-MEC can have higher latency, nonetheless, it is very useful for the applications where TN is not accessible. However, the satellite-MEC may lead to architectural complexity and thereby increased cost. The additional challenges for implementing satellite-MEC may include difficulty in timely maintenance of the servers. Moreover, UAV-assisted MEC operations can also be beneficial by flexibly deploying them in complex terrains or deserts where TN MEC is not available. Such deployment face the challenge of limited payload capacity and energy consumption of the UAVs.  }}
\begin{table*}[h!]
\color{black}
	\caption{\black{NTN integration in MEC.} }
	\centering
	\begin{tabular}{| c || c | c | c |}
		\toprule
		\textbf{Aspect} & \textbf{Terrestrial \cite{mao17,taleb17}} &  \textbf{Aerial \cite{cheng18,zhouWC20}} & \textbf{Space \cite{zhangNetw19,xieNetw20}}\\[0.5ex]
		\hline \hline
		\textbf{Computation Capability} & High & Low & Medium \\
		\hline
		\textbf{Regulatory Hurdles}  & Low  & High &  Moderate \\ 
		\hline
		\textbf{Key Driving Factor} & Low latency & Flexible on-demand deployment & Widespread presence in remote locations\\
		\hline
		\multirow{2}{*}{\textbf{Limiting factors}} & Integration with &Limited payload capacity &Energy consumption\\
    &existing infrastructure &and energy consumption&and high cost\\
		\hline
		\textbf{Architectural complexity and Cost} & Moderate & Low & High\\
		\bottomrule
	\end{tabular}
	\label{tableMEC}
\end{table*}



\section{ML-Empowered NTNs 
} \label{sec:NTN_ML}

In this section, we first briefly elaborate on the ML concept and then we review the related literature. \black{The discussed topics along with the related literature are summarized and categorized in Fig.~\ref{fig:ML_taxonomy}.}

\subsection{Overview on ML
}



Machine learning algorithms are referred to as a collection of tools and algorithms that are used to generate versatile models from large amount of data. According to \cite{domingos2012few}, “Machine learning algorithms can figure out how to perform important tasks by generalizing from examples” and they do that without being explicitly programmed for it \cite{koza1996automated} or without relying on rules-based programming. Machine learning experts recognize it as a rich field with very large themes, patterns and application areas. Understanding such themes will be beneficial to those who wish to apply machine learning to their applied problem. Used in numerous applications, machine learning algorithms are data-driven nor requiring explicit modeling compared with with conventionally engineered algorithms. Often, the existing machine learning algorithms are categorized based on the type of feedback that the learning system has access to. Most machine learning algorithms fall into the categories of \textit{supervised, unsupervised} and \textit{reinforcement learning (RL)} \cite{abu2012learning}, which are explained in the sequel.

\subsubsection{Supervised Learning}
Supervised learning algorithms are identified by the presence of a set of input and output data pairs, often referred to as the training data set or the labeled data set. The task of supervised learning algorithms is to find a proper mapping between the input data (vectors) and the output data. The available labeled data help supervised ML algorithms to fine-tune the parameter values of the parameterized model. The trained model can then be used to map unseen instances of input data (vectors) to output data. Some of the more famous supervised learning algorithms are support vector machine (SVM), linear/logistic regression, decision trees and deep neural networks (DNNs). Examples of supervised learning tasks in (satellite) communication systems include but are not limited to channel equalization \cite{bouchired1998equalisation}, demodulation \cite{meng2018automatic}, decoding \cite{gruber2017deep}, remote sensing \cite{li2020nasa} and network traffic control to mention a few \cite{8088549}. For further readings and insights about supervised learning we would refer the readers to \cite{simeone2018very,simeone2017brief,jiang2016machine} and to \cite{wang2017deep} for deep learning and its applications in wireless communications.

\subsubsection{Unsupervised Learning}
Unsupervised learning (UL) methods are automated algorithms that are designed to learn the existing patterns and features from the available unlabeled raw data. As though, the algorithm must discover the underlying characteristics and features of the available data, such that these inferred features can help to better understand, process or represent our data. Example tasks that are usually targeted by UL techniques are data partitioning/clustering, anomaly detection, latent variable learning, generation of authentic-looking data samples and feature construction. Principal component analysis, together with data clustering are two of the largely deployed classes of UL in satellite/airborne communication systems \cite{hoang2019detection,ziluan2018short,wang2018robust}. These algorithms are often used for feature extraction and dimensionality reduction tasks 
which help to reduce the computational cost of processing the data with the least possible compromise in the performance of the data processing task
. Neural networks have also proven outstandingly helpful in many different types of communication systems tasks, with auto-encoders being utilized in PHY \cite{sadeghi2019physical, felix2018ofdm, jiang2019turbo} and a variety of other deep learning techniques used in remote sensing \cite{al2019audio, osco2021review,opromolla2019airborne} and security \cite{bao2020secrecy} applications of airborne/satellite communications. 
For further readings and insights we would refer the interested readers to explore \cite{simeone2018very,simeone2017brief}.

\subsubsection{Reinforcement Learning}
Although supervised learning algorithms have proven efficient in solving many real world problems, the high cost of obtaining large labeled training and test data sets might become prohibitive factors towards their applicability. Reinforcement learning methods can instead, offer a plethora of different algorithms where we can benefit from a trial and error approach to solve sophisticated sequential decision making problems. Reinforcement learning agents are expected to maximize the expectation of their cumulative future rewards which are obtained through interaction with the environment \cite{sutton2018reinforcement}, trial and error process. Maximization of future rewards is achieved by the agent through the selection of appropriate actions for any interaction with environment given the current(/history of) environment state(s). Environment is usually modelled as a Markov decision process (MDP) or other mathematical models e.g., partially observable MDP (POMDP), and decentralized POMDPs \cite{amato2013decentralized}. These mathematical models of the environment allow RL researchers to design the action selection policy of agents with certain performance guarantees. State of the art RL algorithms has also harnessed the power of deep learning by using DNNs to approximate the action policy function of the agent or to approximate the Q-function (that computes the expected cumulative reward of the agent given current state-action pair) \cite{mnih2015human, lillicrap2015continuous}. After the success that deep reinforcement learning (DRL) has achieved by showing beyond human performance, it has started to be used as a solver of sequential decision making problems in many engineering fields including (satellite) communications \cite{zhouTWC21, deng2019next, yan2020delay, zhao2020deep, hu2020dynamic}. Multiagent reinforcement learning also comes to the picture, when the action vector is jointly selected not by a single agent but by a multitude of entities (agents). This becomes especially helpful when the performing a particular task in a distributed fashion becomes of essence \cite{zhu21, hu2020distributed, hu2020cooperative, wu2020cellular}. For further discussions on the applications of reinforcement learning in telecommunication systems interested readers can see
\cite{feriani2021single,ali20206g}.

\subsubsection{Where to/not to apply machine learning}

A useful set of criteria is offered by \cite{simeone2018very} on the type of problems that can be solved using supervised and unsupervised ML tools. On top of these criteria, there are a number of further limitations concerning the application of RL algorithms in real world problems.  One of the main concerns is the poor performance of the RL during the training phase. Accordingly, it is projected to have a new line of applied research where the focus is limiting the bad performance of the RL agent within the exploration phase \cite{vannella2021remote, nayak2020routing}. Towards this goal, the available literature on safe RL \cite{garcia2015comprehensive}, batch RL \cite{lange2012batch, le2019batch} and off-policy RL \cite{munos2016safe, fujimoto2019off} are of enough significance meriting special attention. 

As mentioned earlier (see Figure \ref{fig:NTNarchitecture}), in general the network architecture includes four entities which are ground/air/space users, BSs, core network, and data network. Depending on the entities engagement the ML applications can be categorized into three levels \cite{samsung6g}: i) local ML, ii) joint ML, and iii) end-to-end (E2E) ML. Local ML is implemented in each entity, for instance optimization of channel coding. A Joint ML refers to a joint operation of two entities such as users and BSs in the optimization of handover based on prediction of future network conditions. E2E ML optimizes the entire communication system through four entities. Accordingly, it becomes possible to identify or anticipate anomalies in network operation and propose corrective actions.
    



\subsection{ML-Empowered Satellite Networks}

In this subsection we review important applications of ML in satellite-based communication and networking. 

\subsubsection{Resource Allocation}

Scarce and expensive satellite resources should be optimized for an improved system performance. The use of ML for such concern is investigated in recent works \cite{jia2020intelligent,deng2019next,ferreira2018multiobjective,nie2019deep,jiang2020reinforcement,yan2020delay,qiu2019deep,ferreira2019reinforcement,lei2020beam,zhao2020deep,cui2020latency,tsuchida2020efficient}. The feasibility and performance of spectrum sharing between satellite and terrestrial networks are studied in \cite{jia2020intelligent}. To improve spectrum efficiency, the authors present intelligent resource management mechanism following sensing, prediction and allocation steps, and given the users priorities and requirements. In this work, the accuracy of spectrum occupancy detection is improved using SVM where convolutional neural network (CNN) is applied to spectrum detection problem. The next generation of heterogeneous satellite systems architectures and intelligent collaboration between satellites are addressed in \cite{deng2019next}. In this work, the potential of DRL in achieving efficient resource allocation is shown. Furthermore, multi-objective RL and artificial neural network based algorithms have been used in \cite{ferreira2018multiobjective} to manage satellite resources and conflicting mission-based targets. 
In \cite{nie2019deep} the multi-band communication in cubesats systems, operating in microwave, mmWave, or THz band, is considered. A multi-objective resource allocation scheme based on DNN is proposed, which takes into account the limited computation and energy budget, and the Doppler effect of mobile CubeSats. A multi-layer heterogeneous satellite network is considered in \cite{jiang2020reinforcement} with intra and inter connectivity links. In order to maximize the network capacity the author propose a Q-learning based algorithm to optimize a long-term utility function. 

\subsubsection{Beam Hopping}
In a conventional satellite network there might exist a mismatch between requested traffic and offered capacity through the satellite beams. Beam hopping technique is a promising solution for this issue to manage asymmetric and variant traffic demands. Using this technique, a subset of active beams is dynamically selected following the time-varying request. The selection of proper subset of beams and the duration that each beam is active, are challenging tasks addressed in several works \cite{hu2020dynamic,lei2020beam}. 
In order to increase the system throughput in satellite-ground network and reduce the transmission delay, beam hopping illumination plan is formulated in \cite{hu2020dynamic}. The authors model the problem as a POMDP and solve it using DRL by taking into account several realistic conditions such as inter-beam interference and spatial-temporal feature of traffic. To optimize the resource allocation efficiently and timely, without violating the system constraints, a joint learning and optimization approach is taken in \cite{lei2020beam}. 

\subsubsection{IoT}
The benefits of satellite integration into IoT networks are elaborated in Section \ref{subsec:SatIOT}. Satellite-assisted ground IoT network in downlink is considered in \cite{yan2020delay}. The authors formulate the problem of optimal power allocation strategy for non-orthogonal multiple access (NOMA) users under minimum rate and delay-QoS requirements for each user. To efficiently allocate the resources for maximum sum effective capacity, DRL technique is adopted. The performance superiority is proved as compared to fixed power allocation strategy and time division multiple access (TDMA) scheme. Using DRL approach an energy efficient channel allocation algorithm in LEO satellite IoT is presented in \cite{zhao2020deep}. The dynamic characteristic of LEO satellites is modeled and novel methods are taken to reduce the size of action set and speed up the learning process. In order to minimize energy cost and latency, affected by several factors including user association and resource allocation for computing and communication, an optimization problem is formulated in \cite{cui2020latency}. To handle the complexity of the problem, DRL technique is partially used to solve joint user association and offloading decision sub-problem. 
To prolong the lifetime of LEO satellites, in \cite{tsuchida2020efficient} the author proposed to deploy adjacent satellites with lower workload to assist the overloaded satellites serving ground users. A Q-learning approach is adopted to dynamically allocate power based on the satellite battery status and traffic volume.



\subsubsection{MEC}
The authors in \cite{wei19} studied a satellite IoT edge computing with deep learning to coordinate with the satellite IoT cloud node and terrestrial cloud center for computation offloading. \black{Edge computing is jointly considered with networking and caching in \cite{qiu2019deep}. In this paper the resource allocation is formulated as a joint optimization problem. To solve this, the problem is described as a Markov decision process for which action and state space along with relevant reward function are defined. Then the authors apply deep Q-learning method to learn the optimal strategies for the resource allocation. The effectiveness of the results are shown through a comparative study with static and individual schemes.} 
\black{\subsubsection{Handover and Interference Management}
As mentioned earlier, HO is one of the challenges in satellite-based systems. In \cite{xu2020qoe}, the problem of handover and user's QoE is addressed. Given the periodic motion of satellites, first the HO factors are anticipated. Then the author applies RL algorithm to optimally make HO decisions. In this work, the spatial correlation between user and satellite movements is modeled to estimate the channel for an improved HO success rate and lower end-to-end delay. 
The problem of interference management is addressed in \cite{henarejos2019deep}. Using IQ samples and by adopting deep learning approach, interference from different ground sources is detected and classified in the presence of satellite systems. In order to control and reduce interference effect of other stations, pointing and tracking of mobile ground terminals served by satellites are studied in \cite{liu2019artificial} using artificial intelligence self learning.}
\black{\subsubsection{Channel Modeling}
Machine learning in combination with satellites can be used for channel modeling as well. The drawback of computationally complex tools such as ray tracing simulator, which makes them improper for real-time coverage optimization, is addressed in \cite{ates2019path} using deep learning. The authors apply deep convolutional neural network to estimate channel parameters such as path loss exponent and shadowing standard deviation from two-dimensional (2D) satellite images, i.e. without the need of 3D model of the region. Same approach in path loss prediction using deep learning is adopted in \cite{thrane2020model, ahmadien2020predicting}. Although such approach is very practical, 2D images may not suffice in some cases such as low altitude and high frequency transmitters in a dense urban areas \cite{ahmadien2020predicting}. In this case in addition to the 2D images the building heights map might be required. Furthermore, for the training phase still the dataset relies on the 3D ray tracing simulators.}



\tikzset{
  level/.style   = { line width=2pt, black },
  level2/.style   = { ultra thick, black},
  connect1/.style = { thick, blue },
  connect2/.style = { thick, dashed, blue },
  notice/.style  = { draw, rectangle callout, callout relative pointer={#1} },
  label/.style   = { text width=2cm }
}

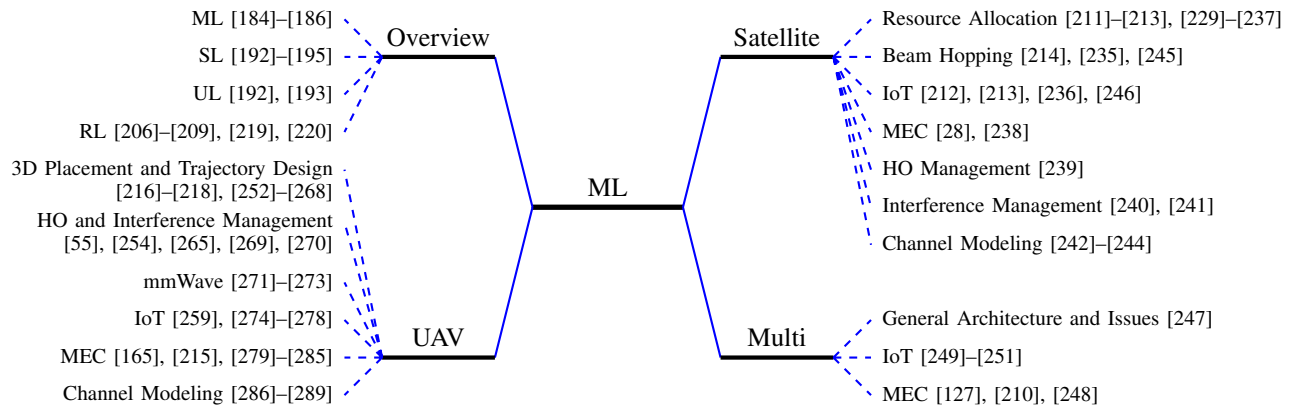
\begin{figure*}
    \centering
\begin{tikzpicture}
  \draw[level] (-1,0) -- node[above] {ML} (1,0);

  \draw[connect1] (-1,0)  -- (-1.5,2) (-1,0) -- (-1.5,-2) (1,0) -- (1.5,2) (1,0) -- (1.5,-2);
  \draw[level2]   (-1.5,2)  -- node[above] {Overview} node[below] {} (-3,2);
  \draw[level2]   (-1.5,-2) -- node[above] {UAV} node[below] {} (-3,-2);
  \draw[level2]   (1.5,2) -- node[above] {Satellite} node[below] {} (3,2);
  \draw[level2]   (1.5,-2) -- node[above] {Multi} node[below] {} (3,-2);

  \draw[connect2] (-3,2) -- (-3.5,2) (-3,2) -- (-3.5,1.5) (-3,2) -- (-3.5,2.5) (-3,2) -- (-3.5,1);
  \draw[level]   (-3.5,2.5)  node[left] {\footnotesize ML \cite{domingos2012few,koza1996automated,abu2012learning}};
  \draw[level]   (-3.5,2)  node[left] {\footnotesize SL \cite{simeone2018very,simeone2017brief,jiang2016machine,wang2017deep}};
  \draw[level]   (-3.5,1.5) node[left] {\footnotesize UL \cite{simeone2018very,simeone2017brief}};
  \draw[level]   (-3.5,1) node[left] {\footnotesize RL \cite{sutton2018reinforcement,amato2013decentralized,mnih2015human, lillicrap2015continuous,feriani2021single,ali20206g}};
  
  \draw[connect2] (3,2) -- (3.5,2.5) (3,2) -- (3.5,2) (3,2) -- (3.5,1.5)  (3,2) -- (3.5,1) (3,2) -- (3.5,0.5) (3,2) -- (3.5,0) (3,2) -- (3.5,-0.5);
  \draw[level]   (3.5,2.5)  node[right] {\footnotesize Resource Allocation \cite{jia2020intelligent,deng2019next,ferreira2018multiobjective,nie2019deep,jiang2020reinforcement,yan2020delay,qiu2019deep,ferreira2019reinforcement,lei2020beam,zhao2020deep,cui2020latency,tsuchida2020efficient}};
  \draw[level]   (3.5,2)  node[right] {\footnotesize Beam Hopping \cite{lei2020beam,hu2019deep,hu2020dynamic}};
  \draw[level]   (3.5,1.5) node[right] {\footnotesize IoT \cite{zhao2020deep,cui2020latency,yan2020delay,wang2019reinforcement}};
  \draw[level]   (3.5,1) node[right] {\footnotesize MEC \cite{xie2020satellite,wei19}};
  \draw[level]   (3.5,0.5) node[right] {\footnotesize HO Management \cite{xu2020qoe}};
  \draw[level]   (3.5,0) node[right] {\footnotesize Interference Management \cite{henarejos2019deep,liu2019artificial}};
  \draw[level]   (3.5,-0.5) node[right] {\footnotesize Channel  Modeling \cite{ates2019path,thrane2020model,ahmadien2020predicting}};

    \draw[connect2] (3,-2) -- (3.5,-1.5) (3,-2) -- (3.5,-2) (3,-2) -- (3.5,-2.5) ;
    \draw[level]   (3.5,-1.5)  node[right] {\footnotesize General Architecture and Issues \cite{kato2019optimizing}};
  \draw[level]   (3.5,-2.5)  node[right] {\footnotesize MEC \cite{zhouTWC21,cheng2019space,liaoIOT21}};
  \draw[level]   (3.5,-2)  node[right] {\footnotesize IoT \cite{michailidis2020ai,zhou2020deep,gu2020coded}};
  
  \draw[connect2] (-3,-2) -- (-3.5,-2) (-3,-2) -- (-3.5,-1.5) (-3,-2) -- (-3.5,-2.5) (-3,-2) -- (-3.5,-1) (-3,-2) -- (-3.5,-0.2) (-3,-2) -- (-3.5,0.5) ;
  \draw[level]    (-3.5,0.5)  node[left] {\footnotesize  3D Placement and Trajectory Design };
  \draw[level]    (-3.5,0.2)  node[left] {\footnotesize  \cite{hu2020reinforcement,liu2019trajectory,challita2019interference,shakoor2020joint,koushik2019deep,hu2020distributed,garg2020machine,mamaghani2020intelligent,bayerlein2020uav,hu2020cooperative,cui2020adaptive,li2020intelligent,wu2020cellular,liu2020machine,susarla2020learning,khamidehi2020double,zeng2019path,liu2019reinforcement,liu2019distributed,huang2019deep}};
  \draw[level]    (-3.5,-0.2)  node[left] {\footnotesize HO and Interference Management};
  \draw[level]    (-3.5,-0.5)  node[left] {\footnotesize \cite{azari2020mobile,arani2020learning,chen2020deep,zeng2019path,challita2019interference}};
  \draw[level]    (-3.5,-1) node[left] {\footnotesize mmWave \cite{li2020millimeter,yuan2020learning,moorthy2020beam}};
  \draw[level]    (-3.5,-1.5) node[left] {\footnotesize IoT \cite{bayerlein2020uav,abedin2020data,yi2020deep,sun2020learning,yang2020multi,cao2019deep}};
  \draw[level]    (-3.5,-2) node[left] {\footnotesize MEC \cite{liu20,sun21,zhangAccess21,zhu21,Cao2020,mostaani2019learning,mostaani2020task,liu120,yangIoT20}};
  \draw[level]    (-3.5,-2.5) node[left] {\footnotesize Channel Modeling \cite{wang2019machine,xia2020generative,zhang2018air,yang2019machine}} ;
;
\end{tikzpicture}
\caption{\black{ML approach for addressing various aspects of NTNs.}} 
    \label{fig:ML_taxonomy}
\end{figure*}


\subsection{ML-Empowered UAV Networks}

In the following we specify the problems that have been dealt using ML approach.

\subsubsection{3D Placement and Trajectory Design}
Different RL based algorithms and corresponding applications for cellular internet of UAVs, i.e. multi-armed bandit learning for user association, Q-learning for trajectory design, actor-critic learning for power management, and deep RL for subchannel allocation, along with their pros and cons are discussed in \cite{hu2020reinforcement}. Trajectory of multiple UAVs along with their transmit power are optimized in the presence of mobile ground users in \cite{liu2019trajectory}. The authors apply RL Q-learning algorithm to dynamically adjust the position of UAVs to maximize the network sum-rate. \black{Optimal path planning of multiple cellular-connected UAVs is considered in \cite{challita2019interference}. The objective is to minimize the interference caused by UAVs and the transmission latency. To optimize the trajectory, a deep RL framework based on echo estate network cells is examined.} 

\subsubsection{\black{HO and Interference Management}}
\begin{figure}
\centering
\begin{subfigure}{\columnwidth}
\centering
\includegraphics[width=\columnwidth]{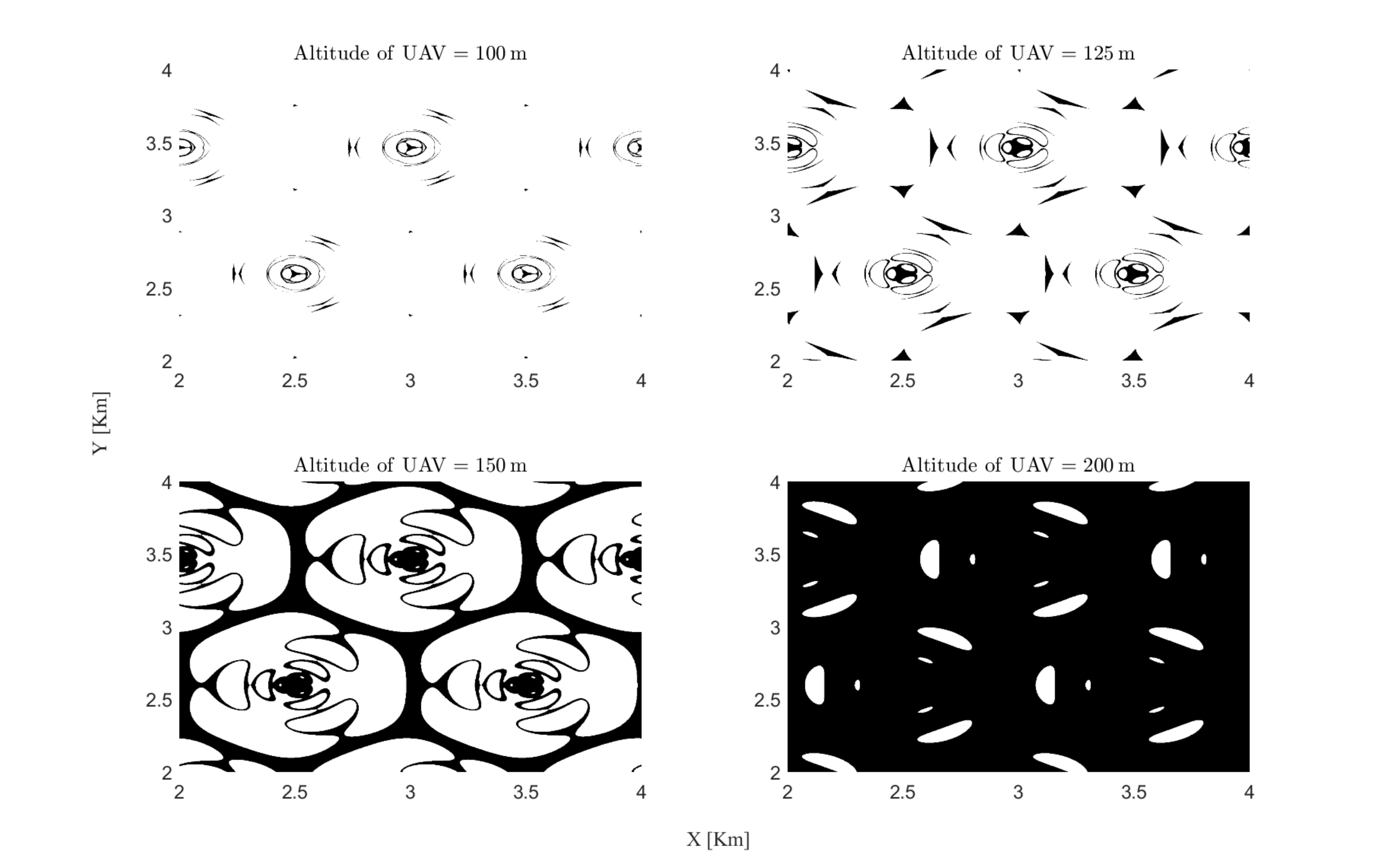}
	\caption{\black{In black regions a target rate of 100\,kbps over one assigned physical resource block can not be satisfied.}}
	\label{CoverageHoles_hUs}
\end{subfigure}
\newline
\begin{subfigure}{\columnwidth}
\centering
\includegraphics[width=\columnwidth]{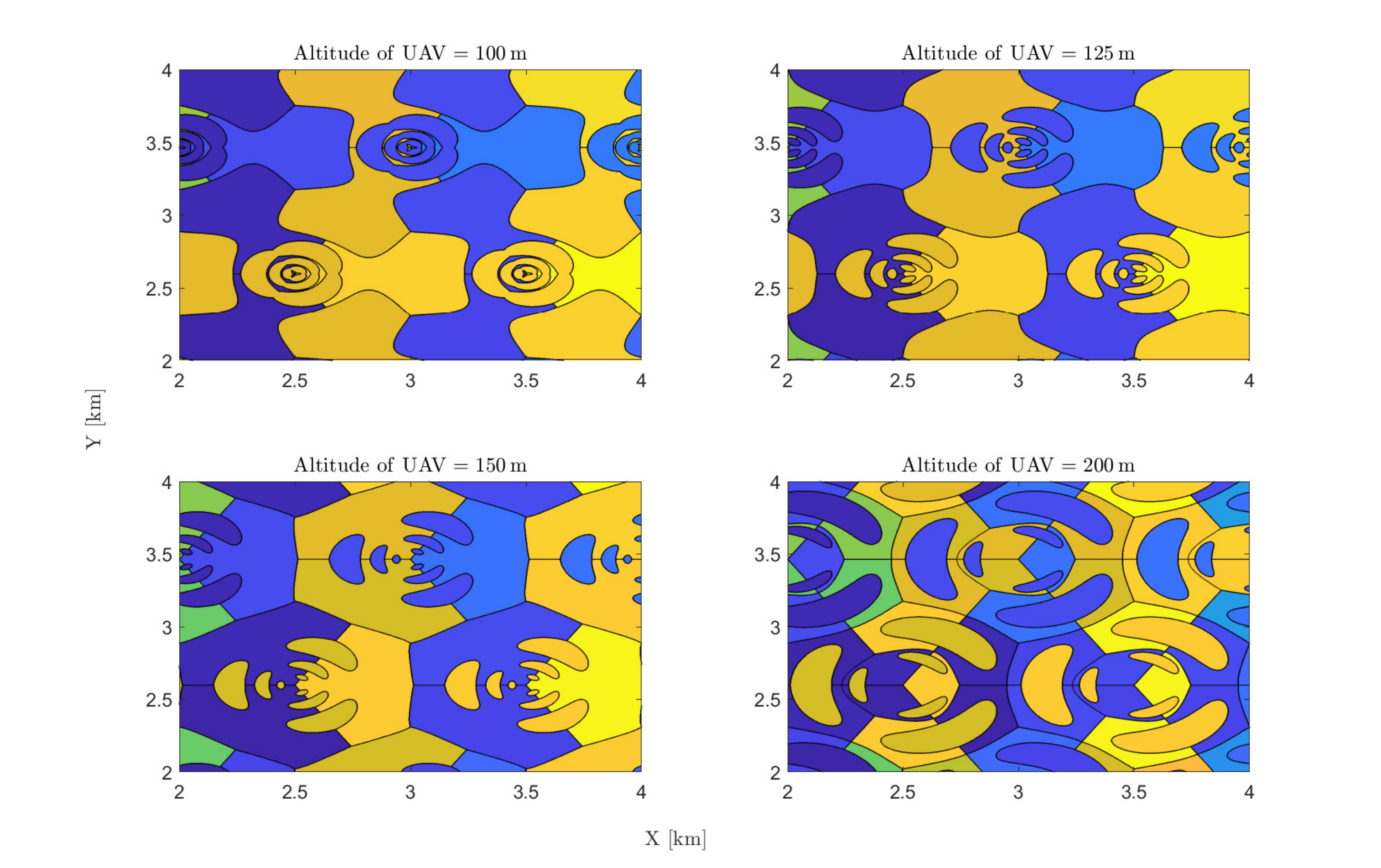}
	\caption{\black{Crossing the lines trigger HO. In general, the lines are denser at higher altitudes resulting in more HOs.}}
	\label{HOlines_hUs}
\end{subfigure}
\caption{\black{Limits illustration of cellular technology for the UAV connectivity \cite{azari2020machine}.}} \label{coverage_HO}
\vspace{-5mm}
\end{figure}

\black{A cellular-connected UAV encounters several specific challenges in addition to the general issues such as limited onboard energy budget and high propulsion energy consumption. Fig.~\ref{CoverageHoles_hUs} highlights the communication link quality at different UAV altitude. As can be seen, a target data rate is harder to achieve at higher altitudes due to significantly higher co-channel LoS interference. Furthermore, Fig. \ref{HOlines_hUs} illustrates the cell association regions at different heights where crossing a line triggers a HO event. Overall, the lines are denser at higher altitudes imposing more frequent HOs which considerably reduce the link reliability due to service interruption and latency in re-connecting. To deal with such issues, authors in \cite{azari2020mobile} propose an RL approach to control the disconnectivity time and HO rate while taking into account the UAV energy consumption (life-time) and time of task accomplishment. A general optimization problem is formulated and transferred into a MAB problem which is solved by adopting a UCB-based learning algorithm. In this paper, the general RL MAB-based learning parameters are tuned based on the importance of each aforementioned metrics which appears in an objective function. Similar approach has been taken in \cite{arani2020learning} for controlling the mutual interference of multiple ground and aerial BSs by positioning the UAVs in optimal 3D locations. The authors show the superiority of the results in terms of overall network throughput as compared to static (strategic) and Q-learning based solutions. A deep Q-learning strategy is adopted in \cite{chen2020deep} to optimize the number of handovers of a mobile cellular-connected UAV. To reduce time of mission completion and disconnectivity time through an optimal path, the authors in \cite{zeng2019path} use RL based algorithms namely temporal-difference (TD) learning.}

\begin{table*}
\color{black}
\centering
\begin{tabular}{| c || c | c |}
    \toprule
        \textbf{ML Techniques} & \textbf{Applications in Space Networks}  & \textbf{References}  \\
    \hline \hline
     SVM   & detection of spectrum occupancy    & \cite{jia2020intelligent}          \\ \hline
     DL, CNN   & spectrum prediction, interference detection and classification,     &      \cite{jia2020intelligent,henarejos2019deep,ates2019path,thrane2020model,ahmadien2020predicting,kato2019optimizing}   \\
        &  channel modeling, routing    &           \\ \hline
     
     Q-Learning   & resource allocation, HO management  & \cite{deng2019next,jiang2020reinforcement,tsuchida2020efficient,qiu2019deep}          \\ \hline
     DRL    & resource management, beam hopping, user association  & \cite{deng2019next,ferreira2018multiobjective,nie2019deep,hu2020dynamic,yan2020delay,zhao2020deep,cui2020latency,qiu2019deep}          \\ 
        \hline
    \addlinespace
    \toprule
        \textbf{ML Techniques} & \textbf{Applications in Aerial Networks}  & \textbf{References}  \\
    \hline \hline
      MAB  & HO rate, connectivity, energy, and velocity management, 3D positioning, user association    & \cite{azari2020machine,arani2020learning,hu2020reinforcement}     \\ \hline
      Q-Learning & trajectory design and path planning, HO management, beamforming and beam-steering   & \cite{liu2019trajectory,chen2020deep,zeng2019path,li2020millimeter,yang2020multi}    \\ \hline
      DRL  & channel allocation, path planning, HO management,  &   \cite{hu2020reinforcement,challita2019interference,chen2020deep,bayerlein2020uav,abedin2020data}  \\ 
        & task scheduling, offloading policy   & \cite{cao2019deep,liu20,liu120,sun21,zhangAccess21,yangIoT20,zhou2020deep,zhouTWC21}       \\ \hline
      RNN, LSTM & predictive beamforming, beam steering, beam tracking, and beam control   & \cite{yuan2020learning, moorthy2020beam}          \\
      GNN, KNN & channel modeling   & \cite{xia2020generative,zhang2018air}          \\
    \addlinespace
    \bottomrule
\end{tabular}
\caption{\black{ML techniques and their applications in air/space networks.}} \label{tab:ML_tech}
    \end{table*}

\subsubsection{mmWave}
The problem of beamforming and beam-steering in a multi-UAV and multi-antenna network serving ground users is considered in \cite{li2020millimeter}. In this network, an efficient beamforming and fast beam-steering is formulated as a joint optimization problem to maximize the sum-rate capacity which is decoupled into two sub-problems of beamforming and beam-steering optimization. Inspired by ML, a hybrid beamforming scheme is proposed. For the second subproblem, a mean field game (MFG) based solution is proposed. However, in order to handle the complexity of the calculation of MFG and to solve the problem of initial sensitivity, the authors adopt a novel RL technique. A higher sum-rate and a faster convergence rate are verified as compared to the algorithm without RL. mmWave UAV communication with ground UEs can significantly be affected by the presence of random wind gusts which imposes inevitable jittering. This indeed results in angle ambiguity and hence frequent beam misalignment. 

To mitigate such issue a deep learning scheme based on multiple long short-term memory (LSTM) layers in a recurrent neural network (RNN) is presented in \cite{yuan2020learning}. In this work, the temporal features of sequential angle data from the past time-slots are exploited to predict the angles ahead in order to promptly establish the UAV-UE link and adapt the beam-steering vectors. The UAV mobility uncertainty is more elaborated and detailed in \cite{moorthy2020beam}, where the impact of wind, engine operation, propeller rotation, and hovering disturbance are experimented under different weather conditions. To control the impact of various angular uncertainties in beam alignment of mmWave/THz links, the authors propose a dynamic prediction of optimal beamwidth of the flying drones using an echo state learning.


\subsubsection{IoT} 
The use of UAV in IoT networks as data collector is considered in \cite{bayerlein2020uav}. To design an efficient path for maximizing the collected data under flying time and obstacle avoidance constraints a double deep Q-network (DDQN) is proposed. Energy-efficient trajectory design of multiple UAV BSs that support IoT devices to improve data freshness is studied in \cite{abedin2020data}. An optimization problem is formulated which is solved using deep RL Q-network algorithm since the state space is too large and the optimization problem is NP-hard. Similar problem is considered in \cite{yi2020deep} where the authors optimize the flight path of a UAV and transmission scheduling of the ground devices in order to minimize the weighted sum of the AoI. In \cite{sun2020learning}, UAVs were deployed as MEC system for Industrial IoT (IIoT) in forest monitoring where learning-based resource allocation solution is proposed to minimize the maximum tasks' processing time. Similarly in \cite{yang2020multi}, deep RL is used for efficient task scheduling among UAVs. In this work, UAVs acting as MEC system, assisted ground IoT devices with limited computing capabilities. In \cite{cao2019deep}, deep RL is used to jointly optimize the power and channel allocation to the IoT devices in uplink, in order to maximize the energy efficiency of the IoT network. 

\subsubsection{MEC} \label{subsubsect: AI-MEC}
A cooperative UAV-enabled edge computing network has been proposed in \cite{liu20}, where UAVs assist the computation of IoT devices as well as neighboring UAVs. DRL algorithms are used for the resource allocation and for optimizing the offloading policies. In \cite{liu120}, UAV is utilized to compute the offloaded tasks from terminal users. To satisfy the QoS of the users, double deep Q-network based DRL algorithm has been proposed. For the application of forest fire monitoring, the authors in \cite{sun21} proposed an architecture in which the data collected by sensors in industrial IoT is offloaded to MEC-enabled UAV for processing. To minimize the maximum response time of forest fire, a learning based cooperative particle swarm optimization algorithm was proposed for optimal UAV resource allocation. In \cite{zhangAccess21}, the authors presented a DRL method for the optimization of UAV's trajectory and offloaded task ratio while maximizing the system stability and minimizing the energy consumption and latency in UAV-MEC system. Zhu \textit{et al.} \cite{zhu21} proposed multi-agent reinforcement learning algorithms for UAV-MEC systems to obtain an offloading policy of task and bandwidth allocations which minimizes the average response time for the computations. In \cite{yangIoT20}, the authors considered a multi-UAV assisted MEC system which targets the load balancing and effective task scheduling for UAVs deployment. Also, a DRL based scheduling algorithm has been proposed to improve the task execution efficiency at the UAV. 

\textcolor{black}{ One major challenge in MEC is how to jointly design the MEC's local action policy with communications policy to balance the exploration-exploitation tradeoff. The authors of \cite{Cao2020} has shown the potential of such joint design for industrial Internet-of-things (IIoT) systems, in which edge-device acts as a machine-type agent (MTA). The MTAs collaboratively learn optimal policy for channel access and task offloading in multi sub-carrier D2D environment. At every time slot, each MTA determines to compute the task locally or off-loads the task to the MEC server. By modelling the state space including off-loading decisions, channel access status and computation task, the authors proposed to use multiagent deep deterministic policy gradients algorithm on actor-critic network at each MTA.  then the MTAs exchange their locally trained model to generalize the global model. The benefit of multiagent MEC system presented in \cite{Cao2020} is based on an assumption that all the edge nodes can perfectly communicate to each other during the training phase. In many practical cases, such assumption rarely occurs due to imperfect communications among the edge nodes and the highly dynamic network topology. This asks for novel joint designs of source/channel coding at IoT devices and computational processes at the MEC device \cite{mostaani2019learning,mostaani2020task}.} 

\subsubsection{Channel Modeling}
A new generative neural network mmWave channel modeling for UAV communication is studied in \cite{xia2020generative}. In this paper, the channel is modeled in two steps: first the state of the channel is predicted, being in line-of-sight (LoS), non-LoS (NLoS), or outage; then path loss, delay, and angles of arrival and departure for different propagation paths are obtained. Random Forest and KNN based ML algorithms are employed in \cite{zhang2018air} to predict the path loss between two aerial nodes. It is shown that machine learning algorithms are able to provide accurate predictions with acceptable computational efficiency as compared to empirical modelings. \black{To evaluate the accuracy of the predicted models metrics of mean absolute error (MAE) and root mean square error (RMSE) are adopted where it is shown that random forest method presents smallest prediction error. Both afore-mentioned reports use ray tracing software to provide sufficiently large datasets for training and test phases.}


\subsection{ML-Empowered Multi-Segment Networks}


Several important challenges of multi-segment networking that significantly influence the performance of overall network have been recognized in \cite{kato2019optimizing}, including network control, spectrum management, energy consumption, routing design and handover management, and security issue. Then, the AI-based solutions are discussed. Specifically, the authors consider routing problem and apply deep learning schemes to improve the integrated network performance. The efficiency of the proposed method has been shown in terms of network throughput and packet loss rate. The use of LEO satellites along with mobile UAVs to assist communication between two faraway ground terminals has been studied in \cite{lee2020integrating}. The goal is to maximize the end-to-end throughput by optimally selecting one satellite from an orbiting constellation and designing the location of one flying UAV. In order to tackle the complexity of the problem due to the high number of satellites and the time-varying network topology, the authors adopt deep RL with a new technique for action dimension reduction.

\subsubsection{IoT}
In \cite{michailidis2020ai}, the potential use of UAV and satellite for industrial IoT scenarios has been discussed. Then, the authors shed some lights on different roles of ML techniques for corresponding challenges including latency, energy consumption, and resource allocation. Design of computing task scheduling for delay sensitive IoT devices is considered in \cite{zhou2020deep}. In this study, IoT computing tasks are collected by a UAV and the UAV makes an online decision to either process the tasks itself or offload the data to ground BS or LEO satellites. The objective is to minimize the computing delay under UAV energy consumption constraints and the optimal policy is obtained using a deep risk-sensitive RL algorithm. Dynamic structure of integrated ground-air-space networks may result in connection error and slow down the computation capabilities in distributed mechanisms. For this, a reliable storage and flexible computation offloading through a novel integrated architecture is proposed in \cite{gu2020coded} to speed up distributed learning algorithms such as traffic prediction and resource allocation. The proposed system is examined in IoT scenarios, where the ground terminals' computing tasks are offloaded to UAVs and HAPs acting as MEC servers. The superiority of the performance in terms of retrieval and offloading delay is shown under unreliable network conditions.

\subsubsection{MEC}
An integrated GAS network has been considered for MEC-assisted offloading in \cite{zhouTWC21} where a UAV gathers the task from the IoT devices to process it locally or offload it to a satellite/ground edge server.
The dynamic scheduling problem has been solved with DRL-based method. 
In \cite{cheng2019space} the problem of computation delay and energy consumption of remote IoT devices has been addressed using a flexible and dynamic architecture of an integrated ground-air-space network. In this architecture, the UAVs act as near-user low delay mobile edge computing platforms, while the satellite segment grant access to the cloud computing with higher delay. To provide an efficient edge and cloud computing services, a flexible joint communication and computation framework has been formulated. Then, an optimal offloading policy for minimum energy consumption, delay, and cost of server usage has been proposed using RL strategy.
The authors in \cite{liaoIOT21} proposed a learning-based task offloading with queue awareness and resource allocation strategy in integrated GAS network for IoT applications.

Table \ref{tab:ML_tech} summarizes different ML techniques applied to various problems of aerial and space networks with the corresponding references.\\

\noindent\textit{\black{\textbf{Key Takeaways} -- Tremendous complexity and lack of decent models motivate machine learning (ML) applications to several problems that were previously perceived highly challenging. Particularly, ML has received significant attention in handling path planning, resource allocations, interference detection and classification, beamforming and beam tracking, mobility management, user association, and ground-air-space channel modeling. Local ML algorithm might need to adapt to NTs limitations such as UAVs limited onboard computation power and energy budget. In addition, while the majority of references have applied RL techniques, poor performance of such optimization method during the training phase is to be well understood and mitigated. To comprehensively capture the benefits of ML-based approach for integrated ground-air-space networks, furthermore, E2E network comprehension and corrective actions are required.}}

\section{Higher Layer Advancements
} \label{sec:NTN_higherlayer}


In recent years, there has been a trend to provide communication networks with greater "flexibility". This term is defined in \cite{Yazar_2020} as ``the capability of making suitable choices out of variable options depending on the internal and external changes of the communication systems, and eventually evolves with an increasing number of new options''. 
\black{This term takes on more relevance when we refer to NTN's. Unlike TNs, the NTNs present a highly dynamic behavior, both in their topological configuration and in their radio channel. Therefore, the integration of these networks must be supported by a high reconfiguration capability, in the topological part of traffic engineering and its physical layer. The NTNs must be} highly reconfigurable, dynamically adaptable to a changing network conditions, intelligence provisioned and ultimately with cost reduction in the deployment and network operation. This NTNs capability has been acquired by the gradual implementation of the paradigm known as SDN. This includes the integration of NTNs into the broad spectrum of 5G and beyond networks and services. Its main key principles such as the separation of the control and user plane, centralized control and programmability, position SDN as a technological enabler in order to face the high integration requirements in highly heterogeneous environments \cite{Yazar_2020}. 
 
 In this context, several efforts have recently emerged that invite the incorporation of SDN into NTN. Although the first efforts mainly focused its utilization in satellite networks, the use in aerial networks is increasingly being considered. In this regard, several of these works basically focus on architectural solutions for implementation and exploitation of main characteristics such as global vision and centralized control. Given the remarkable differences among the NTN technologies (e.g., satellites, HAPs and UAVs) regarding processing capacity, data transmission and network topologies, etc., there is a significant difference in the way of approaching solutions for SDN implementations. For satellite networks, many of the recent works point towards the need of control sets and management functions as well as compatible interfaces (APIs and/or network protocols).  
 This is for a full end-to-end (E2E) networking realization where the whole satellite-terrestrial network behavior can be programmed in a consistent and inter-operable manner \cite{Mendoza_2019}. Other works, aimed at the development of platforms and architectures have been presented in \cite{Akyildiz_2015,toufik_2018}, generally, positioning SDN controllers in the satellites synchronized with SDN controllers in the terrestrial networks.  

\black{Thus far, we can see that the NTNs integration process into TNs probably will be take place first at an architecture level and then at physical level \cite{open5gcore}.} In this context, the seamless satellite-terrestrial integration and SDN capabilities are expected to offer new or improved services/technologies such as:

\subsubsection{Network Virtualization}
Network virtualization is defined in \cite{Prakash2013} as the ability to manage and prioritize traffic in portions of a network that might be shared among different external networks. Network slicing is a more particular case of network virtualization, defined in \cite{Ordonez2017} as an E2E logical network/cloud running on a common underlying (physical or virtual) infrastructure, mutually isolated, with independent control and management that can be created on demand. 

Some projects, such as 5G-VINNI \cite{5G-VINNI}, directly address satellite integration in 5G networks from the point of view of highly dynamic and flexible network architectures, service deployment and testing, to create new technical and commercial service deployment models, enabling virtualized functions from the network and service layer.

\subsubsection{C-RAN Architecture and Cloud/Edge Computing} 
Satellite networks have faced difficulties in their integration with terrestrial networks mainly because of the lack of common interfaces and lack of common management and control \cite{toufik_2018}. Assuming the implementation of SDN and network function virtualization (NFV) technologies, the cloud RAN (C-RAN) concept (very popular in the terrestrial architectures) can be extrapolated to aerial/satellite networks. Instead of having multiple and expensive ground component segments, C-RAN decouples network functions from the actual equipment therefore reducing the deployment and maintenance cost. Furthermore, the centralized cloud architecture enables the so-called cloud computing, with a large and scalable computing capabilities that are shared on-demand. Essentially, the cloud server is in charge of higher layer functionalities and represent the link with the core network \cite{Khalili_2019}. To not compromise the C-RAN resources and in order to reduce communications to/from the cloud, edge computing has emerged to exploit the distributed computing resources in close proximity to users. With C-RAN and edge computing, aerial/satellite networks can adhere to the 3GPP 5G cellular access network standard \cite{Liolis_2019} which seems to be a key factor for the final integration into 5G \cite{Bacco_2019}.
   
\subsubsection{Transmission Control Protocol (TCP)}
In order to benefit the most from the multiple radio interfaces/data paths provided by an aerial-supported wireless network, dynamic selection of the most suitable path for a given data unit is of key importance. From a transport layer point of view, multi-path transmission control protocol (MP-TCP) appears as a valid approach \cite{ford_2020}, by improving resource usage and user experience through simultaneous use of these multiple paths. However, general TCP protocol was not conceived to work on long-propagation links such as the satellite links. TCP's main goal is to avoid congestion before it happens. TCP may interpret the long-delay of a packet as a loss event and consider that the link is congested. There are a number of ways to deal with the conventional TCP limitations. For example, one can tune the TCP parameters of the end systems to match the environment. However, the most widely used technique is to use performance enhancing proxies (PEPs) \cite{Peng_2012,Border_2001}. PEPs exploit topology awareness by basically acting as a ``TCP-splitting'', where standard TCP is considered on the terrestrial legs, while an optimized TCP is considered for the long-round-trip-delay aerial connections. Unfortunately, PEPs also exhibit some weaknesses, e.g. spoofing and spitting break the semantics of TCP, thus causing interoperability issues \cite{deCola_2017}.
   
   \subsubsection{Smart Gateway Diversity and Aerial Links}
    For GEO satellites, the trend of moving feeder links to higher and higher bands combined with the increased high-data rates has resulted in the need of multiple gateway (GW) links for nominal service. In the NGSO and UAV case, usually a fully meshed aerial network architecture is considered, generally combined with a complex ground network with a significant number of gateways. In both cases, the ground segment has become a  source of growing concern. Proper GW management with quick adaptation to potential outage is of key importance \cite{Kyrgiazos_2014}.
    Having ``at least'' one GW within the coverage area of each aerial device is not always possible. For instance, in remote or oceanic areas, or regions where it is not safe to set up a GW for security aspects. In those cases, and as an alternative to reduce the complexity of the ground segment and improve the network’s security, aerial links can be considered \cite{Chen_2021}. The main research challenges of such links are the need for on-board routing and network management mechanisms capable of dealing with the motion of NGSO aerial systems, together with the non-uniformly congested and dynamic network traffic with different QoS classes \cite{Radhakrishnan_2016}.\\

\noindent\textit{\black{\textbf{Key Takeaways} --  It is highly probable that NTN-TN integration will take place first at an architecture level and later at the PHY level. Following this, the benefits of SDN technology for the highly dynamic NTNs integration into 5G networks and beyond is surveyed in this section.}}



\section{NTNs Field Trials and Industrial Efforts 
}\label{sec:NTN_FieldTrials}

\begin{table*}[]
\color{black}
\caption{\black{NTNs field trials and industrial efforts.}}
\centering
\begin{tabular}{|c||c|c|}
\toprule
\textbf{Field Trials}                                                                                                    & \textbf{Space Networks}                                                   & \textbf{Aerial Networks}                                                                                                           \\ \hline \hline
\multicolumn{1}{|c||}{\multirow{7}{*}{\textbf{Industry led efforts}}}                                                              & GEO satellite-based broadband across globe \cite{viasat,hughesnet}                   & \begin{tabular}[c]{@{}c@{}}High altitude unmanned aircraft for wide \\ coverage \cite{facebook,google}\end{tabular} \\ \cline{2-3} 
\multicolumn{1}{|c||}{}                                                                                                   & \begin{tabular}[c]{@{}c@{}}LEO satellite constellation for high-speed data\\   \cite{oneweb,eutelsat,starlink,kuiper}\end{tabular}  & \begin{tabular}[c]{@{}c@{}}Cellular connected UAVs \cite{nokia,lte,Ericsson,Quantum,Nokia1,vodafone}\end{tabular}                                                                      \\ \cline{2-3} 
\multicolumn{1}{|c||}{}                                                                                                   & \begin{tabular}[c]{@{}c@{}}To launch next generation MEO satellites for beyond \\   urban reach \cite{ses}\end{tabular} & \begin{tabular}[c]{@{}c@{}}Time-based conformance monitoring for UAVs \cite{Nasa}\end{tabular}                                                                                                                               \\ \cline{2-3} 
\multicolumn{1}{|c||}{}                                                                                                   & \begin{tabular}[c]{@{}c@{}}Role of high-altitude pseudo-satellites\\ for telecommunication \cite{HAPS-TELEO} \end{tabular}  & \begin{tabular}[c]{@{}c@{}}Stratospheric platforms for reliable 5G service \cite{stratosphere}\end{tabular}                                                                                                                              \\ \midrule
\multicolumn{1}{|c||}{\multirow{8}{*}{\begin{tabular}[c]{@{}c@{}}\textbf{Industry-academia}\\ \textbf{consortia led efforts}\end{tabular}}} & \begin{tabular}[c]{@{}c@{}} Satellite and 5G convergence \cite{METEORS1,fraunhofer,sat5g1,5Ggenesis,5g-allstar}\end{tabular}                                                                     & \begin{tabular}[c]{@{}c@{}} Integrated cellular-satellite system\\ for command control of UAS  \cite{droc2om} \end{tabular}                                                                                                                              \\ \cline{2-3} 
\multicolumn{1}{|c||}{}                                                                                                   &        \begin{tabular}[c]{@{}c@{}} Satellite-based 5G backhaul \cite{ThalesKT} and\\HAP for emergency and backhauling \cite{HAPPIEST}\end{tabular}                                                                & \begin{tabular}[c]{@{}c@{}} Validation of 5G KPIs for eMBB, URLLC,\\ and mMTC \cite{5GDrone1} \end{tabular}                                                                                                                              \\ \cline{2-3} 
\multicolumn{1}{|c||}{}                                                                                                   & \begin{tabular}[c]{@{}c@{}} Various ESA projects for different aspects of\\ NTN integration in 5G \cite{satis51,EdgeSAT,CLOUDSAT1,5G-Goa}\end{tabular}                                                                      & \begin{tabular}[c]{@{}c@{}} End-to-end 5G trials for industry 4.0 and \\autonomous drone scout \cite{5GDIVE} \end{tabular}                                                                                                                              \\ \cline{2-3} 
\multicolumn{1}{|c||}{}                                                                                                   & \begin{tabular}[c]{@{}c@{}} ML solutions for satellite communications \cite{SATAI,MLSAT,ATRIA}\end{tabular}                                                                     & \begin{tabular}[c]{@{}c@{}} Integration of 5G and drones for increased\\ connectivity and coverage \cite{AERPAW} \end{tabular}                                                                                                                                                                                                                          \\ \midrule
\multicolumn{1}{|c||}{\multirow{3}{*}{\textbf{Academia led efforts}}}                                                              & \begin{tabular}[c]{@{}c@{}} Dynamic integrated satellite-terrestrial backhaul\\  supported by mmWave\cite{artigaNetw18} \end{tabular}                                                                    & \begin{tabular}[c]{@{}c@{}} UAV relay systems \cite{dixonJSAC12,guoCSNDSP14,johansenGC14} \\  UAV-aided localization \cite{liuHAWK14,gongTVT17}\end{tabular}                                                                                                                                \\ \cline{2-3} 
\multicolumn{1}{|c||}{}                                                                                                   & \begin{tabular}[c]{@{}c@{}} Satellite backhaul operation for the \\5G cellular access network\cite{liolis5GWF19} \end{tabular}                                                                        &  \begin{tabular}[c]{@{}c@{}} Path planning and resource allocation \cite{diConf15,FadlullahNetw16,wigardPIMRC17,amorimGC17} \\UAV for flying ad hoc network \cite{BekmezciRAST15} and agriculture \cite{PopescuConf19} \end{tabular}                                                                                                                          
                         \\ \bottomrule
\end{tabular}
\label{tableField}
\end{table*}


\black{The ongoing 5G and beyond 5G vision encapsulates a broad range of application scenarios including ubiquitous connectivity in remote and under-served areas to propel an all-inclusive growth. The current pandemic has ensued the vitality of communication technologies which enabled the government, people, and other industry verticals stay connected virtually.} This section highlights the major field trials and experimental efforts by various industries and academic institutions to showcase the viability and prospects of NTN integration in the future networks design. \black{ A summary of these efforts is given in Table~\ref{tableField}.}
\subsection{Space Networks}
To address the proliferating consumer demands and massive connectivity requirements in 5G and beyond, space networks will play a pivotal role due to their unique attributes such as ubiquity, mobility, broadcast/multicast, and security etc. Extensive research across various domains are being pursued to facilitate the seamless integration of space networks in 5G ecosystem. 
\subsubsection{Industry led efforts}
Substantial industrial developments have been focusing on the satellite broadband and internet services for ubiquitous global connectivity. Viasat \cite{viasat} and Hughesnet \cite{hughesnet} are among the popular satellite-based broadband internet service providers. While Hughesnet operates only in America, Viasat has a widespread presence across the globe. For the internet service, both providers use GEO satellite for the wider coverage, however, Viasat is also developing small LEO satellites for specific operations such as military communications and  for crosslink with GEO satellite. Oneweb \cite{oneweb} has been building a system to provide end-to-end solution and supply the high-speed data in every part of the world using the constellation of LEO satellites. It expects to provide the coverage everywhere above 50 degrees north by end of Q4 2021 and global coverage by the end of 2022. GEO satellites usually suffer with high latency problem, whereas, LEO satellites cover a small portion of earth due to its short distance with earth. Capitalizing on this, SES \cite{ses} plans to launch its next generation MEO satellites network O3b mPOWER in Q3 2021 with the aim of accelerating 5G reach beyond urban centres to cater high throughput and low latency. 
In addition to its GEO satellite services, Eutelsat has been planning to develop the LEO fleet towards the narrowband connectivity for IoT \cite{eutelsat}. It aims to offer the widespread satellite links while complementing the LPWAN IoT terrestrial networks. 
Another project HAPS-TELEO \cite{HAPS-TELEO} had focused on studying the role of high-altitude pseudo-satellites for telecommunication and complementary space applications.
Further, Starlink \cite{starlink} has been providing beta-stage broadband internet service with moderate latency of 20 ms-40 ms and speed of 50 Mbps-150 Mbps. It aims to launch more and more LEO satellites and also enhance the software capabilities to improve the speed and latency performance in the near future. Amazon \cite{kuiper} also follows the suit by announcing a project Kuiper which will build a constellation of LEO satellites to provide  reliable and affordable broadband internet services around the world.
\subsubsection{Industry-academia consortia led efforts}
Some industry-academic consortia are also striving to augment the capabilities of 5G networks using the cooperation from space networks. For instance, in compliance with the objectives in 3GPP RAN and 3GPP SA, 5G METEORS project \cite{METEORS1} focuses on the investigation and prototyping in the 5G satellite convergence. Recently, certain extensions of 5G NR have been demonstrated to support the NTN integration over GEO satellites \cite{fraunhofer, METEORS1}. A research project SAT5G \cite{sat5g1} aims to enable the virtualisation of satellite functions to make them compatible with 5G SDN and NFV which can eventually lead to integration of satellite and mobile network elements. Other key features include extending 5G capabilities into satellites, integrated network management and orchestration, and caching and multicast for content delivery and NFV distribution. Under the umbrella of 5GENESIS \cite{5Ggenesis}, the Limassol 5G platform aims to employ NFV-/SDN-enables satellite communications to achieve throughput enhancement via air interface aggregation, dynamic spectrum allocation among satellite and terrestrial networks, and multi-radio slicing. 5G-ALLSTAR \cite{5g-allstar} is an international collaborative project which will demonstrate the multi-connectivity support for cellular and satellite access and also explore new radio feasibility to offer broadband and reliable 5G services based on satellite access. More recently, Thales Alenia Space and KT SAT are working jointly to experiment the satellite-based 5G backhaul to expand the global outreach particularly in the areas where terrestrial infrastructure is difficult to construct \cite{ThalesKT}.
Furthermore, European space agency (ESA) has been governing several research projects including SATis5 \cite{satis51}, EdgeSAT \cite{EdgeSAT}, CLOUDSAT \cite{CLOUDSAT1}, and 5G-GOA \cite{5G-Goa} leading different aspects of NTN integration into 5G architecture.
HAPPIEST \cite{HAPPIEST} focused on studying the application of high altitude pseudo-satellites for communication and complementary services. It explored two  reference scenarios viz., HAPS as telecommunication back-up system in the case of emergencies and optical backhauling to space.
 To investigate the applications of AI for satellite communications, ESA also governed two  dedicated ML-based projects \cite{SATAI,MLSAT}. Based on ML solutions, three proof of concepts were developed in \cite{SATAI} i.e., automatic interference detection, flexible payload configuration in the presence of interferers, and user demands prediction. In a similar line, a Horizon Europe supported project ATRIA recently kicked off \cite{ATRIA}. It aims to exploit the AI tools to autonomously optimize the configuration of the available satellite resources as per the service requests.
\subsubsection{Academia led efforts}
A couple of academic research works have demonstrated the experimental outcomes of major projects. The authors in \cite{artigaNetw18} reported a dynamic integrated satellite-terrestrial backhaul solution supported by mmWave band that can overcome the limitations of the conventional fixed backhaul. Specifically, in such an integrated design, a terrestrial network can reconfigure its topology based on the traffic demands and frequency reuse. The primary technology enables to achieve these goals are smart antennas and software defined intelligent hybrid network management. In \cite{liolis5GWF19}, the authors demonstrated an over-the-air test where a satellite is integrated with 3GPP release 15 5G core network. It allowed satellite backhaul to operate as a 5G cellular access network to showcase the efficient edge content delivery.

\subsection{Aerial Networks}





The development of aerial networks is still in progressive phase and yet to discover its full potential. Over the past few years, significant research attention has been driven towards this domain from both industries and academia. 
\subsubsection{Industry led efforts}
Some early industrial efforts include Facebook’s Aquila \cite{facebook} and Google’s loon \cite{google}, which were initiated to provide coverage over a large remote area from the sky using unmanned aircraft at high altitudes. These projects aimed to prolong the life span of aerial vehicles by using solar power in Aquila and exploiting wind currents instead of propellers in Loon. In 2016, Nokia bell labs developed an experimental flying-cell (F-cell) technology to eliminate the costly backhaul wires enabling flexible small-cell deployment \cite{nokia}. The architecture of F-cell supported NLoS wireless transmission by using time division duplex (TDD) or frequency division duplex (FDD) mode and allowed the system throughput of around 1 Gbps over the existing LTE networks.  Further, in 2017, Qualcomm conducted various field trials of the LTE-supported UAS \cite{lte}. Flights were operated at different altitudes using commercial LTE bands and eventually an array of data logs was obtained to characterize the performance. Based on these trials, complemented by the simulations, the feasibility of drone operation had been illustrated. Ericsson conducted a successful proof-of-concept trial for mobile network on a drone \cite{Ericsson}. It deployed a small cellular network of about 150 gram weight on a drone to provide the mission-critical voice and video connectivity over an area with poor or no coverage. Together with Quantum Systems \cite{Quantum}, Ericsson successfully completed the initial test for implementation of 5G technology in the drone. The test included measurement of various transmission metrics, for example, data throughput, delay, and signal quality. Nokia \cite{Nokia1} also offers an end-to-end solution for drone networks which can provide services in critical applications such as public safety. The solution enables the drones for an automated mission by providing them connection over a private network which can remain unaffected due to congestion in public networks. Together with Sendai city in Japan, they also tested a potential use of drones in the event of Tsunami or other disasters for the preventive and management efforts using a private LTE network. Vodafone has been advancing the capabilities of cellular-connected drones for various applications \cite{vodafone}. During various trials, they have demonstrated the capabilities of dynamic no-fly zones detection and geofencing, BVLoS operation, and drone interference mitigation. NASA has also been involved in advancing efforts to facilitate several UAV applications \cite{Nasa}. For example, a field trial has been conducted to investigate a time-based conformance monitoring (TBCM). TBCM allows to monitor whether the UAV flights are following the planned trajectory. Furthermore, stratospheric platforms are also being explored to provide fast and reliable 5G technology from high altitudes \cite{stratosphere}.
\subsubsection{Industry-academia consortia led efforts}
In addition to the industry-led developments, various industry-academic consortia are also thriving to advance drone-aided communication technologies. In particular, the DroC2om project \cite{droc2om}, which started in 2017 and ran for two years, targeted the datalink of UAS to enable the airspace sharing between manned and unmanned systems. Specifically, the project evaluated an integrated cellular-satellite system design for command and control to support the safe and reliable UAS operation based on the real drone measurements and modelling. Moreover, the 5G!Drones project \cite{5GDrone1} by EU H2020 aims to design, implement, and run trials for various use cases on the 5G infrastructure. It intends to trial several use cases such as eMBB, URLLC, and mMTC and to validate 5G KPIs for supporting these use cases. Another research project by EU H2020 called 5G-DIVE \cite{5GDIVE} targets the end-to-end 5G trials to highlight the merits and business value proposition of 5G technologies. It primarily focuses on two use cases viz., industry 4.0 and autonomous drone scout. The latter case involves the drone fleet navigation for better piloting the drone swarm and intelligent image processing on the drones to aid the automation in the drone scouting. Recently, an NSF-funded research project, AERPAW comprising a large-scale testbed for experimentation with advanced wireless technology and systems involving UASs, is being developed \cite{AERPAW}. It plans to integrate drones and 5G for mutual benefits. Drones will support the 5G by providing increased connectivity and coverage, while 5G will support the drones by dispensing the location data and improved signals. The use cases aimed under the project include package delivery, smart agriculture, traffic control, etc.
\subsubsection{Academia led efforts}
Besides the major industry and academic-led research projects, several small-scale testbed developments and field trials also showcase the viability of drone-based networks. For instance, an experimental work in \cite{dixonJSAC12} presented a mobility control algorithm for a UAV relay-aided end-to-end communication chain. Herein, several flight experiments were conducted to validate the performance of the algorithms. Flight tests were performed with a UAV which measures the signal strength of 802.11 b/g communication links from the multiple nodes on the ground placed at the known locations. In \cite{liuHAWK14}, a fully functional and portable mini UAV system is introduced which is programmable for localization tasks. A software controller is developed in this work to implement the waypoint functionality based on the PI-control laws. The authors in \cite{guoCSNDSP14}  analyzed the potential of UAVs as relay to support the cellular network. They presented the field-test results from the observation in both urban and rural areas. Further, in \cite{johansenGC14}, the authors demonstrated a field experiment wherein a UAV operates as a wireless relay for an autonomous underwater vehicle at the ocean surface. They presented the design of relay payload, optimal flight conditions, network configuration, and experimental results. Using the real drone flight measurements, path planning for the UAV is presented in  \cite{diConf15} to minimize energy consumption while satisfying other requirements such as coverage and image resolution. In \cite{BekmezciRAST15}, testbed implementation of flying ad hoc networks has been presented. This work employed an 802.11 connection and Raspberry Pi module to establish the communication link between multiple Ar.Drone 2.0 as UAVs. Authors in \cite{FadlullahNetw16} proposed a dynamic algorithm to control the trajectory in UAV-based networks for improving the delay and throughput. They also conducted field experiments to substantiate the effectiveness of multihop communication and to analyze the effect of separation between users and UAVs. Furthermore, authors in \cite{gongTVT17} proposed a drone-assisted localization framework for wireless networks. Based on this framework, field experiments were also performed, which demonstrated promising results. Research works in \cite{wigardPIMRC17} and \cite{amorimGC17} discussed different methodologies to distinguish between the ground and aerial users, which is crucial for optimizing interference and mobility management. For these purposes, LTE radio measurements were utilized. Authors in \cite{PopescuConf19} introduced a collaborative approach utilizing the UAV and sensor network to improve the precision and ecological agriculture performance. \\

\noindent\textit{\textbf{Key Takeaways} -- 
For the space networks, the industrial efforts are primarily driven towards addressing the large propagation delays from the communication using the GEO satellites. To this end, various companies are planning to launch LEO or MEO satellite constellations for offering the low latency services. Joint effort from industry and academia is trying to explore the effective solutions such as NFV, radio slicing, ML, etc., for integrating the space networks with TNs. Meanwhile, experiments from the academia have focused on the integrated terrestrial-satellite backhaul solutions.} \\
\indent For the aerial networks, industries have demonstrated the operation and feasibility of cellular connected UAVs to identify new applications and challenges. Several industry-academia consortia have been trying to explore the new applications and services that can be enabled by UAVs' integration. The field trials from academia have been developing various aspects of UAVs such as mobility management, localization, aerial relay, flying ad hoc network, trajectory planning, interference management, etc. Despite the benefits of NTN, there are several challenges which are needed to be addressed for the seamless integration of NTN with TN. For instance, channel modelling incorporating the Doppler effects, admission control by satellites, storage limitations in air and space, flexible addressing and routing, mobility and constellation management, and spectrum co-existence are some of the crucial design challenges.

\section{NTNs Integration in 6G 
} \label{sec:NTN_6G}

NTNs integration into 6G will further advance the key features of 5G through eMBB+, mMTC+, and uRLLC+ next to the new features by including new spectrum and AI. In what follows, we discuss the prospective use cases, new architectures, technological enablers, and the higher layer aspects related to the new space/air paradigm in 6G and beyond.

\subsection{Prospective Use Cases}

\begin{figure*}[!t]

\centering
\includegraphics[width=0.8\textwidth]{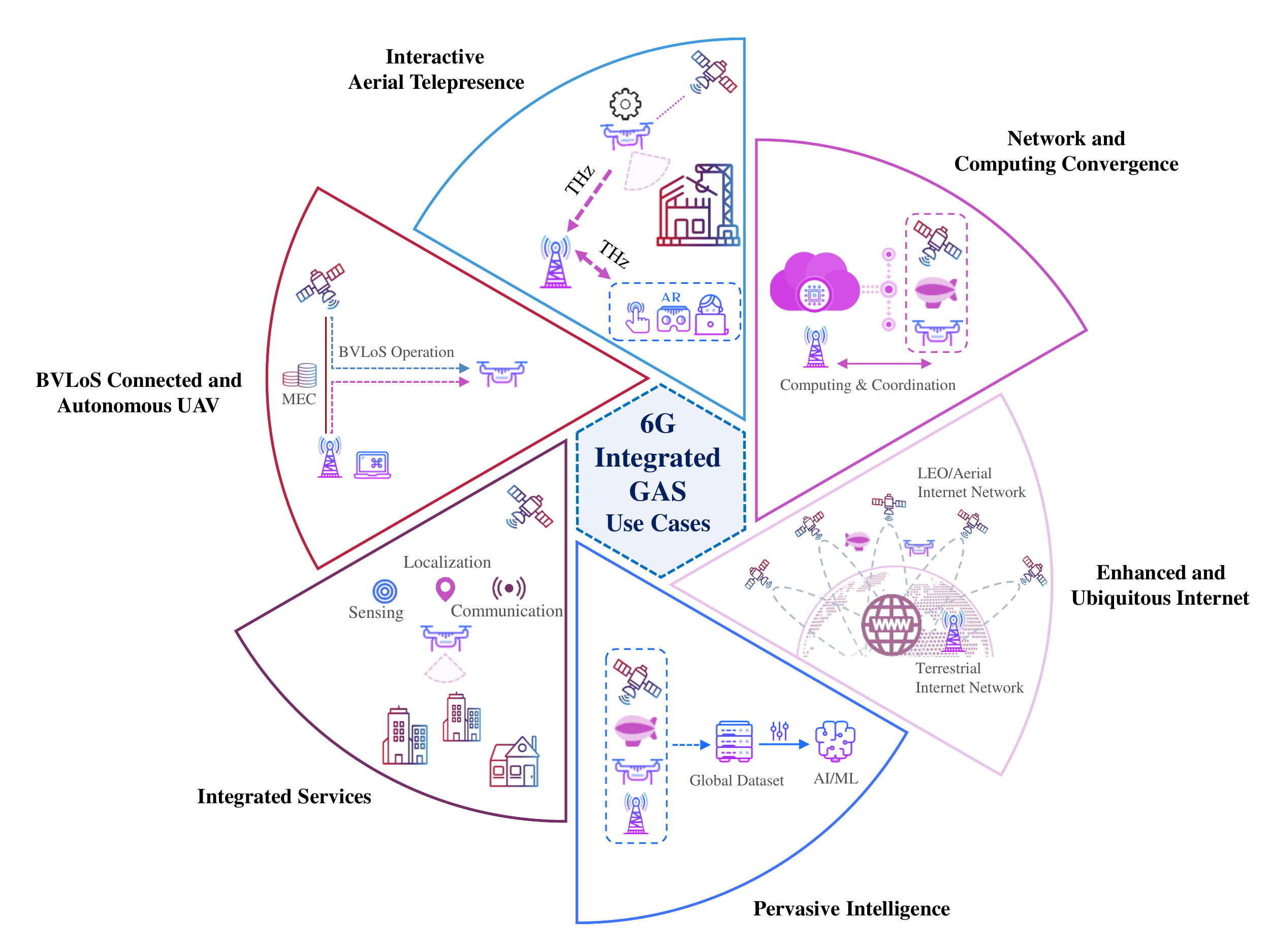}
\caption{\black{6G integrated ground-air-space (GAS) prospective use cases.}}
\label{fig:6g_usecases}
\end{figure*}

A harmonic integration of UAVs and satellites supplement the coverage of terrestrial networks and hence enables important 6G use cases, some of them are listed blow. \black{A graphical abstract of the prospective use cases is also presented in Fig.~\ref{fig:6g_usecases}.}
    
\subsubsection{Network and Computing Convergence \cite{samsung6g}}  
The computing capability of a network edge site is typically limited and may not be expanded easily. This fact leads to the use of multiple network edge sites for which network coordination is needed. Depending on the computing requirement UAVs and satellites may support the ground infrastructure in both computing and coordination. For instance, for an intensive computation task, ubiquitous availability of satellites enable a fair distribution of the computation load, yet they may assist the network with coordinating the tasks over the available resources. Accordingly a full integration of GAS networks enables the computing-aware networking in 6G and beyond.
    
\subsubsection{Enhanced and Ubiquitous Internet}
Leveraging from the inter-connected LEO satellites and aerial access points, a parallel Internet network comparable with its terrestrial counterpart is envisioned in 6G and beyond. Accordingly, Internet services can be available everywhere such as oceans, deserts, and on plane and ships. The social impact of ubiquitous internet is significant for instance by offering the education or medical consultations to everyone. Furthermore, a better data delivery can be provided by using satellite paths compared to the terrestrial Internet routing paths. 
    
    
    
\subsubsection{Pervasive Intelligence} 
The effectiveness of AI applications in wireless networks heavily depends on reliable data at the network disposal. The integrated GAS networks and global availability of satellites enable a holistic data integration anywhere and anytime, which increases the effectiveness of ML-based solutions. Furthermore, a global rich data-set provided by satellites and aerial platforms allows to globally adjust the ML parameters and therefore improve the performance. Nevertheless to say that air/space networks can provide the local storage and computing sites for the edge AI based solutions in order to fulfill the AI-based decisions and network management \cite{akyildiz20206g}. 

\subsubsection{Integrated Services}
In 6G networks, communication, localization, and sensing services are envisioned to coexist, sharing the same time-frequency-spatial resources. This coexistence will remarkably boost the spectrum efficiency and avert current regulatory hurdles. Substantial mutual benefits lie with the cooperation between the three services. On the one hand, dense communication devices can aid the simultaneous localization and mapping (SLAM) service, and on the other hand, location information can enable context-aware communication \cite{de2021convergent}. While the dense communication devices in 6G networks will provide rich measurements for SLAM, the achieved accuracy will be subject to the environment's multipath and LoS characteristics. The integration of NTN will not only improve the SLAM measurement's accuracy but will also enable SLAM with 3D location and orientation information \cite{de2021convergent}. Applications in 6G networks will also benefit from the context-awareness of joint communication, localization, and sensing in fully integrated GAS networks. Here, the integration of NTN will allow fine-grained context-awareness, enabling multi-modal communication and quality of service based on the current location or context. For instance, with an accurate context-awareness, one can select the subset of terrestrial, aerial, and space elements based on the application requirements \cite{xiao2020overview}.


\subsubsection{BVLoS Connected and Autonomous UAV Operation} 
One of the inhibiting factors which restricts the full potential of UAV technology is the inability to command and control the UAV beyond a visual sight or out of the dictated range of inherent communication channel. Effective UTM, which will essentially control the air-traffic for UAV, can ensure the safe BVLoS operation by using the support from either terrestrial or satellite communication links \cite{geraci2021will}. A 3D ultra-reliable low-latency network can be provided in 6G through the deployment of various technological enablers such as cell-free and IRS deployments. Such important feature enables reliable autonomous operation of drones. Further, MEC capability in GAS networks would allow the crowdsourcing of flight data from other UAVs fleet and information about the no-fly zones, which eventually can be processed at the nearby edge server to design the optimal trajectory and thereby facilitating the autonomous operations \cite{commag22}. MEC support can also bolster the UTM when combined with ML applications and various optimization algorithms. Finally, accurate 3D localization is of high importance to enable a safe operation of autonomous UAVs. 6G and beyond through the deployment of higher frequencies and extremely massive antenna arrays may provide advanced sensing solutions with finer range and higher angular resolution \cite{geraci2021will}.

\subsubsection{Interactive Aerial Telepresence} 
To do unsafe, costly, or time-critical tasks for humans, UAVs can be deployed and controlled remotely while having humans in the loop, the so-called \textit{aerial telepresence}. The concept can be further enriched when combined with augmented reality which provides 3D visual feedback and real-time teleinteraction with the target environment \cite{commag22}. Such haptic guidance, indeed, enhances the UAV capability, creates new applications, and more importantly brings experts to the scene from anywhere at anytime. For this application scenario, particularly for the real-time interactions where the AR can not be compressed \cite{giordani2020toward}, the data rate may reach the Gbps and hence THz deployment plays an important role. Further to the communication rate and latency requirement, UAVs THz sensors can offer highly accurate environmental cognition and instantaneous perception of the environment being crucial for precision \textit{interactive tasks and immersive experience}.

In the following, we review several relevant aspects of 6G and beyond networks with mutual impact on GAS networking. This comprises novel architectures, technological enablers, and the higher layer aspects. 




\subsection{Architectures}

\subsubsection{Open-RAN} 
During the last few years, the telecommunication domain has shown a tremendous interest towards Open RAN (ORAN) architectures. The movement of ORAN actively promotes disaggregated RAN architectures enabled by standardized communication and control interfaces among the constituent components. The motivation is threefold 1) empowered innovation, 2) enhanced security, 3) higher sustainability. In terms of innovation, the ORAN philosophy can enrich the vendor ecosystem with smaller highly-innovative players who focus on specialized components instead of a few highly-integrated global vendors. In terms of security, open interfaces can safeguard the information exchanged between different components and move the design and control decision-making from the highly-integrated vendors towards the operators. In terms of sustainability, the disaggregated architecture can enable the continuous system integration with the latest features instead of its complete replacement with new generation equipment every decade. Based on the above discussion, it becomes obvious that integrated GAS networks could greatly benefit from the ORAN movement. However, the adoption of the terrestrial ORAN designs, components, interfaces and controllers would not be straightforward due to the particularities of the GAS networks. The first and foremost challenge is the new control/communication interfaces needed to interconnect the RAN which is part of the network control center (NCC) with the air-space control center (SCC), which is responsible for the orchestration of the UAV and/or satellite assets. From an algorithmic point of view, this motivates new ORAN intelligent controllers which would be responsible for the communication control co-design. 

 \subsubsection{Multi-Segment 3D Networks}
The GAS networks are inherently multi-segment. In conventional 5G architectures, the terminals are mostly located on the ground and in some cases on UAVs. However in the context of 6G, the terminals might be located in higher orbits, for example on VLEO or LEO nanosat constellations meant for observation of data collection missions. This enlarged view of multi-segment networks completely changes the targeted architecture, since the higher layers are no longer designed purely for backhauling but they might as well generate traffic. In this context, there is a wealth of challenges to be addressed. Firstly, low-mass low-power antennas would be needed so that nanosats can effectively communicate directly with the large LEO space internet providers. In parallel, radio access for the LEO space internet satellites would have to be redesigned taking into account all possible terminals located on ground, air or space and their heterogeneous requirements in terms of link budgets and relative speeds. In parallel, the intra- and inter-layer backhauling network will have to be densified to allow uninterrupted connectivity. In this direction, THz and free space optics will play a crucial role to decongest the lower frequencies which are more suitable for radio access. As always, the development of low-mass low-power transceivers is a crucial challenge given the energy limitations of UAVs and the launch mass limitations of satellites.

\subsubsection{3D Cell-Free}
Cell-free communication is considered a key ingredient for any 6G network, where the elimination of cell boundaries can result ideally in interference-free communication in scenario with many access points or antennas. The concept is not that different from distributed MIMO or coordinated multipoint \cite{akyildiz20206g}, except for the fact that there are no cell boundaries considered when allocating users to access points. They result in much higher spectral efficiency compared to small cells, and are more robust to interference and other non-idealities. Interesting opportunities can arise when considering cell-free concepts for aerial access points, and in \cite{cellfreeUAV} it was indeed shown that cell-free communication results in superior performance compared to traditional massive MIMO  networks, also for UAV users. 
When scaling up the number of satellite base-stations, each location on the ground will be served naturally by multiple satellite beams from different satellites; as a consequence, cell-free satellite communication networks are also a key ingredient in high-throughput, dense satellite communication systems.  Spatial multiplexing for cell-free communication requires first angular resolvability, which means that the beams from the multiple satellites could be resolved by exploiting different phase gradients on the receiving array. Exploiting the high altitude of those networks, it is easy to serve a small location on the ground from multiple distinct directions that can be resolved by arrays of moderate size.  But, given the high altitude, we also have a lot of vertical degrees of freedom go create multiple layers of satellites or UAVs. This could potentially also make it possible to resolve information from multiple layers by exploiting amplitude information.

 
 


\subsubsection{Mega LEO Constellation}

Recent significant technology advances related to satellites not only sharply reduced the cost of satellite building, launching, and operating, but also enabled a faster and more flexible deployments being essential for the large deployments of LEO types. Accordingly, the vision of truly global coverage and broadband service \textit{anywhere, anytime, for anything,} is expected to be in place by 2030. The use of higher frequencies in Ku-band and Ka-band and larger available bandwidth allow LEO satellites to offer higher data rates and boost the system traffic capacity. Therefore, the integration of space into 6G networks are becoming more effective than before. Nevertheless, there are yet several challenges to be addressed. The presence of thousands LEO satellites cause significant adjacent satellite interference (ASI) where other orbital constellations generate signal interference \cite{mahonen}. The use of higher frequencies and the co-existence with the ground networks operating at the same frequency might be even more consequential due to LoS interference received from the sky towards so many ground networks.
Therefore, an extremely careful network management is required to avoid interference coming from different layers and orbital constellations under various propagation delay characteristics. 
MegaLEO \cite{MEGALEO} considers a self-organized LEO mega constellations management where satellite and network operation configurations are decided and executed in space. In addition, debris is an important issue while deploying large LEO constellation which makes the near space activities difficult. For this, commercial players such as OneWeb and SpaceX have been collaborating to mitigate the danger caused by space congestion and debris.

\subsection{Technological Enablers}
The expected technological enablers are highlighted below:
\subsubsection{X-Communication Co-Design}
\begin{itemize}
    \item Control-Communication Co-Design:
    Looking into the envisaged architectures for the 6G GAS networks, it becomes obvious that communication design is inherently interconnected with the control of the ground and space assets. This is mainly due to two reasons. Firstly, the antennas are most often firmly attached to the body of the flying asset. As a result, any change in the asset attitude directly affects the orientation of the antenna and thus the experienced communication channel. Secondly, the experienced end-to-end user connectivity depends on a number of hops, some of which are relayed by flying assets. Therefore, any configuration of the system radio resource management has to be applied simultaneously across ground and flying components. Especially for the flying components, these instructions are considered as critical signalling information that has to be orchestrated and relayed over separate control channels e.g., telemetry and telecommand (TT\&C). In this context, one of the great challenges in the 6G GAS vision is to develop efficient and resilient algorithms for the co-design of communication and control parameters. These algorithms should ideally be deployed in a distributed and autonomous fashion to avoid single points of failure. 
    \item Sensing-Communication Co-Design:
    Joint Sensing and Communication is believed to be a key driver for 6G systems as well. As antenna densities, bandwidths and frequencies increase, spatial resolution of RF sensing becomes very high. In \cite{imecradar} a 10 GHz radar with center frequency of 145 GHz is shown to enable a range resolution of 30 mm. Extending this to cell-free systems \cite{cellfreeradar}, one can show that exploiting the spatial resolution allows to achieve good sensing with lower bandwidths as well \cite{needBW}. For joint communication and sensing, bandwidth is however much desirable and waveform design for joint communication and sensing is still an open issue. Most solutions for joint sensing and radar dynamically allocate resources to the communication or sensing problems \cite{jointsensing}, but rarely send data and to sensing with the same waveform as done in \cite{Ali}. This survey paper gives an overview of the main challenges in joint communication and sensing (JCAS) for 6G, and how it is different from the more traditional RadCom, where radar hardware is reused to also achieve sensing \cite{NokiaJCAS}. 
\end{itemize}


 \subsubsection{Intelligent Reconfigurable Surfaces}
Intelligent reconfigurable surfaces (IRSs) based on metamaterials are extensively studied as one of the promising 6G enablers in terrestrial communications. Most of the existing developments have focused on the terrestrial deployment of the IRSs such as on the facades of the buildings. However, such terrestrial deployment may be hindered by appropriate site selection, service access which can be limited to only half of the space, and scattering in undesired direction in urban areas \cite{luTWC21}. To counteract on these shortcomings of the terrestrial IRS, aerial deployment of IRSs on UAVs has been recently explored \cite{srbTVT22,srbOjComs22}. Deploying aerial-IRSs are primarily driven by the fact that they enjoy more flexibility and can establish direct LoS links with the terrestrial users which, in turn, avoid a large power loss. Furthermore, UAVs can usually support the limited payloads and, therefore, may not be able to carry heavy RF transceivers required in BSs or relays. Also, active nodes (e.g., BSs/relays) would consume a lot of energy that can limit the endurance and operational life-time of the UAVs. Thus, employing passive IRS on the UAVs can save some energy cost.
Nevertheless, there are many challenges associated with the aerial deployment of IRSs \cite{alfattaniICM21}. The main challenge would be to incorporate IRSs in antennas attached to flying assets, where the main objectives are low-mass, low-power, and large range flexibility. 
Another challenge would be to implement effective controllers for the surface configuration given that the channel might be changing aggressively, while the propagation distance/delay between the flying BS and the surface would be considerable. More recently, the possibility of exploiting IRS-aided cooperation for inter-satellite THz links in LEO constellation has also been explored \cite{kursatArxiv21}. 

\subsubsection{Multi-Mode Communication}
6G devices will support a number of heterogeneous radios operating at a range of frequencies \cite{yang20196g}. This opens new multi-connectivity opportunities to connect aerial nodes to multiple ground or aerial access points, even operating at multiple frequencies. Such multi-connectivity techniques can extend the current boundaries of cells, resulting in cell-free operation as nodes are not contained to a single cell or even network. Such cell-free operation will significantly impact handovers and user scheduling, challenging for mobile node scenarios. Satellite cells, giving extremely large coverage areas, could serve as megacells controlling resources and assisting handovers. The devices should be able to seamlessly transition among different heterogeneous links (e.g., sub-6 GHz, mmWave, THz, or VLC) as function of availability of terrestrial or aerial nodes. 


\subsubsection{Dynamic Spectrum Access}
When it is not possible to harvest new frequency ranges, coexistence with terrestrial networks is mandatory.  
Interference to and from the aerial platforms increases significantly with altitude, until of course at very high altitude when the signals from space become very weak. Going to higher altitude platforms, \cite{satDSA} proposed a dynamic spectrum sharing method for LEO and GEO satellites.  A survey on data-based aided spectrum sharing for satellite networks can be found in \cite{databasesat}. Future satellite systems can largely benefit from the ability to access more spectrum bands beyond the limited dedicated licensed spectrum bands, especially when satellite systems at different orbits will coexist. A main conclusion from the paper is the fact that the non geostationary systems should adapt to the geostationary ones, by changing frequency, tilting their antenna or adapt their transmit power to avoid interference.
\black{In addition, the earlier discussions on dynamic spectrum access (DSA) were primarily based on the classical microwave band and mainly driven for the terrestrial networks. In contrast, the DSA will continue to be developed for the new emerging bands such as mmWave and THz band. As a consequence, the application of DSA in these emerging bands will be a potential technological enabler for the future NTN.}

\subsubsection{THz Communication}
THz communication has been attracted researcher attentions over the past few years. This technology is expected to support the Tbps links with moderate and viable spectral efficiency, which can revolutionize the 6G wireless networks \cite{cYiJSAC2021,commag22}. 
\black{Despite its advantages, the THz communication is still in its nascent stage and dedicated research efforts are required to make this technology a reality in practice. For instance, the channel modelling for THz propagation is yet to be fully understood. As opposed to the conventional microwave or mmWave propagation, THz waves are comparable to the size of rain,  dust,  or  snow due to extremely small wavelengths. As a result, attenuation caused by molecular absorption such as oxygen ($\textrm{O}_2$) and water ($\textrm{H}_2$O) molecules is remarkable in the THz band. Moreover, THz waves are also susceptible to  diffuse  scattering,  specular reflections, and  diffraction  etc. These characteristics can limit the communication range significantly at THz band, especially, in the dense urban scenarios and hence the LoS propagation are more desirable.
Nonetheless, the transmission loss by these impediments  can be mitigated to some extent by using the large antenna arrays \cite{commag22}.} Notably, aerial networks can greatly benefit from THz supported transmissions \cite{commag22}. Specifically, due to their flexible deployment, the UAVs can establish LoS links with the terrestrial users to counteract the high path-loss and absorption, prevalent at high frequencies. Moreover, THz links can also facilitate the intra-satellite communication operations such as communication within a constellation or between LEO and GEO satellites \cite{TekbiyikCM2020}. The feasibility of THz based transmission for space networks is also  supported by the fact that the path-loss due to molecular absorption can be non-significant in the space. 

\subsubsection{AI-empowered Communication and Networking} 
In the presence of 3D multi-segments and mega LEO constellations, 6G networks become super heterogeneous even in vertical dimension, where a ground terminal may have access to several communication segments and may operate in multi-frequencies for various purposes. While this appears as an opportunity from an end-user perspective, the network management becomes a more complex task. Therefore, AI-based algorithms can assist to provide solutions with reduced response time and operating costs. Several other challenges listed in Section \ref{sec:NTN_ML} are relevant issues to be addressed for an efficient integration using AI-based solutions. In particular, an E2E learning-based corrective actions are required to provide a harmonized integration of air/space networks into 6G ecosystem \cite{samsung6g}.

\subsubsection{Task-Oriented Communication}
Whenever a constellation of satellites cooperate toward a unique goal, they might largely benefit from task-oriented communication (TOC) designs, refer to \cite{7462495} for more on cooperative satellite networks. TOC design approaches follow the aim of making communications between machines more efficient by considering the value of communicated bits towards the goal for which those bits are required to be communicated. This design perspective is gaining more momentum as we move towards 6G, since we will have an accelerated increase in the number of M2M communications, see \cite{mostaani2019learning, mostaani2020task, mostaani2021task, tung2021effective} for further readings on task-oriented design. Be it a cooperative earth observation, remote sensing, or a deep space observation task, the M2M communications between satellites can be optimized while maintaining the performance of the constellation in attaining its objective. All the tasks in satellite-assisted telecommunication networks which require extensive coordination between satellites are the application fields where M2M inter-satellite communications can be efficiently designed according to TOC schemes. Example tasks include but are not limited to the very challenging issue of handover management in rapidly moving LEO constellations \cite{giordani2020non},  cooperative load balancing \cite{yang2020multi,jiang2015cooperative} and cooperative routing \cite{tang2018multipath}.

Take the cooperative earth observation task as an example, where multiple satellites capture prohibitively large images from different locations on earth and after fusing their data in space, they estimate a particular environmental parameter e.g., traffic parameters of a wide area road network. The potential advantages of task-oriented communication design in similar scenarios are multi-fold: (i) reducing the size of (or energy consumed for) communication bits among satellites, (ii) reducing the complexity of the parameter estimation algorithms (iii) facilitating satellite on-board data processing due to the reduced complexity of computations, and (iv) reducing the overall delay in accomplishing the computing task - as the satellite constellation acts as a distributed processing system at the edge of the network. \textcolor{black}{The importance of task-oriented communications in LEO satellite networks is specifically pronounced as LEO satellites are run under very tight energy budgets. By utilizing task-oriented communications, the power allocated to transmit and process data will be remarkably reduced. Accordingly, task-oriented communications provide promising means for LEO edge computing as well as inter-LEO and LEO-ground station communications. }


\subsubsection{Quantum Satellite Networks}
Quantum communication, or quantum key distribution (QKD), provides a super secure hacker-proof means of information sharing between two parties located far away from each other. In this context, the limitation of terrestrial optical networks such as significant signal attenuation over long distances and difficulty of intercontinental communications can be overcome by exploiting satellites \cite{picchi2020towards}. Study cases show that integrating the fibre and free-space QKD links can increase the range of communication from several hundred to a total distance of 4600 km \cite{chen2021integrated}. In the time of growing cybersecurity threats such enhanced QKD mechanism, leading to a global quantum network, revolutionizes sharing sensitive data and protecting information. Nevertheless, there are some underlying challenges for realizing the quantum communication in space. For instance, quantum signals through free space are traversed by various noise sources such as atmospheric turbulence, background noise from stray light, diffraction, etc. In addition, the development of quantum technologies is necessary that can endure and withstand the harsh space weather.

\begin{figure*}[!t]
\centering
\includegraphics[width=0.9\textwidth]{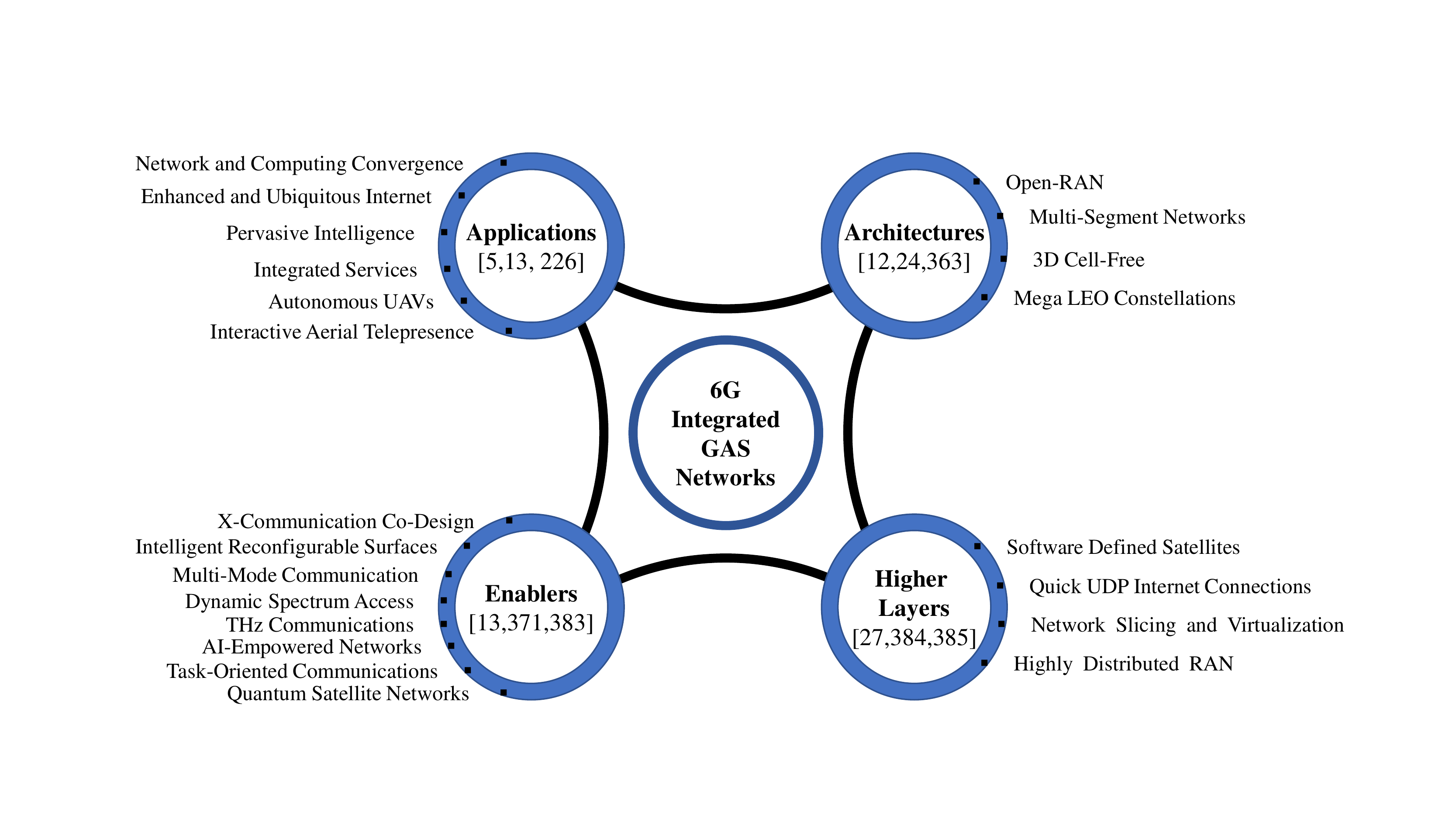}
\caption{6G integrated ground-air-space (GAS) networks.}
\label{fig:6G}
\end{figure*}


\subsection{ Higher Layer Aspects
}

The upcoming 6G networks will accommodate an extensive range of different technologies, thus posing some challenges on the management of such heterogeneous networks in higher layers \cite{oubbati2020softwarization,zhang2017software}.

\subsubsection{Software-Defined Satellites}
 A fully integrated NTN into the highly heterogeneous 6G ecosystem, entails the need for novel network architectures to facilitate the harmonization of all network elements, as well as the execution of algorithms, strategies, protocols, schemes, etc., to efficiently extend all the services provided by terrestrial networks to the aerial/space domain. 
 For example, in 6G networks is expected a greater extent of AI-based technologies with sophisticated algorithms for different purposes, such as smart structures, smart industry, network slicing, self-organizing networks, etc. This open up new development scenarios for architectures, protocols, traffic routing schemes, etc., that meet these new demands. 
  
 \subsubsection{Quick  UDP  Internet  Connections (QUIC)}
 As alternative to TCP, QUIC has emerged as a new mechanism implementing TCP-like properties over UDP transport. QUIC was conceived as an improved TCP over the following aspects \cite{QUIC1,QUIC2,Cui2017,Langley2017}: (i) combines the transport and crypto to  minimize  the  connection  latency (i.e., zero Round-trip time (RTT) connection establishment); (ii) independent streams multiplexed in a single connection, thus avoiding the so-called ``Head-of-line blocking'' occurring when one lost packet blocks the rest of the data; (iii) QUIC authenticates all of its headers and encrypts all data, including its signalling.
   
\subsubsection{Network Slicing and Virtualization}
6G networks are expected to further boost the use of virtualization schemes. For NTN, the implementation of the virtualization feature must address at least two major challenges; (1) Highly dynamic networks. These networks requires the dynamic VNs implementations schemes in order to adapt efficiently the VNs configurations according to the changing network conditions,  and; (2) Network "awareness". In the context of network slicing implementations, the development of increasingly efficient advanced Virtual Network Embedding algorithms (e.g., AI-based), requires real-time and detailed network state information. This information can range from traffic load, processing capacity, energy consumption, and even topology and capacity in case of dynamic networks (e.g., Non-GEO satellite constellations). While the adoption of SDN-based solutions is expected to fulfill these tasks, most SDN-based solutions are based at the level of the packet-oriented Layers 2 and 3 (e.g., Ethernet, IP/MPLS). This opens the opportunity of developing new architectures, protocols and APIs in order to fully integrate the NTN elements for a global network orchestration schemes in the context of network slicing implementations. 

\subsubsection{Highly Distributed RAN} 
As the latency requirements becomes more critical and the amount of traffic does not cease to increase, the cloud-RAN 
architecture suffers from congestion caused by the fact that a substantial amount of data has to go through the core. Edge clouds and edge computing is gaining popularity for low-latency services. However, 6G will require not only the decentralized computing and storage concept, but a highly decentralized network architecture where each node is equipped with sufficient intelligence and self-reconfiguration capabilities. For instance, each node is expected to intelligently route the data packets to the suitable network slice according to its requirements. The latter is only possible if each node is aware of the network status. 
  

\noindent \textit{\textbf{Key Takeaways} -- It is evident from the combined efforts across industries and academia that the integration of NTN with terrestrial architecture is perceived to play a significant role in advancing the embryonic 6G efforts. The new added capabilities offered by NTNs can bring a boost to meet the stringent requirements of 6G wireless networks. 6G integrated GAS networks provide new application opportunities such as reliable BVLoS control of drones. To this end, enabling technologies such as ML and new spectrum in THz are required to address key challenges related to networking complexity, accurate localization, co-existence, and integrated services, to name a few. However, to unlock the potential of both TNs and NTNs when working together, it is essential to improve the entire network programmability facilitating the network operation, maintenance, and scalability. Such higher network programmability will grant the flexibility of network in order to dynamically modify system parameters, configurations, and routing schemes. Figure~\ref{fig:6G} summarizes various components of 6G integrated GAS networks including prospected use cases, candidate architectures, technological enablers, and higher layer aspects elaborated in this section.}

\section{Conclusion} \label{sec:conclusion}

NTNs are perceived to play a significant role in the forthcoming generations of wireless networks, thanks to their lower cost and widespread reach. The unique characteristic of NTNs in providing global and on-demand coverage can facilitate a diverse range of new applications and services that are accessible to anything, anywhere, at anytime. With the steady progression towards the advanced wireless ecosystem, additional tasks are being delegated to NTNs justifying their broader integration into the TNs.

Considering the increasingly essential roles of NTNs, we provided an extensive review study on partially and fully integrated GAS networks from 5G to 6G by discussing the remarkable techniques ranging from new services (e.g., IoT and MEC), to new spectrum bands  (e.g., mmWave and THz), to new approaches (e.g., ML). 
It is noted that, although NTNs will be rapidly adopted to complement the existing terrestrial infrastructure, several challenges arising from the technical peculiarities such as distinguished characteristics of the NT channel, Doppler effect, handover, seamless vertical integration, etc., need to be addressed. UAVs mobility pose critical challenges due to blockage and beam misalignment, particularly at higher frequencies viz., mmWave and THz. Satellite integration with cellular networks using mmWave links seemed to be an important research domain, however, large propagation delays, Doppler effects, and co-channel interference need to be mitigated. For IoT networks, NTN integration is quite promising, especially, where terrestrial infrastructure does not exist. Moreover, MEC integration in NTN can overcome the computation limitations by offloading the complex processes while opportunistically accessing the different network segments. For instance, UAV's onboard computation limitation demands offloading of heavy tasks to ground or space. Notably, ML-based methods can provide effective solutions to address several challenges such as network deployment optimization, channel estimation, mobility management, etc. However, more attention is required for an E2E ML-based corrective actions particularly in multi-segment GAS networking. Further, for a reconfigurable and dynamically adaptable NTNs, the adoption of SDN is found to be a potential paradigm. It is also concluded that NTN-TN integration will most likely happen first at an architecture level and later at the PHY level.

Our review, further presented industry's position in 5G NTNs integration efforts. Finally, we detailed the NTNs integration in 6G ecosystem from various perspectives entailing new use cases, supporting architectures, key technological enablers, and higher layer aspects. 


\section*{Acknowledgment}

This work was supported by the Luxembourg National Research Fund (FNR)-5G-Sky Project, ref. FNR/C19/IS/13713801/5G-Sky, and the SMC funding program through the Micro5G and IRANATA projects. The work of Hazem Sallouha was funded by the Research Foundation – Flanders (FWO) Postdoc Fellowship.

\bibliographystyle{IEEEtran}
\balance
\bibliography{main}

\end{document}